\documentclass[a4paper,11pt,oneside]{mythesis_arxiv}

\providecommand{\bysame}{\leavevmode\hbox to3em{\hrulefill}\thinspace}
\providecommand{\MR}{\relax\ifhmode\unskip\space\fi MR }
\begin{document}

\title{Poisson sigma models\\on surfaces with boundary:\\
\vskip 0.2cm
classical and quantum aspects}
\author{Iv\'an Calvo Rubio}
\degree{
%{\rm PhD thesis}
}
\degreedate{
%March 2006
}

%%\maketitle

%%\cleardoublepage
\pagestyle{empty}

\begin{figure}[h]
\begin{center}
\resizebox{1.5in}{!}{\includegraphics{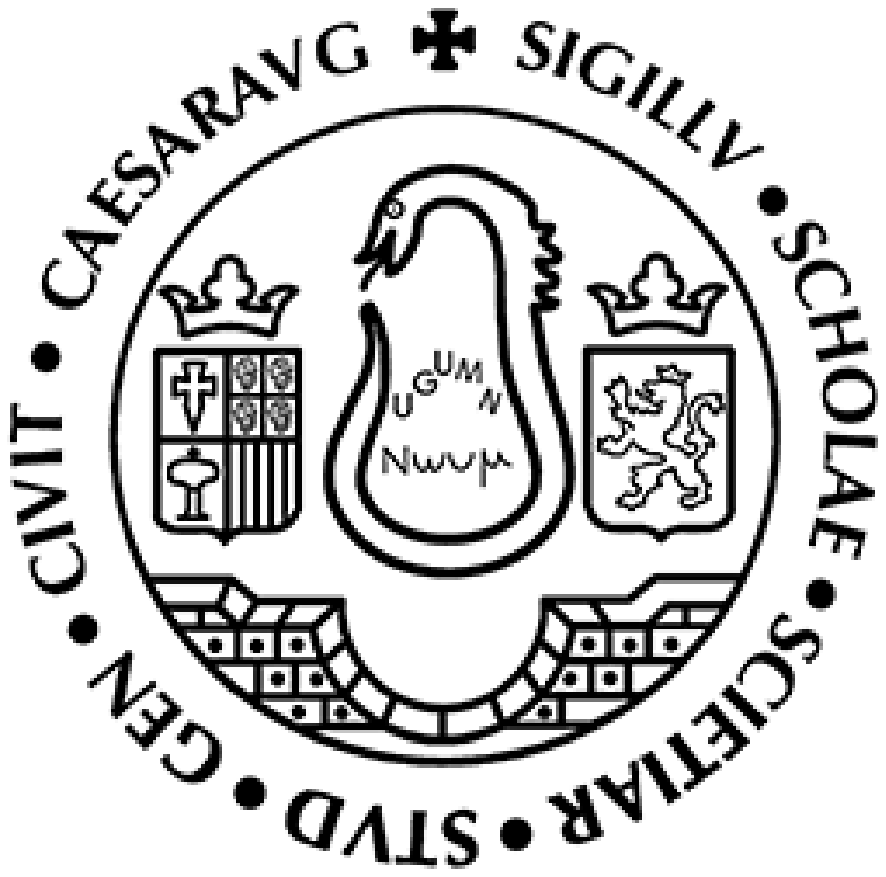}}
\end{center}
\end{figure}

\vskip 0.5cm

\begin{center}
{\bf{\Large POISSON SIGMA MODELS\\
\vskip 0.1cm
ON SURFACES WITH BOUNDARY:\\
\vskip 0.6cm
CLASSICAL AND QUANTUM ASPECTS}}
\end{center}

\vskip 5cm

{\large
\begin{center}
Memoria presentada por\\\vskip 0.2cm
{\bf Iván Calvo Rubio}\\\vskip 0.2cm
en la Facultad de Ciencias de la Universidad de Zaragoza\\
para optar al grado de Doctor. Junio de 2006.
\end{center}

\vskip 1cm

\begin{center}
Director de la Tesis: Fernando Falceto Blecua, Departamento de Física
Teórica, Universidad de Zaragoza.
\end{center}
}

\cleardoublepage
\pagenumbering{roman}
\pagestyle{plain}
\tableofcontents
\cleardoublepage

\newpage
\pagestyle{empty}

\chapter{Introduction}
\pagenumbering{arabic}
\thispagestyle{empty}
\pagestyle{empty}
\cleardoublepage

\pagestyle{headings}
\renewcommand*{\chaptermark}[1]{\markboth{\small \scshape  \thechapter.\ #1}{}}
\renewcommand*{\sectionmark}[1]{\markright{\small \scshape \thesection. \ #1}}

The understanding of Yang-Mills and instanton equations led to major
mathematical advances in topology and geometry which proved to be
connected to the quantum theory. It is the study of such relations
what has been called {\it topological quantum field theory}.

The first examples of topological field theories appear in the works
of Schwarz (\cite{Sch}) and Witten
(\cite{Wit821},\cite{Wit822}). Their different constructions include
all known topological field theories up to now and one usually talks
about topological field theories of the Schwarz or Witten type. The
importance of Witten's approach has been enormous. In \cite{Wit881} he
gave a field theory description of Donaldson's work on
four-manifolds. In \cite{Wit89} he gave a three-dimensional
interpretation of the polynomial invariant of knots previously found
by Jones (\cite{Jones}). Different examples of topological field
theories were provided also by Witten in \cite{Wit882},\cite{Wit883}
in relation to the construction of topological invariants of complex
manifolds.

All these models share common features which allow to give a
definition of a topological field theory. Roughly speaking, they do
not depend on a metric tensor. More precisely, a topological (quantum)
field theory consists of:
\begin{enumerate}
\item[(i)]A set of (Grassmann graded) fields $\varphi$ defined on a
Riemannian manifold $(\Sigma,g)$.
\vspace{-0.25cm}
\item[(ii)]An odd nilpotent operator $Q$.
\vspace{-0.25cm}
\item[(iii)]An energy momentum tensor which is $Q$-exact. That is,
there exist some functionals $F_{\mu\nu}$ of the fields and the metric
tensor such that $T_{\mu\nu}=\{Q,F_{\mu\nu}(\varphi,g)\}$.
\end{enumerate}

Property (ii) allows to define a cohomology on the Hilbert space so
that physical states are $Q$-cohomology classes, whereas (iii) ensures
that expectation values of $Q$-closed (and metric independent)
observables do not depend on the metric tensor, i.e. they are
topological invariants.

As mentioned above, all known topological field theories are
classified as being of either Witten or Schwarz type
(\cite{BirRakTho}). Witten type theories are those in which the
complete quantum action is $Q$-exact. On the other hand, theories of
the Schwarz type are those in which one starts from a classical action
$S_c(\varphi)$ which does not depend on the metric and the quantization
procedure adds terms which are $Q$-exact, so that the full quantum
action is of the form $S_q(\varphi,g)=S_c(\varphi)+\{Q,V(\varphi,g)\}$.

This dissertation is devoted to a thorough study of a two-dimensional
topological field theory of Schwarz type known as {\it Poisson sigma
model} (PSM). It was introduced in \cite{SchStr94} as a generalization
of the mathematical structure underlying a number of two-dimensional
gauge theories such as 2D Yang-Mills, gauged WZW and models of 2D
gravity. The mathematical object common to these theories is a Poisson
structure $\Pi$ on the target.

The fields of the PSM are given by a bundle map $(X,\eta)$ from the
tangent bundle of a surface $\Sigma$ to the cotangent bundle of a
Poisson manifold $(M,\Pi)$. Here $X$ stands for the base map and
$\eta$ for the fiber map. Apart from its original motivation, the
interest of the PSM resides in the fact that it naturally encodes the
Poisson geometry of the target and allows to unravel it by means of
techniques from Classical and Quantum Field Theory. This became
manifest for the first time in 1999 with the work of Cattaneo and
Felder \cite{CatFel99}, who used the PSM defined on the disk to give a
field theoretical interpretation of Kontsevich's formula for
deformation quantization in terms of Feynman diagrams (see also
\cite{CatFel01}).

The same authors showed in \cite{CatFel00} the connection between the
structure of the phase space of the PSM and the symplectic groupoid
(when the latter exists) integrating the target Poisson
manifold. Their ideas inspired to a large extent the crucial works
\cite{CraFer2},\cite{CraFer} of Crainic and Fernandes on the
integrability of Lie algebroids.

In these works the boundary conditions considered for the PSM are
such that $X$ is free at $\partial\Sigma$ and $\eta$ vanishes on
vectors tangent to $\partial\Sigma$.

The study of more general boundary conditions for the PSM was started
in \cite{CatFel03}. The question is what submanifolds (branes)
$N\subset M$ are admissible in the sense that the field $X$ can be
consistently restricted to $N$ at the boundary. In \cite{CatFel03} it
was shown that perturbative branes, i.e. branes admitting a
perturbative quantization around $\Pi=0$, coincide with coisotropic
submanifolds of $M$.

In this dissertation we perform a general study of the PSM defined on
a surface $\Sigma$ with boundary at both the classical and quantum
level and show that much more general branes are allowed.

In Chapter \ref{ch:PoissonGeometry} we give a self-contained
introduction to Poisson geometry with emphasis in the reduction of
Poisson manifolds. We introduce Dirac structures, beautiful objects
living on $TM\oplus T^*M$ and generalizing simultaneously the notions
of Poisson and presymplectic manifolds. Then, we present some original
results dealing with the reduction of Dirac structures. The chapter
ends with the statement of the problem of deformation quantization of
Poisson manifolds and Kontsevich's solution.

In Chapter \ref{ch:branesPSM} we present the PSM and perform the
general study on a surface with boundary. We identify the classically
admissible branes and show that the (pre-)symplectic structure on the
phase space is related to the Poisson bracket induced on the brane. In
this chapter we also recall the results of Cattaneo and Felder
regarding the symplectic groupoid integrating the Poisson manifold $M$
by means of the PSM on the strip with free boundary conditions. Then,
we see how everything is modified for more general branes.

On the light of the results of Chapter \ref{ch:branesPSM}, it is
tempting to conjecture that the perturbative quantization of the PSM
with a general brane is related to the deformation quantization of the
brane with the induced Poisson bracket. In Chapter \ref{ch:quantumPSM}
we show that this actually holds, although the suitable approach is
quite different from that used in \cite{CatFel99},\cite{CatFel03}. We
shall see that non-coisotropic branes are non-perturbative in a sense
and require a redefinition of the perturbative series.

Chapters \ref{ch:Liebial} and \ref{PLSM} are devoted to an interesting
particular case of the PSM, in which the target is a Lie group and the
Poisson structure is compatible with the product on the group, i.e. a
{\it Poisson-Lie group}. In Chapter \ref{ch:Liebial} we briefly
introduce the results on Lie bialgebras and Poisson-Lie groups that
will be needed in Chapter \ref{PLSM}. Poisson-Lie groups come in dual
pairs $G,G^*$. In Chapter \ref{PLSM}, we solve the models with target
$G$ and $G^*$ for any simple, connected and simply connected
Poisson-Lie group $G$. Poisson-Lie structures on simple Lie groups are
either factorizable or triangular. In the first case the PSM over
$G^*$ is known to be locally equivalent to $G/G$ theory. We show that
in the second case it is equivalent to BF-theory. We end Chapter
\ref{PLSM} by finding a family of branes which preserve the duality
between the models.

Chapter \ref{ch:SusyWZP} deals with the relation between extended
supersymmetry in two-dimensional first order sigma models and
generalized complex geometry from a phase space point of view. The
particular case in which the metric in the target vanishes coincides
with the PSM or more generally, if a WZ term is included, with the
so-called twisted or WZ-Poisson sigma model. We generalize the
definitions of Chapter \ref{ch:PoissonGeometry} concerning the
geometry of $TM\oplus T^*M$ to the twisted case. Then, we work out the
conditions for extended supersymmetry in the WZ-Poisson sigma model
and give a simple geometrical interpretation.

\newpage

\pagestyle{empty}
\chapter{Poisson geometry} \label{ch:PoissonGeometry}
\thispagestyle{empty}
\pagestyle{empty}
\cleardoublepage

\pagestyle{headings} \renewcommand*{\chaptermark}[1]{\markboth{\small
\scshape \thechapter.\ #1}{}}
\renewcommand*{\sectionmark}[1]{\markright{\small \scshape
\thesection. \ #1}}

In any course on analytical mechanics the Poisson bracket of two
functions on phase space, in coordinates $(q^i,p_i),i=1,\dots,n$, is
introduced:
\begin{equation}\notag
\{f,g\}=\sum_{i=1}^n\left(\frac{\partial f}{\partial q^i}\frac{\partial
g}{\partial p_i}-\frac{\partial f}{\partial p_i}\frac{\partial
g}{\partial q^i}\right).
\end{equation}

In the study of some systems, in particular systems with constraints,
more general brackets and then a generalization of the notion of
Poisson bracket, are needed. The general theory of Poisson manifolds,
i.e. manifolds whose set of smooth functions is equipped with a
Poisson bracket, has been developed since the 1970s. The importance of
Poisson geometry from the point of view of Physics resides not only on
a geometrical formulation of classical mechanics, but also on the
issue of quantization. It is a well-known fact that the starting point
in the (canonical) quantization of a classical theory is precisely the
Poisson bracket.

For the purposes of this dissertation some knowledge on Poisson
geometry is obviously needed. This chapter provides the necessary
geometrical background to carry out the general study of the PSM on
surfaces with boundary in Chapters \ref{ch:branesPSM} and
\ref{ch:quantumPSM}. We review some basic facts on Poisson geometry
and we develop the theory of reduction of Poisson manifolds. Then we
introduce more advanced topics related to Dirac structures, which
generalize and contain as particular cases symplectic and Poisson
structures. Finally, we state the problem of deformation quantization
of Poisson manifolds and present the solution given by Kontsevich.

\section{Poisson algebras and Poisson manifolds}

Let $\cal A$ be an associative commutative algebra with unit over the
real or complex numbers equipped with a bilinear bracket
$\{\cdot,\cdot\}:\cal A\rightarrow\cal A$. We say that $\cal A$ is a
{\it Poisson algebra} if
\begin{enumerate}
\item[(i)]$\{f,g\}=-\{g,f\}$ \hfill(antisymmetry)
\vspace{-0.25cm}
\item[(ii)]$\{f,\{g,h\}\}+\{g,\{h,f\}\}+\{h,\{f,g\}\}=0$ \hfill(Jacobi)
\vspace{-0.25cm}
\item[(iii)]$\{f,gh\}=g\{f,h\}+\{f,g\}h$ \hfill(Leibniz)
\vskip 0.1cm
\end{enumerate}
\vspace{-0.2cm}$\forall f,g,h\in\cal A$.

\vskip 0.1cm

We shall be mainly interested in the case in which ${\cal A}$ is the
algebra of smooth functions on an $m$-dimensional manifold $M$, which
is then called a {\it Poisson manifold}. A Poisson algebra structure
on $C^\infty(M)$ determines in a unique way a bivector field
$\Pi\in\Gamma(\wedge^2TM)$ such that the Poisson bracket of two
functions is given by:
\begin{equation} \label{definitionPB}
\{f,g\}(p)=\iota(\Pi_p) (\dd f\wedge \dd g)_{p},\ p \in M.
\end{equation}

Taking local coordinates $(x^1,\dots,x^m)$ on $M$ the components of
$\Pi$ are $\Pi^{ij}(x)=\{x^i,x^j\}$. The Jacobi identity for the
Poisson bracket reads in terms of $\Pi^{ij}$:
\begin{equation} \label{JacobiIdentity}
\Pi^{ji} \partial_{i}\Pi^{kl} + \Pi^{ki} \partial_{i}\Pi^{lj} +
\Pi^{li} \partial_{i}\Pi^{jk} =0
\end{equation}
where summation over repeated indices is understood and
$\partial_i:=\frac{\partial}{\partial x^i}$. A bivector field
satisfying (\ref{JacobiIdentity}) is called a {\it Poisson tensor} or
a {\it Poisson structure} on $M$. It is obvious that a Poisson tensor
defines a Poisson bracket on $C^\infty(M)$ through formula
(\ref{definitionPB}).

Some definitions are in order at this point:

\begin{definition}
Let $V(M)=\{V_p(M),p\in M\}$ be a set of linear subspaces of the
tangent space $T_pM$ at each point of $M$. $V(M)$ is called a {\it
general differentiable distribution} if for every $p\in M$, there
exist vector fields $X_1,\dots,X_k\in V(M)$ such that
$V_p(M)=\mbox{span}\{X_1(p),\dots,X_k(p)\}$. If $\mbox{dim}(V_p(M))$
is constant, $V(M)$ is a differentiable distribution in the usual
sense.
\end{definition}

\begin{definition}
A function $f:M\rightarrow {\mathbb R}$ is {\it upper semicontinuous}
at $p_0$ if for every $\varepsilon > 0$ there exists a neighborhood
$\cal U$ of $p_0$ such that
$$f(p)<f(p_0)+\varepsilon,\ \forall p\in {\cal U}.$$
\end{definition}

\begin{definition}
A function $f:M\rightarrow {\mathbb R}$ is {\it lower semicontinuous}
at $p_0$ if for every $\varepsilon > 0$ there exists a neighborhood
$\cal U$ of $p_0$ such that
$$f(p)>f(p_0)-\varepsilon,\ \forall p\in {\cal U}.$$ A function $f$ is
continuous if and only if it is upper and lower semicontinuos.
\end{definition}

\begin{remark}
The dimension of a general differentiable distribution is a lower
semicontinuous function on $M$.
\end{remark}

\vskip 4mm

Now define $\Pi^\sharp:T^*M\rightarrow TM$ by
\begin{equation} \label{sharpmap}
\beta(\Pi^\sharp(\alpha))=\iota(\Pi) (\alpha \wedge \beta),\
\alpha,\beta \in T^{*}M.
\end{equation}
For any $f\in C^\infty(M)$ we call $X_f:=\Pi^\sharp\dd f$ the {\it
Hamiltonian vector field} of $f$. Clearly,
\begin{equation}\notag
\{f,g\}=X_f(g),\ \forall f,g\in C^\infty(M).
\end{equation}

By virtue of Jacobi identity the image of $\Pi^{\sharp}$,
$${\rm Im}(\Pi^{\sharp}):=\bigcup_{p\in M}{\rm Im}(\Pi^{\sharp}_{p})$$
is a completely integrable general differential distribution and $M$
admits a generalized foliation. $M$ is foliated into leaves which may
have varying dimensions. The Poisson structure can be consistently
restricted to a leaf and this restriction defines a non-degenerate
Poisson structure on it. That is why we shall also refer to the leaves
as {\it symplectic leaves} and to the foliation as the {\it symplectic
foliation} of $M$. This result comes from a generalization of the
classical Frobenius theorem for regular distributions.

Next we give some examples of Poisson manifolds and Poisson algebras.

\vskip 0.3cm

\begin{example}
$\Pi=0$ is the trivial Poisson structure.
\end{example}

\vskip 0.3cm

\begin{example} \label{exLinPoi}
Take $M={\gl g}^{*}$, where ${\gl g}$ is a Lie
algebra. Linear functions on ${\gl g}^*$ can be viewed as elements of
${\gl g}$. The Poisson bracket of two such functions is given by the
Lie bracket on $\gl g$:
\begin{equation} \label{KostantKirillov}
\{f,g\}=[f,g],\  f,g \in {\gl g}
\end{equation}
and is extended to $C^\infty({\gl g}^{*})$ by the Leibniz rule. This
defines the so-called {\it Kostant-Kirillov Poisson structure}.

The symplectic leaves in this case correspond to the orbits under the
coadjoint representation of any connected Lie group $G$ with Lie
algebra $\gl g$ and have, in general, varying dimensions; in
particular, the origin is always a symplectic leaf since the Poisson
structure vanishes.
\end{example}

\vskip 0.3cm

\begin{example}
Take $M={\mathbb R}^3$. Then,
\begin{equation} \label{exnonlinPoisson}
\Pi^{12}(x)=-(x^3)^2+\frac{1}{4},\ \Pi^{23}(x)=x^1,\ \Pi^{31}(x)=x^2
\end{equation}
defines a Poisson structure invariant under rotations around the $x^3$
axis. This example will appear in Chapter \ref{ch:branesPSM} in the
context of two-dimensional $R^2$-gravity.
\end{example}

\vskip 0.3cm

\begin{example} \label{expresymp}
A presymplectic structure $\omega$ on $M$ is a closed 2-form on
$M$. We can define a Poisson algebra ${\cal A}_\omega$ consisting of
functions that possess a Hamiltonian vector field, i.e.  those
functions $f\in{C}^\infty(M)$ for which the equation
\begin{equation}\notag
\iota(X_f)\omega=-\dd f
\end{equation}
has a solution ${X_f}\in{\gl X}(M)$. Given $f,g\in{\cal A}_\omega$
with Hamiltonian vector fields ${X}_f,{X}_g$ respectively, $fg$ has
Hamiltonian vector field $f{X}_g+g{X}_f$ and then ${\cal A}_\omega$ is
a subalgebra of ${C}^\infty(M)$. The Poisson bracket is defined by
$$\{f,g\}=\omega({X}_f,{X}_g).$$

Note that in general the Hamiltonian vector field ${X}_f$ for $f\in
{\cal A}_\omega$ is not uniquely defined but the ambiguities are in
the kernel of $\omega$ and then it leads to a well defined Poisson
bracket.

Due to the closedness of $\omega$, $\{f,g\}\in{\cal A}_\omega,\
\forall f,g\in{\cal A}_\omega$ and the Jacobi identity is satisfied,
so that ${\cal A}_\omega$ is actually a Poisson algebra. It is worth
mentioning that in this case the center of ${\cal A}_\omega$ (Casimir
functions) is the set of constant functions on $M$.

If $\omega^\flat:TM\rightarrow T^*M$ is invertible at every point of
$M$, $\omega$ is called symplectic. A presymplectic structure $\omega$
defines a Poisson structure on $M$ (equivalently, ${\cal A}_\omega =
C^\infty(M)$) if and only if it is symplectic.
\end{example}

\vskip 0.4cm

We end this section by giving some definitions.

\begin{definition}
Given two Poisson algebras $({\cal A}_1,\{\cdot,\cdot\}_1)$, and
$({\cal A}_2,\{\cdot,\cdot\}_2)$ and a homomorphism of abelian
associative algebras, $\Psi:{\cal A}_1\rightarrow{\cal A}_2$, we say
that $\Psi$ is a {\it Poisson homomorphism} if
$$\Psi(\{f,g\}_1)=\{{\Psi}(f),{\Psi}(g)\}_2,\ \forall f,g\in {\cal A}_1.$$

Analogously, we say that ${\Psi}$ is an {\it anti-Poisson homomorphism}
if
$${\Psi}(\{f,g\}_1)=-\{{\Psi}(f),{\Psi}(g)\}_2,\ \forall f,g\in {\cal A}_1.$$
\end{definition}

\begin{definition}
Let $(M_1,\Pi_1)$ and $(M_2,\Pi_2)$ be two Poisson manifolds. A smooth
map $F:M_1 \rightarrow M_2$ is called a {\it Poisson map} if
$$\{f,g\}_2 \circ F = \{f\circ F,g\circ F\}_1,\ \forall f,g \in
C^{\infty}(M_2).$$
and an {\it anti-Poisson map} if
$$\{f,g\}_2 \circ F = -\{f\circ F,g\circ F\}_1,\ \forall f,g \in
C^{\infty}(M_2).$$
That is, $F^*:C^{\infty}(M_2)\rightarrow C^{\infty}(M_1)$ is a
(anti-)Poisson homomorphism.
\end{definition}

\begin{definition}
Let $N\subset M$ be a submanifold of $M$. $N$ is called a {\it Poisson
submanifold} if the inclusion $i:N\hookrightarrow M$ is a Poisson map.
\end{definition}

\section{Reduction of Poisson manifolds} \label{reduction}

Let $N$ be a closed submanifold of $(M,\Pi)$. Can we define in a
natural way a Poisson structure on $N$? The answer is negative, in
general. What we can always achieve is to endow a certain subset of
$C^{\infty}(N)$ with a Poisson algebra structure. The canonical
procedure below follows in spirit reference \cite{Kim}, although we
present some additional, new results.

We adopt the notation ${\cal A} = C^{\infty}(M)$ and take the ideal 
(with respect to the pointwise product of functions in ${\cal A}$. 
We will use the term Poisson ideal when we refer to an ideal 
with respect to the Poisson bracket),
$${\cal I} =\{f \in {\cal A} \ \vert \ f(p) = 0,\ \forall p \in N\}.$$

Define ${\cal F}\subset{\cal A}$ as the
set of {\it first-class functions}, also called the {\it normalizer} of ${\cal I}$,
$${\cal F}=\{f\in{\cal A} \ \vert \ \{f,{\cal I}\}\subset{\cal I}\}.$$
Note that due to the Jacobi identity and the Leibniz rule ${\cal F}$
is a Poisson subalgebra of ${\cal A}$ and ${\cal F}\cap{\cal I}$ is a
Poisson ideal of ${\cal F}$. Then, we have canonically defined a
Poisson bracket in the quotient ${\cal F}/({\cal F}\cap{\cal I})$.
However, this is not what we want, as our problem was to find a
Poisson bracket on $C^{\infty}(N)\cong {\cal A}/{\cal I}$ (or, at
least, in a subset of it). To that end we define an injective map
\begin{eqnarray}\label{mapphi}
 \begin{matrix}\phi:&{\cal F}/({\cal F}\cap{\cal I})&\longrightarrow&{\cal A}/{\cal I}\cr
&f+{\cal F}\cap{\cal I}&\longmapsto&f+{\cal I}
\end{matrix}
\end{eqnarray}
which is a homomorpism of abelian associative algebras with unit and
then induces a Poisson algebra structure $\{\cdot,\cdot\}_N$ in the
image, that will be denoted by ${\cal C}(\Pi,M,N)$, i.e.:
\begin{eqnarray}\label{ourDirac}
\{ f_1+{\cal I}, f_2+{\cal I}\}_{_N}=
\{ f_1, f_2\}+{\cal I}.\qquad f_1,f_2\in {\cal F}.
\end{eqnarray}

\vspace{0.3 cm}

\begin{remark}
{\it Poisson reduction} is a generalization of the symplectic
reduction in the following sense: If the original Poisson structure is
non-degenerate, it induces a symplectic structure $\omega$ in $M$.
Then, we may canonically define on $N$ the closed two-form
$i^*\omega$, where $i:N\hookrightarrow M$ is the inclusion map.  As
described before, this presymplectic two-form on $N$ defines a Poisson
algebra for a certain subset of ${C}^\infty(N)\cong {\cal A}/{\cal
I}$. The Poisson algebra obtained this way coincides with the one
defined above.
\end{remark}

\begin{remark}
The elements of ${\cal F}\cap{\cal I}$ are, in the language of
physicists, the generators of {\it gauge transformations} or, in
Dirac's terminology, the {\it first-class constraints}. They are
constraints whose Hamiltonian vector fields are tangent to the
submanifold $N$.
\end{remark}

\vspace{0.3 cm}

The problem is that in general $\phi$ is not onto and $N$
cannot be made a Poisson manifold. The goal now is to use the
geometric data of the original Poisson structure to interpret the
algebraic obstructions.

Let $TN^0$ (or ${\rm Ann}(TN)$) be the conormal bundle of $N$ (or
annihilator of $TN$), i.e. the subbundle of the pull-back
$i^{*}(T^{*}M)$ consisting of covectors that kill all vectors in $TN$.
One has the following

\vspace{0.3 cm}

\begin{theorem}
Assume that:

a) ${\rm dim}(\Pi^\sharp_p(T_pN^0)+ T_pN)= {\rm const.},\ \forall p \in N$, and
\vskip0.5mm
b) $\Pi^\sharp_p(T_pN^0)\cap T_pN=\{0\},\ \forall p \in N.$
\vskip 1mm
Then, the map $\phi$ of (\ref{mapphi}) is an isomorphism of
associative commutative algebras with unit.
\end{theorem}

\vspace{0.3 cm}

{\it Proof:} Condition b) implies that
$$T^*_pM = {\rm Ann}(\Pi^\sharp_p(T_pN^0)\cap
T_pN)=T_pN^0+\Pi^{\sharp-1}_p(T_pN)$$
and, then, 
$\Pi_p^{\sharp}(T^*_pM)\subseteq \Pi^\sharp_p(T_pN^0)+ T_pN, \quad
\forall p \in N$.

Now define a smooth bundle map:
$$\Upsilon:TN\oplus TN^0\longrightarrow i^*TM$$
that maps $(v_p,\alpha_p)\in T_pN\oplus T_pN^0$ to 
$v_p+\Pi^\sharp_p \alpha_p$. Due to condition 
a) the map is of constant rank and then every 
smooth section of its image has a smooth preimage.

Take $f\in {\cal A}$.  As shown before $\Pi^\sharp_p (\dd f)_p\in
\Pi^\sharp_p(T_pN^0)+ T_pN$ for any $p\in N$. Then, the restriction to
$N$ of $\Pi^\sharp \dd f$ is a smooth section of the image of
$\Upsilon$. Let $(v,\alpha)$ be a smooth section of $TN\oplus TN^0$ with
$\Upsilon(v,\alpha)_p=\Pi_p^\sharp (\dd f)_p$ for $p \in N$. Now for any
section $\alpha$ of $TN^0$ there exists a function $g\in{\cal I}$ such
that $\alpha_p = (\dd g)_p$ for any $p\in N$.

Hence, one has that $\tilde f=f-g\in {\cal F}$ and 
$\phi(\tilde f+ {\cal F}\cap{\cal I})= f+ {\cal I}$.
$\,$\hfill$\Box$\break

Condition a) of Theorem \ref{thm:phionto} is not necessary as shown by
the following

\begin{example}
Take $M={\gl sl}(2)^*$. In coordinates $(x_1,x_2,x_3)$ the linear
Poisson bracket is given by $\{x_i,x_j\}=\epsilon^{ijk}x_k$. Now
define $N$ by the constraints: $x_1=0, x_2=0$. Clearly,
$${\rm dim}(\Pi^\sharp_p(T_pN^0)+ T_pN)=
\begin{cases}
3\quad {\rm for}\ p \not= 0\cr
1\quad {\rm for}\ p = 0
\end{cases} $$
and for any $f\in {C}^\infty(M)$ we may define $\tilde
f=f-x_1\partial_1 f -x_2\partial_2 f\in{\cal F}$ such that
$\phi(\tilde f+{\cal F}\cap{\cal I})=f+{\cal I}$, i.e. $\phi$ is
onto. The Poisson structure induced in this case is, of course, zero.
\end{example}
\vspace{0.5 cm}

Condition b), however, is necessary:

\begin{theorem} \label{thm:phionto}
{\it \noindent 
If the map $\phi$  of (\ref{mapphi})
is onto then  $\Pi^\sharp_p(T_pN^0)\cap T_pN=\{0\}$.}
\end{theorem}

\vspace{0.3 cm}

{\it Proof:} Assume that $\exists v_p\not=0, v_p \in
\Pi^\sharp_p(T_pN^0)\cap T_pN$.  It is enough to take a function
$f\in{\cal A}$ such that its derivative at $p$ in the direction of
$v_p$ does not vanish. Then $f+{\cal I}$ is not in the image of
$\phi$.  $\,$\hfill$\Box$\break

\vspace{0.5 cm}

This result tells us that when $\Pi^\sharp_p(T_pN^0)\cap
T_pN\not=\{0\}$ one cannot endow $N$ with a Poisson structure. The only
functions on $N$ that have got a well-defined Poisson bracket
(i.e. the physical observables) are those in the image of $\phi$. 
On the other hand, it is easy to see that all functions in the image of $\phi$
belong to the subalgebra of gauge invariant functions
$${\cal A}_{inv}=\{ f\in{\cal A} | \{f,{\cal F}\cap{\cal I}\}\subset{\cal I}\}.$$
One may wonder when the physical observables are precisely
the gauge invariant functions.
A sufficent condition is given by the following

\vspace{0.3 cm}

\begin{theorem} \label{imagegaugeinv}
If ${\rm dim}(\Pi^\sharp_p(T_pN^0)+ T_pN)= {\rm const.},\ \forall p\in N$, then \newline
$\phi ({{\cal F}}/{{\cal F}}\cap{{\cal I}})={\cal A}_{inv}/{\cal I}$.
\end{theorem}

\vspace{0.3 cm}

Before proving the theorem we shall establish a Lemma that will be
useful in the following.

\vspace{0.3 cm}

\begin{lemma} \label{lemmaPoiRed}
{\it \parindent 0pt
The following two statements are equivalent:

a) ${\rm dim}(\Pi^\sharp_p(T_pN^0)+ T_pN)={\rm const.},\ \forall p \in N$
\vskip 0.5mm
b) $\Pi_p^{\sharp-1}(T_pN)\cap T_pN^0=\{({\rm d} g)_p | \ g\in{\cal F}\cap{\cal I}\},\
\forall p \in N$.
}
\end{lemma}

\vspace{0.3 cm}

{\it Proof:}

a) $\Rightarrow$ b): Assume that ${\rm dim}(\Pi^\sharp_p(T_pN^0)+
T_pN)$ is constant on $N$. Then, ${\rm Ann}_p(\Pi^\sharp(TN^0)+
TN)=\Pi_p^{\sharp-1}(T_pN)\cap T_pN^0$ is also of constant dimension
and $\Pi^{\sharp-1}(TN)\cap TN^0$ is a subbundle of $TN^0$ whose fiber
at every point of the base is spanned by a set of sections. For every
section $\alpha$ of this subbundle there exists $g\in{\cal I}$ such
that $\alpha_p=(\dd g)_p$.  But since $(\dd g)_p\in \Pi_p^{\sharp-1}(T_pN)$,
it follows that $g\in{\cal F}\cap{\cal I}$.

The other inclusion is trivial as differential of first-class
constraints are in $TN^0$ (because they are constraints) and their
Hamiltonian vector fields transform constraints into constraints
(because they are first-class) so their restrictions to $N$ are in $TN$.

b) $\Rightarrow$ a): Assuming b) one has that ${\rm
dim}(\Pi_p^{\sharp-1}(T_pN)\cap T_pN^0)$ is a lower semicontinuous
function on $N$ because the fiber of $\Pi_p^{\sharp-1}(T_pN)\cap
T_pN^0$ at every point is spanned by local sections (see
\cite{Vai94}). For the same reason, ${\rm dim}(\Pi^\sharp_p(T_pN^0)+
T_pN)$ is also lower semicontinuous. But from the relation
$$\Pi^\sharp_p(T_pN^0)+ T_pN= {\rm Ann}_p(\Pi^{\sharp-1}(TN)\cap TN^0)$$
we infer that ${\rm dim}(\Pi^\sharp_p(T_pN^0)+ T_pN)$ is upper
semicontinuous, so it is continuous and, being integer valued it is
indeed constant.  $\,$\hfill$\Box$\break

\vspace{0.5 cm}

{\it Proof of Theorem \ref{imagegaugeinv}:} First note that $f\in{\cal
A}_{inv}$ implies that $\Pi_p^\sharp(\dd f)_p\in {\rm Ann}_p(\{\dd g|
g\in{\cal F}\cap{\cal I}\})$.  But from the previous Lemma we have
that the latter is equal to $\Pi^\sharp_p(T_pN^0)+ T_pN$.

Then, $\forall f\in{\cal A}_{inv}$ one has $\Pi_p^\sharp(\dd f)_p\in
\Pi^\sharp_p(T_pN^0)+ T_pN$.  And from here on the proof is like that
of Theorem \ref{thm:phionto}.  $\,$\hfill$\Box$\break

\vspace{0.5 cm}

At first sight we might expect a result analogous to Theorem
\ref{thm:phionto} for the case with gauge transformations in the
constrained submanifold, namely that a necessary condition for $\phi$
mapping onto the space of gauge invariant functions on $N$ is that the
space of Hamiltonian vector fields of first-class constraints at every
point coincides with $T_pN\cap \Pi_p^\sharp(T_pN^0)$. This is not true,
however, as shown by the following example in which the spaces above
differ in some points whereas the image of map $\phi$ of
(\ref{mapphi}) is ${\cal A}_{inv}/{\cal I}$.

\begin{example}
Take $M={\mathbb R}^6=\{(q_1,q_2,q_3,p_1,p_2,p_3)\}$ with the standard
Poisson bracket $\{p_i,q_j\}=\delta_{ij}$. Now consider the
constraints 
$$g_i:=p_i-q_iq_{\sigma(i)}\quad i=1,2,3$$ 
with $\sigma$ the cyclic permutation 
of $\{1,2,3\}$ s. t. $\sigma(1)=2$.
In this case
$${\rm dim}(\Pi^\sharp_x(T_xN^0)\cap T_xN)=
\begin{cases}1\quad {\rm for}\ x \not= 0\cr
3\quad {\rm for}\ x = 0
\end{cases}
$$
while the gauge transformations 
are restrictions to $N$ of Hamiltonian vector fields  
of $fg$ with $f\in{C}^\infty(M)$ and 
$g=q_2g_1+q_3g_2+q_1g_3$.
It implies that at $x=0$ the gauge transformations 
vanish and, hence, they do not fill
$\Pi^\sharp_x(T_xN^0)\cap T_xN$.

We will show that the image of the map $\phi$ of (\ref{mapphi}) is
${\cal A}_{inv}/{\cal I}$.  In every class of ${\cal A}_{inv}/{\cal
I}$ we may take the only representative independent of the $p_i$'s.
Gauge invariant functions $f(q_1,q_2,q_3)$ are then characterized by:
$$(q_2\partial_{q_1}+q_3\partial_{q_2}+q_1\partial_{q_3})f=0,$$
and for any of them we may define
$$\tilde f=f+\sum_i a_i g_i$$
with $a_i$ smooth, given by
\begin{eqnarray}\notag
a_1(q_1,q_2,q_3)&=&\frac{1}{q_1}
[\partial_{q_2}f(q_1,q_2,q_3)-\partial_{q_2}f(0,q_2,q_3)],\cr
a_2(q_1,q_2,q_3)&=&\frac{1}{q_1}
[\partial_{q_1}f(q_1,q_2,q_3)-\partial_{q_1}f(0,q_2,q_3)],\cr
a_3(q_1,q_2,q_3)&=&
\frac{1}{q_2}
\partial_{q_2}f(0,q_2,q_3)= -\frac{1}{q_3}
\partial_{q_1}f(0,q_2,q_3).
\end{eqnarray}
Now $\tilde f$ is first class and $\phi(\tilde f+{\cal F}\cap{\cal
I})=f+{\cal I}.$ This shows that in this case the image of $\phi$
fills ${\cal A}_{inv}/{\cal I}$.
\end{example}

\vspace{0.5 cm}

For later purposes it will be interesting to introduce coordinates
adapted to the submanifold and to the Poisson structure. This is not
possible for an arbitrary submanifold $N$, but be must require some
regularity conditions on it. The sufficient condition of Theorem
\ref{imagegaugeinv},
\begin{eqnarray} \label{regularity}
{\rm dim}(\Pi^\sharp_p(T_pN^0)+ T_pN)=k+\dim(N),\quad \forall p\in N.
\end{eqnarray}
will be called {\it the strong regularity condition} (here $k$ is a
non-negative constant). If it is satisfied, we can choose in a
neighborhood ${\cal U}\subset M$ of every $p\in N$ adapted local
coordinates on $M$, $(x^a,x^\mu,x^{A})$, with $a=1,\dots,\dim(N)$,
$\mu=\dim(N)+1,\dots,\dim(M)-k$ and $A=\dim(M)-k+1,\dots,\dim(M)$,
verifying:
\begin{enumerate}
\item[(i)]$N\cap{\cal U}$ is defined by $x^\mu=x^{A}=0$.
\vspace{-0.1cm}
\item[(ii)]$\{x^{\mu},x^{\nu}\}\vert_{N\cap{\cal
U}}=\{x^{\mu},x^A\}\vert_{N\cap{\cal U}}=0$,
i.e. $x^{\mu}$ are {\it first-class constraints}.
\vspace{-0.1cm}
\item[(iii)]${\rm det}(\{x^A,x^B\}(p))
\neq 0,\quad \forall p\in{N\cap{\cal U}}$, 
i.e. $x^A$ are {\it second-class constraints}.
\end{enumerate}

It is clear that in these adapted coordinates the Poisson structure
satisfies:
\begin{eqnarray}\notag
\Pi^{\mu\nu}\vert_{N\cap{\cal U}} = 0,\quad
\Pi^{A\mu}\vert_{N\cap{\cal U}} = 0,\quad {\rm
det}(\Pi^{AB})\vert_{N\cap{\cal U}} \neq 0.
\end{eqnarray}

\vskip 0.2 cm

Recall that by Lemma \ref{lemmaPoiRed} the strong regularity condition
is equivalent to
\begin{eqnarray}\label{equivalence}
\Pi_p^{\sharp-1}(T_pN)\cap T_pN^0=\{(\dd f)_p | f\in{\cal F}\cap{\cal
I}\},\quad \forall p\in N.
\end{eqnarray}

\begin{definition}
$N\subset M$ is called {\it coisotropic} if
$\Pi_p^\sharp(T_pN^0)\subseteq T_pN,\ \forall p\in N$.
\end{definition}

For a coisotropic submanifold $N$ the strong regularity condition
(\ref{regularity}) is obviously satisfied and every constraint is
first-class. In addition, (\ref{ourDirac}) is the original bracket on
$M$ restricted to the gauge-invariant functions on $N$. The other
extreme case is given by the absence of first-class constraints:
\begin{definition}
We say that $N$ is a {\it second-class submanifold} if
$$\Pi_p^{\sharp-1}(T_pN)\cap T_pN^0=\{0\},\ \forall p\in N.$$
\end{definition}

Equivalently, $N$ is a second-class submanifold if every constraint
defining it is second-class. In the adapted coordinates defined above
there are no Greek indices and the strong regularity condition is
fulfilled. In this case the matrix of the Poisson brackets of the
constraints $\Pi^{AB}=\{x^A,x^B\}$ is invertible on $N$. Defining on
$N$ the matrix $\omega_{AB}$ by $\omega_{AB}\Pi^{BC}=\delta_A^C$ the
Poisson bracket (\ref{ourDirac}) can be written locally:
\begin{eqnarray}\label{Dirac}
\{f +{\cal I},f'+{\cal I}\}_{_N}=\{f,f'\}-\{f,x^A\}\omega_{AB}
\{x^B,f'\}+{\cal I}
\end{eqnarray}
which is the usual definition of the {\it Dirac bracket} restricted to
$N$. In this case every function on $M$ is trivially gauge-invariant
(since ${\cal F}\cap{\cal I}=0$), the image of $\phi$ is $C^\infty(N)$
and we get a Poisson structure on $N$. In adapted coordinates the
components of the canonical 
Poisson tensor on $N$ corresponding to (\ref{Dirac})
are given by:
\begin{eqnarray} \label{DiracComponents}
\Pi^{ab}_{\cal D}=\Pi^{ab}-\Pi^{aA}\omega_{AB}\Pi^{Bb}
\end{eqnarray}
where the subscript $\cal D$ stands for Dirac.

\vskip 0.2 cm

When first-class constraints are present one can still use formula
(\ref{DiracComponents}). Given a choice of adapted coordinates $(x^a,
x^\mu, x^A)$ the expression
\begin{eqnarray} \label{DiracCompFirst}
\Pi^{pq}_{\cal D}=\Pi^{pq}-\Pi^{pA}\omega_{AB}\Pi^{Bq}
\end{eqnarray}
with the indices $p,q=1,\dots,\dim(M)-k$ running over $a$ and $\mu$
values, defines a Poisson bracket on the submanifold $N'$ on which the
second-class constraints vanish (if we assume $\det(\Pi^{AB})\not=0$
on $N'$).  The submanifold $N'$ is not uniquely defined, as it depends
on the concrete choice of the set of second-class constraints.  $N$ is
now a coisotropic submanifold of $N'$ and the Poisson algebra induced
by $\Pi_{\cal D}$ on the gauge invariant functions on $N$ is indeed
canonical, independent of the choice of $N'$, and equals the one given
by (\ref{ourDirac}).

One can also extend the Poisson tensor to a tubular neighborhood of
$N'$ by taking $\Pi_{\cal D}^{Ap}=\Pi_{\cal D}^{AB}=0$.  If one
considers the tubular neighborhood equipped with the Dirac bracket,
$N$ is coisotropic in it and $N'$ is a Poisson submanifold.

\vskip 0.2cm

Later on in this dissertation, it will be important to consider the
following {\it weak regularity condition}:
\begin{eqnarray} \label{weakregularity}
\dim\{(\dd f)_p | f\in{\cal F}\cap{\cal I}\}=\dim(M)-\dim(N)-k,\quad
\forall p\in N
\end{eqnarray}
for some non-negative constant $k$.

That the strong regularity condition (\ref{regularity}) implies the
weak one with the same value for the constant $k$ is clear from
(\ref{equivalence}). The weak regularity condition is equivalent to
the existence of local coordinates on a tubular neighborhood of every
patch of $N$ with a maximal (and constant) number of coordinates which
are first-class constraints. In other words, (\ref{weakregularity})
holds if and only if there exist local coordinates satisfying (i),
(ii) as above and

\vskip .2cm
(iii)$'$ ${\rm det}(\{x^A,x^B\}(p)) \neq 0\quad \text{for
} p \text{ in an open dense subset of } {N\cap{\cal U}}$.
\vskip .2cm
\noindent However, the weak regularity condition is not enough to
guarantee that $\phi({\cal F}/{{\cal F}\cap{\cal I}})={\cal
A}_{inv}/{\cal I}$.

In general, if ${\rm dim}(\Pi^\sharp_p(T_pN^0)+ T_pN)$ is not constant
on $N$ we cannot define a Poisson bracket on the set of gauge
invariant functions. The only thing we can assert is that we have a
Poisson algebra on the subset of $C^{\infty}(N)$ given by the image of
$\phi$.  However, an efficient description of the functions in the
image (the space of {\it observables}) is not available in the general
case.

\subsection{Poisson-Dirac submanifolds}

In this section we would like to make contact between our previous
results and terminology in absence of gauge transformations and those
appearing in two papers by Crainic and Fernandes \cite{CraFer} and
Vaisman \cite{Vai02}.

In \cite{CraFer}, the interpretation of a Poisson structure in terms
of Dirac structures (see Section \ref{sec:Diracstructures}) is adopted
to carry out the reduction procedure. This geometrical approach is
slightly more general than our algebraic methods as we shall see
below. A submanifold $N\subset M$ satisfying
\begin{equation}\label{PoiDir}
\Pi^\sharp_p(T_pN^0)\cap T_pN=\{0\},\ \forall p \in N
\end{equation}
is called {\it pointwise Poisson-Dirac} in \cite{CraFer}. Recall that
this is condition b) of Theorem \ref{thm:phionto}.

In this case, for any $p\in N$ there exists a unique map
$\hat\Pi^\sharp_p$ that makes the following diagram
\begin{equation}\notag
\begin{aligned}
\xymatrix{
  T^*_pN\ar[rr]^{\hat\Pi^\sharp_p}
    &&T_pN\ar[dd]^{{i_{*p}}}\\
  \\
  \Pi_p^{\sharp-1}(T_pN)\ar[rr]_{\Pi_p^\sharp}\ar[uu]^{i^*_{p}}&&T_pM
}
\end{aligned}
\end{equation}
commutative. If the maps $\hat\Pi^\sharp_p:T_p^*N\rightarrow T_pN$
define a smooth bundle map, the latter gives a Poisson structure on
$N$, which is then called a {\it Poisson-Dirac submanifold}. With these
definitions, Poisson-Dirac submanifolds are the most general
submanifolds which inherit in a canonical way a Poisson structure from
$(M,\Pi)$. It is clear that $\phi$ onto implies that $N$ is a
Poisson-Dirac submanifold. The following is an example in which $\phi$
is not onto while $N$ is still a Poisson-Dirac submanifold, being
possible to endow it with a Poisson structure.

\begin{example}
Consider $M={\mathbb R}^4=\{(q_1,q_2,p_1,p_2)\}$ with Poisson structure
$\{p_i,q_j\}=\delta_{ij}q_{i}{\rm exp}(-1/q_{i}^{2})$ smoothly
extended to $q_i=0$ and $N$ defined by the constraints $g_1 = p_1 -
{q_2^{2}}/{2}$, $g_2 = p_2 + {q_1^{2}}/{2}$. We can take $\sigma_i :=
q_i$ as coordinates on $N$.

$\Pi^\sharp_x(T_xN^0)\cap T_xN=\{0\}$ on $N$ but $\phi$ is not
onto. For instance, take $f_i = q_i \in C^{\infty}(M)$. If we try to
find a first-class function in the class $f_1 + {\cal I}$ (its pre-image by
$\phi$) we obtain for $q_i\neq 0$
$$\tilde{f_1}:= f_1-\frac{q_1{\rm exp}(-1/q_{1}^{2})}{q_1^{2}{\rm exp}(-1/q_{1}^{2})+q_2^{2}{\rm exp}(-1/q_{2}^{2})}$$
which fails to extend continuously to $q_i = 0$. Then, $f_1$ does not belong to the image of $\phi$. However, the hamiltonian vector field associated to this singular $\tilde{f_1}$ is smooth and we can define a Poisson structure on $N$:
$$\Pi_{_N}^{12}(\sigma_1,\sigma_2) =
\{\tilde{f_1},f_2\}(\sigma_1,\sigma_2,\frac{1}{2}\sigma_2^2,-\frac{1}{2}\sigma_1^2)
= \frac{\sigma_1\sigma_2}{\sigma_1^2{\rm
exp}(1/{\sigma_2^2})+\sigma_2^2{\rm exp}(1/{\sigma_1^2})}.$$
\end{example}

\vspace{0.5 cm}

If $\phi$ is onto and, in addition, ${\rm dim}(\Pi^\sharp_p(T_pN^0)+
T_pN)$ is constant on $N$ (i.e. the situation of Theorem
\ref{thm:phionto}), we have what is called in \cite{CraFer} a {\it
constant rank Poisson-Dirac submanifold}.

\vspace{0.5 cm}

The surjectivity of $\phi$ is equivalent to the existence of an
algebraically compatible normal bundle in the language of
\cite{Vai02}. Following the latter paper, define a {\it normalization}
of $N$ by a {\it normal bundle} $\nu N$ as a splitting
$TM\vert_{N}=TN\oplus\nu N$. For every $p\in N$ there exists a
neighborhood ${\cal U}$ where we can choose adapted coordinates
$(g^A,y^{a})$ such that, locally, $g^A\vert_{N\cap {\cal U}}=0$ and
$y^{a}$ are coordinates on $N\cap {\cal U}$. Vaisman calls $\nu N$
{\it algebraically} $\Pi${\it -compatible} if, in these coordinates,
$\Pi^{A a}\vert_{N}=0$. Then, we have the following
\begin{theorem}
\noindent $\phi$ {\it is onto iff there exists an algebraically}
$\Pi${\it-compatible normal bundle.}
\end{theorem}

{\it Proof:}

Only local properties in a neighborhood of each point of $N$ matter
for this proof.

$\Rightarrow )$ Let $(g^A,z^a)$ be local coordinates such that
$N$ is locally defined by $g^A=0$ and $z^a$ are coordinates on
$N$. Take the pre-image by $\phi$ of the coordinate functions
$z^a$ and denote them by $y^a$. $(g^A,y^a)$ are local
coordinates such that $\Pi^{Aa}\vert_{N}=0$.

$\Leftarrow )$ For any $f(g^A,y^a) \in C^\infty({\cal U})$,
$\tilde{f}(g^A,y^a)=f(0,y^a)\in {\cal F}$ and $\tilde{f}-f \in {\cal
I}$. Then, $\phi$ is onto.$\,$\hfill$\Box$\break

\section{Dirac structures}\label{sec:Diracstructures}

As stressed before the structure underlying the reduction of Poisson
or symplectic manifolds is a Poisson algebra. Such an algebraic
structure can be encoded in geometric terms through the concept of
{\it Dirac structure}, which generalizes the Poisson and presymplectic
geometries by embedding them in the context of the geometry of
$TM\oplus T^*M$. Dirac structures were introduced in the remarkable
paper by T. Courant \cite{Cou}.  Therein, they are related to the
Marsden and Weinstein \cite{MarWei} reduction and to the Dirac bracket
\cite{Dir} on a submanifold of a Poisson manifold.  This simple but
powerful construction allows to deal with mechanical situations in
which we have both gauge symmetries and Casimir functions.

Dirac structures (more precisely their twisted counterparts) will play
an essential role in Chapter \ref{ch:SusyWZP} when working out the
conditions for extended supersymmetry in the WZ-PSM. In addition, the
fact that Dirac structures can be seen as a generalization of Poisson
structures suggests generalizations of the PSM (see Dirac sigma models
in Section \ref{sec:generalizationsPSM}) and of the deformation
quantization scheme (\cite{Sev}). Below we describe Dirac structures
and give some original results on their relation to Poisson algebras
and on reduction issues.

It will be useful for later purposes to introduce the notion of Lie
algebroid, of which a Dirac structure is a particular case. A {\it Lie
algebroid} over a smooth manifold $M$ is a vector bundle $E$ with the
following additional structure:
\begin{enumerate}
\item[(i)] a Lie bracket $[\cdot , \cdot]$ on sections of $E$.
\vspace{-0.25cm}
\item[(ii)] A bundle map $\rho:E\rightarrow TM$, called the {\it anchor},
inducing a Lie algebra homomorphism on sections.
\vspace{-0.25cm}
\item[(iii)] $[\alpha,f\beta]=f[\alpha,\beta]+(\rho(\alpha)f)\beta$, where
$f\in C^\infty(M)$ and $\alpha,\beta\in\Gamma(E)$.
\end{enumerate}

Let us provide some elementary examples of Lie algebroids:

\begin{example}
$M$ a point and $\rho$ trivial. Then, $E$ is a Lie algebra.
\end{example}

\begin{example}
$M$ arbitrary and $\rho =0$. Then, $E$ is a bundle of Lie algebras.
\end{example}

\begin{example}
$E=TM$ endowed with the ordinary bracket and $\rho = \mbox{id}_{TM}$.
\end{example}

\begin{example}
Let $(M,\Pi)$ be a Poisson manifold. $E= T^*M$ is a Lie algebroid with
anchor $\rho=\Pi^\sharp$ and bracket defined on exact elements by
$[\dd f,\dd g]=\dd \{f,g\}$ and extended by the Leibniz rule to all
1-forms.
\end{example}

\vskip 0.5cm

Dirac structures are special types of Lie algebroids contained in the
vector bundle $TM \oplus T^*M$. We shall use the notation $\rho_1:TM
\oplus T^*M\rightarrow TM$, $\rho_2:TM \oplus T^*M\rightarrow T^*M$
for the canonical projections.

The {\it Courant bracket} on sections of $TM \oplus T^*M$ is defined
by\footnote{In Chapter \ref{ch:SusyWZP} we shall define the twisted
version of the Courant bracket by a closed 3-form and the twisted
analogues of the rest of objects introduced in this section.}
\begin{equation}\notag
[(X,\xi),(X',\xi')]=([X,Y], \iota(X)\dd\xi'-\iota(X')\dd\xi +{1\over
2}\dd(\iota(X)\xi'-\iota(X')\xi)).
\end{equation}
It is bilinear and antisymmetric but not a Lie bracket, since it does
not satisfy the Jacobi identity in general.

Consider the natural symmetric bilinear form on $TM \oplus T^*M$,
\begin{equation}\label{sympairing}
\langle (X,\xi),(X',\xi')\rangle = \iota(X)\xi'+\iota(X')\xi.
\end{equation}

Following \cite{Cou} we shall define a {\it Dirac subbundle} as a
subbundle $D\subset TM \oplus T^*M$ maximally isotropic with respect
to (\ref{sympairing}). Maximal isotropy implies that $D^\perp = D$,
where $D^\perp$ stands for the orthogonal of $D$. In particular,
$\mbox{dim}(D)=\mbox{dim}(M)$. Dirac subbundles have been recently
considered (see \cite{BlaRat}) in connection to the singular reduction
of implicit Hamiltonian systems.

A {\it Dirac structure} is a Dirac subbundle $D$ such that its sections
close under the Courant bracket. In this case, as shown in \cite{Cou},
the restriction to $D$ of the Courant bracket fulfills the Jacobi
identity and $D$ with anchor $\rho_1\vert_D$ is a Lie algebroid.

\begin{example}
For a 2-form $\omega$, the graph $L_\omega$ of
$\omega^\flat:TM\rightarrow T^*M$ is a Dirac structure in $TM \oplus
T^*M$ if and only if $\dd\omega =0$. This type of Dirac structure is
characterized by the fact that $\rho_1(L_\omega)=TM$ at every point of
$M$.
\end{example}

\begin{example}
Let $\Pi$ be a bivector field on $M$. The graph $L_\Pi$ of the map
$\Pi^\sharp:T^*M\rightarrow TM$ is a Dirac structure if and only if
$\Pi$ is a Poisson structure. In this case $\rho_2$ maps $L_\Pi$ onto
$T^*M$.
\end{example}

The definition of a symmetry of a Dirac structure is a straightforward
generalization of the definitions in the presymplectic and Poisson
cases:
\begin{definition}\label{def:symmetryDirStruct}
A vector field $Y$ is a {\it symmetry of the Dirac structure}
$D\subset TM \oplus T^*M$ if $({\cal L}_Y X,{\cal L}_Y\xi)\in
\Gamma(D)$ for every $(X,\xi)\in\Gamma(D)$.
\end{definition}

\subsection{Dirac structures and Poisson algebras}

Let $D$ be a Dirac structure on $M$. We say that $f\in C^\infty(M)$ is
an {\it admissible function} if there exists a vector field $X_f$ such
that $(X_f,\dd f)$ is a section of $D$. It is not difficult to see that
the admissible functions form an algebra with the pointwise
product. In addition, given two admissible functions $f$ and $g$,
\begin{equation}\notag
\{f,g\}=X_f(g)
\end{equation}
defines a Poisson bracket. Thus the set of admissible functions is a
Poisson algebra. The converse, i.e. whether given a Poisson algebra we
may define a Dirac structure $D$ associated to it, is a more subtle
issue. The situation is as follows.

Let ${\cal A}\subset C^\infty(M)$ be a Poisson algebra and let us
define its associated Dirac structure $D$. For any point $p\in M$,
$(X_p,\xi_p)\in D_p$ if and only if
\vskip 2mm

a) There exists $f\in {\cal A}$ such that $\xi_p=(\dd f)_p$.

b) $\iota(X_p) (\dd g)_p=\{f,g\}(p)$ for every $g\in \cal A$.
\vskip 2mm

We are tacitly assuming that $\forall f\in{\cal A}$ there exists a
vector field satisfying condition b). Equivalently, we assume that
$\rho_2(D_p)=\{(\dd f)_p| f\in{\cal A}\}$. This is mandatory if we
want to define a meaningful Dirac structure associated to the Poisson
algebra ${\cal A}$.

Notice that $\rho_2(D_p)=\{(\dd f)_p| f\in{\cal A}\}$ if and only if
$(\dd g)_p=0,g\in{\cal A}\Rightarrow \{f,g\}(p)=0,\forall f\in{\cal
A}$. This requirement is not met for an arbitrary Poisson
algebra of functions, as shown by the next example.

\begin{example}
Take $M={\mathbb R}^2$ with coordinates $(x,y)$ and ${\cal A}$
consisting of functions of the form $f(x^2,y^2)$. We endow ${\cal A}$
with the Poisson bracket $\{x^2,y^2\}=1$. The function $x^2\in{\cal
A}$ has vanishing differential at the origin and non-zero Poisson
bracket with $y^2\in{\cal A}$. Consequently, the vector field $X$ of
condition b) above does not exist for every $f\in{\cal A}$.
\end{example}

\vskip 3mm

However, it is not difficult to show that for Poisson algebras of
functions obtained by the reduction procedure of Section
\ref{reduction}, the aforementioned condition $\rho_2(D_p)=\{(\dd
f)_p| f\in{\cal A}\}$ actually holds. To see this, consider the
Poisson algebra ${\cal C}(\Pi,M,N)$ with Poisson bracket given by
(\ref{ourDirac}). Take $g+{\cal I}\in {\cal C}(\Pi,M,N)$ with
$g\in{\cal F}$ such that $(\dd g)_p\vert_{TN}=0$ at a point $p\in
N$. For any $f+{\cal I}\in {\cal C}(\Pi,M,N)$ with $f\in{\cal F}$,
formula (\ref{ourDirac}) obviously implies that $\{f,g\}_N=0$. Then,
conditions a), b) above define a Dirac structure associated to ${\cal
C}(\Pi,M,N)$.

\vskip 2mm

{}From now on we assume that $\rho_2(D_p)=\{(\dd f)_p| f\in{\cal A}\}$.

\vskip 0.5cm

It is clear that $D_p$ is a linear subspace. Given $(X_p,\xi_p)$,
$(X'_p,\xi'_p)\in D_p$ corresponding to $f$ and $f'$, we have that
$(aX_p+a'X'_p,a\xi_p+a'\xi'_p)\in D_p,\forall a,a'\in{\mathbb R}$ and
corresponds to $af+a'f'\in{\cal A}$.

$D_p$ is isotropic with respect to (\ref{sympairing}) as a consequence
of the antisymmetry of the Poisson bracket i.e. given $(X_p,\xi_p),
(X'_p,\xi'_p)\in D_p$ associated to functions $f$ and $f'$ we have
\begin{eqnarray}\notag
\langle(X_p,\xi_p),(X'_p,\xi'_p)\rangle&=&
\iota(X_p) (\dd f')_p+\iota(X'_p) (\dd f)_p\cr&=&\{f,f'\}(p)+\{f',f\}(p)=0.
\end{eqnarray}

We can show that $D_p=(D_p)^\perp$ by a dimensional argument. Since
$$\rho_2(D_p)=\{(\dd f)_p| f\in{\cal A}\}$$
and
$$\Ker\rho_2\vert_{D_p}=\{X_p| \ \iota(X_p)(\dd f)_p=0,\ \forall
f\in{\cal A}\}$$
it is clear that we have the relation
$$(\Ker \rho_2\vert_{D_p})^0 = (\Im \rho_2\vert_{D_p})$$
and  then 
$$\dim (D_p) = \dim (\Im \rho_2\vert_{D_p})+\dim(\Ker \rho_2\vert_{D_p})=
\dim(M).$$ 
This proves that $D_p$ is maximally isotropic.  

One can also show that sections of $TM\oplus T^*M$ with values in
$D_p$ as defined before close under the Courant bracket.  Firstly,
consider sections of the form $(X_f,\dd f)$ with $f\in{\cal A}$; their
bracket reads
\newpage
\begin{eqnarray}\notag
[ (X_f,\dd f) , (X_{f'},\dd f') ] &=& ([X_f,X_{f'}], {1\over 2}
\dd(X(f')-X'(f)))\cr &=&([X_f,X_{f'}], \dd\{f,f'\})
\end{eqnarray}
and due to the Jacobi identity for the Poisson bracket,
$$[X_f,X_{f'}](g) = \{\{f,f'\},g\},\ \forall g\in{\cal A}$$
so that exact sections close under the Courant bracket. Using now that
for two orthogonal sections of $TM\oplus T^*M$ and $h\in C^\infty (M)$
we have
$$[(h X, h\xi),(X',\xi')]=h [(X, \xi),(X',\xi')] - X'(h)(X,\xi)$$ the
closedness relation can be extended to arbitrary sections. Hence, if
the fibers $D_p$ define a subbundle $D$ of $TM\oplus T^*M$, then $D$
is a Dirac structure. Although the fibers have constant dimension, in
general they do not define a subbundle. This can be shown in the
following simple example.

\begin{example}
Consider $M={\mathbb R}^3$ and the algebra $\cal A$ of smooth functions 
in $M$ invariant under the flow of the vector field: 
$$x_2\partial_{x_1}+x_3\partial_{x_2}+x_1\partial_{x_3}.$$  
This is a Poisson algebra if we endow it with the zero 
Poisson bracket.

Now for $(x_1,x_2,x_3) \not= 0$ we have
$$ D_{(x_1,x_2,x_3)}={\rm
  span}\{(x_2\partial_{x_1}+x_3\partial_{x_2}+x_1\partial_{x_3}, 0), 
(0,\xi_1),(0,\xi_2)\}$$
where $\xi_1$, $\xi_2$ are two independent covectors in $T^*_{(x_1,x_2,x_3)}M$
that kill the vector $x_2\partial_{x_1}+x_3\partial_{x_2}+x_1\partial_{x_3}$.
However, as the differential of any function in $\cal A$ vanishes
in $(0,0,0)$ we have
$$D_{(0,0,0)}=T_{(0,0,0)}M,$$
which shows that the bunch of fibers is not a subbundle of $TM\oplus
T^*M$.
\end{example}
\vskip 2mm

Of course, when the Poisson algebra is induced by a Dirac structure
the latter is recovered form the former as sketched above.

\subsection{Induced Dirac structures}

In this paragraph we shall introduce a method to obtain a Dirac
structure from another one. The method is very simple but it has
important applications like the generalization of the Dirac bracket,
reduction of Dirac structures to submanifolds or the Marsden and Ratiu
reduction of Poisson manifolds \cite{MarRat}.
 
\vskip 2mm

Take a subbundle $S\subset TM\oplus T^*M$ isotropic with respect to
(\ref{sympairing}), i.e. $S\subset S^{\perp}$. It is easy to show that
given a Dirac subbundle $D$ we may obtain another one $D^S= (D\cap
S^\perp) + S$ provided that it is a subbundle. We must show that it is
maximally isotropic, but this is immediate:
\begin{eqnarray}\notag
(D^S)^\perp &=& (D^\perp + S)\cap S^\perp\cr
&=& (D\cap S^\perp)+S =D^S,
\end{eqnarray}  
where in the last line we have used that $D$ is maximally isotropic
and $S$ is a subset of $S^\perp$. In a sense $D^S$ is the Dirac
subbundle closest to $D$ among those containing $S$, as stated in the
following
\begin{theorem}
Let $D,S$ and $D^S$ be as above and let $D'$ be a Dirac subbundle such
that $S\subset D'$. Then, $ D'\cap D\subset D^S\cap D$. In addition,
$D'\cap D =D^S\cap D$ if and only if $D'=D^S$.
\end{theorem}

\vskip 0.3cm

{\it Proof:}

{}From the isotropy of $D'$ and given that $S\subset D'$ we deduce
 that $D'\subset S^{\perp}$. Hence,
$$D'\cap D \subset S^\perp\cap D = D^S\cap D.$$

If the equality $D'\cap D =D^S\cap D$ holds, then $D'\supset D'\cap D
= S^\perp\cap D$. Since $S\subset D'$, we deduce that $D^S=(D\cap
S^\perp) + S\subset D'$. But $D^S$ and $D'$ have the same dimension
and hence they are equal. $\,$\hfill$\Box$\break

In general even when $D^S$ is a maximally isotropic smooth subbundle
its sections do not close under the Courant bracket. We discuss now
some particularly interesting examples and applications of this
construction.

\vskip 5mm

{\bf Dirac bracket (or Dirac Dirac structure):} Consider an integrable
distribution ${\Phi}\subset TM$ and take ${{S}}={\Phi}^0\subset
T^*M$. Then, for any Dirac structure $D$ on $M$, $D^{{S}}$ (if smooth)
is a Dirac structure such that $\rho_1(D^{{S}})$ is everywhere tangent
to the foliation. That is,
$$(D^{{S}})_p=\{(X_p,\xi_p+\nu_p)| (X_p,\xi_p)\in D_p, X_p\in{\Phi}_p,
\nu_p\in {{S}}_p\}.$$

Let us see that sections of $D^{{S}}$ close under the Courant
bracket:
\begin{eqnarray}\label{CourDirac}
&&\hspace{-1cm}[(X,\xi+\nu),(X',\xi'+\nu')]=\cr
&&=([X,X'], \iota(X)\dd\xi'-\iota(X')\dd\xi+
{1\over 2} \dd(\iota(X)\xi'-\iota(X')\xi))+\cr
&&+(0,\iota(X)\dd\nu'-\iota(X')\dd\nu).
\end{eqnarray}  
$(X,\xi), (X',\xi')$ are sections of the Dirac structure and then so
it is the first line of the right hand side of (\ref {CourDirac}).
$X, X'$ are vector fields in the integrable distribution $\Phi$ and
then $[X,X']$ is also a vector field of $\Phi$.  Finally, we have to
show that the third line of (\ref {CourDirac}) is a section of
$\Phi^0$. Contracting with a vector field $Y$ in $\Phi$ we obtain:
\begin{eqnarray}\notag
\iota(Y)(\iota(X)\dd\nu')=
X(\iota(Y)\nu')-Y(\iota(X)\nu')-\iota([X,Y])\nu'=0
\end{eqnarray}
where we have used that $\nu'$ is a section of ${{S}}={\Phi}^0$
and $X$, $Y$ are sections of $\Phi$, which is an integrable
distribution. Therefore, we have proven that $D^{{S}}$ is a Dirac
structure on $M$ (assuming it is a subbundle).

$D^{{S}}$ can be considered as a generalization of the Dirac bracket
extended to the whole manifold for the case of Dirac
structures. Recall that the Dirac structure $D^{{S}}$ is the graph of
a bivector field (which is a Poisson structure due to involutivity) if
and only if $\rho_2(D^{{S}})=T^*M$ at every point of $M$. This is
equivalent to
\begin{eqnarray} \label{DiracfromPoisson}
D^{{S}}+TM=TM\oplus T^*M.
\end{eqnarray} 
Taking the orthogonal of (\ref{DiracfromPoisson}) we get the more
familiar (and equivalent) condition
\begin{eqnarray}\label{eq:condPoissongen}
(D+{\Phi}^0)\cap{\Phi}=\{0\}.
\end{eqnarray} 
If this holds, $D^{{S}}$ corresponds to a Poisson bracket which should
be called Dirac bracket. Indeed, it coincides with the standard Dirac
bracket if $D$ itself comes from a Poisson structure $\Pi$. In this
case (\ref{eq:condPoissongen}) can be rewritten as
$\Pi^\sharp({\Phi}^0)\cap{\Phi}=\{0\}$.

Once we have defined the Dirac structure $D^{{S}}$ we can restrict
it to any leaf $N$ of the foliation induced by $\Phi$. Let
$i:N\rightarrow M$ be the inclusion. The image of the bundle map
$$i_*^{-1} \oplus i^* : D^{{S}}\vert_N \rightarrow TN\oplus T^*N$$
defines a Dirac $D^{{S}}_N$ structure in $TN\oplus T^*N$. The isotropy
of $D^{{S}}_N$ is obviously inherited from the isotropy of $D^S$. Now,
using that $\mbox{Ker}(i_*^{-1} \oplus i^*)={\Phi}^0\vert_N$ and that
$\mbox{dim}(D^{{S}}_N)=\mbox{dim}(D^{S})-\mbox{Ker}(i_*^{-1} \oplus
i^*)$ we deduce that $D^{{S}}_N$ is maximally isotropic in $TN\oplus
T^*N$. The proof of the closedness of the Courant bracket in
$D^{{S}}_N$ is just the proof of the closedness in $D^{{S}}$ given
above (recall that we are assuming that $D^{{S}}$ is a subbundle.

Notice that $D^{{S}}_N$ corresponds to a Poisson structure if and only
if (\ref{eq:condPoissongen}) is satisfied on $N$, i.e. if
\begin{equation}\label{eq:condequivDB}
(D\vert_N+TN^0)\cap TN =\{0\}.
\end{equation}
If $D$ is the graph of a bivector field $\Pi$, (\ref{eq:condequivDB})
is equivalent to requiring $\Pi^\sharp(TN^0)\cap TN=\{0\}$ at every
point of $N$. That is, $N\subset M$ is a pointwise Poisson-Dirac
submanifold (recall (\ref{PoiDir})). The smoothness of the induced
bivector field is ensured by our regularity asumptions on $D^S$ so
that $N$ is a Poisson-Dirac submanifold and is equipped with a Poisson
bracket.

\vskip 5mm

{\bf Projection along an integrable distribution:} Now, let $S\subset
TM$ be an integrable distribution. Assuming that both $D\cap S^\perp$
and $D^S$ are subbundles, any section of $D^S$ can be written as
$(X+Y,\xi)$ where $(X,\xi)$ is a section of $D\cap(TM\oplus S^0)$ and
$Y$ is a section of $S$.

$D^S$ is not Courant involutive in general. This is not strange since
one would expect to be able to define a Dirac structure only on $M/S$,
the space of leaves of the foliation defined by $S$ (assuming $M/S$ is
a manifold). Objects on $M$ which descend suitably to $M/S$ will be
called {\it projectable along} $S$. Functions on $M/S$,
$C^\infty(M)_{\mbox{pr}}$, can be viewed as the set
\begin{equation}\notag
C^\infty(M)_{\mbox{pr}}=\{f\in C^\infty(M)\ \vert \ X(f)=0,\
\forall X\in S\}.
\end{equation}

Vector fields on $M/S$ are then defined as derivations on
$C^\infty(M)_{\mbox{pr}}$, i.e.  ${\mathfrak
X}(M)_{\mbox{pr}}=\{X\in{\mathfrak X}(M)\ \vert \
X(C^\infty(M)_{\mbox{pr}})\subset C^\infty(M)_{\mbox{pr}}\}$. Notice
that $X$ belongs to ${\mathfrak X}(M)_{\mbox{pr}}$ if and only if
$Z(X(f))=0,\ \forall f\in C^\infty(M)_{\mbox{pr}},\ \forall
Z\in\Gamma(S)$. Or equivalently, $[Z,X](f)=({\cal L}_Z X)(f)=0$ using
that $Z(f)=0$. Hence, we obtain the more useful characterization
\begin{equation}\notag
{\mathfrak X}(M)_{\mbox{pr}}=\{X\in{\mathfrak X}(M)\ \vert \ {\cal
L}_Z X \in \Gamma(S),\ \forall Z\in \Gamma(S) \}.
\end{equation}

Analogously, 1-forms on $M/S$, $\Omega^1(M)_{\mbox{pr}}$, are linear
maps from ${\mathfrak X}(M)_{\mbox{pr}}$ to
$C^\infty(M)_{\mbox{pr}}$. An analogous argument to that followed for
vector fields yields:
\begin{equation}\notag
\Omega^1(M)_{\mbox{pr}}=\{\xi\in \Gamma(S^0)\ \vert \ {\cal L}_Z\xi
=0,\ \forall Z\in\Gamma(S)\}.
\end{equation}

\vskip 5mm

We are now ready to prove that sections $(X+Y,\xi)$ of $D^S$ which are
projectable along $S$, i.e.
\begin{enumerate}
\item[a)]$\xi$ is a section of $S^{0}$,
\vspace{-2.5mm}
\item[b)]$\forall Z\in\Gamma(S),\ ({\cal L}_Z(X+Y),{\cal
L}_Z\xi)=(Z',0)$, with $Z'\in\Gamma(S)$
\end{enumerate}
are Courant involutive. Notice that the integrability of $S$ implies
that ${\cal L}_Z X=[Z,X]$ in condition b) must be a section of
$S$. Hence, if $(X+Y,\xi)$ is a projectable section, $X$ itself must
be a projectable vector field.

Take two such sections $(X+Y,\xi)$ and $(X'+Y',\xi')$. First notice
that
$$[X+Y,X'+Y']-[X,X']\in\Gamma(S)$$ because $[X,Y']$, $[Y,X']$ and
$[Y,Y']$ are sections of $S$ due to condition b) and the integrability
of $S$. Therefore,
\begin{equation}\label{eq:commprojsec}
[X+Y,X'+Y']=[X,X']+W,\ W\in\Gamma(S).
\end{equation}
We must prove that ${\cal L}_Z[X+Y,X'+Y']=[Z,[X+Y,X'+Y']]$ is a
section of $S$ for any $Z\in\Gamma(S)$. Using (\ref{eq:commprojsec}),
this amounts to prove that $[Z,[X,X']]$ belongs to $\Gamma(S)$. By the
Jacobi identity,
\begin{equation}
[Z,[X,X']]=[[Z,X],X']-[[Z,X'],X]
\end{equation}
which is a section of $S$ because $X$ and $X'$ are prjectable vector
fields.

For the cotangent part of the Courant bracket of projectable sections
we have
\begin{eqnarray}\label{Courproy}
&&\dd(\iota(X+Y)\xi')=\dd(\iota(X)\xi'),\cr\cr
&&\iota(X+Y)\dd\xi'=\iota(X)\dd\xi'+{\cal
L}_Y\xi'-\dd\iota(Y)\xi'=\iota(X)\dd\xi'
\end{eqnarray}
where we have used conditions a) and b) above. We want to show that
the right-hand side of both equations in (\ref{Courproy}) are
projectable sections, i.e. both are in $\Gamma(S^0)$ and have
vanishing Lie derivative along any section $Z\in\Gamma(S)$. This is
not difficult:
\begin{eqnarray} \notag
&&\iota(Z)(\iota(X){\dd}\xi')=-\iota(X)({\cal
L}_Z\xi'-{\dd}(\iota(Z)\xi'))=0,\cr\cr &&\iota(Z){\dd}(\iota(X)\xi')=
{\cal L}_Z(\iota(X)\xi')=\iota({\cal L}_Z X)\xi'+ \iota(X){\cal
L}_Z\xi'=0,\cr\cr &&{\cal L}_Z(\iota(X){\dd}\xi')=\iota({\cal L}_Z
X){\dd}\xi'+\iota(X){\cal L}_Z {\dd}\xi'= \cr&&\hskip 2.03cm= {\cal
L}_{Z'}\xi'-{\dd}(\iota(Z')\xi')+\iota(X){\dd}({\cal
L}_Z\xi')-\iota(X){\dd}(\iota(Z)\xi')=0,\cr\cr &&{\cal
L}_Z({\dd}(\iota(X)\xi'))={\dd}(\iota(Z){\dd}(\iota(X)\xi'))=0,
\end{eqnarray}
where $Z'={\cal L}_Z X=[Z,X]\in\Gamma(S)$.

If for any point $p\in M$ and any $(X_p,\xi_p)\in (D^S)_p$ there
exists a (local) projectable section of $D^S$ such that it coincides
with $(X_p,\xi_p)$ at $p$, then $S$ is a symmetry of $D^S$ in the
sense of Definition \ref{def:symmetryDirStruct}. To see this, denote
by $\{{\cal X}^i\}_{i=1}^m$, $m=\mbox{dim}(M)$, a basis of
$\Gamma(D^S)$ formed by projectable sections. Any local section of
$D^S$ is written as
\begin{equation}\notag
{\cal X} =\sum_{i=1}^m
f^i{\cal X}_i
\end{equation}
for some smooth functions $f^i$.  Taking the Lie derivative with
respect to $Z\in\Gamma(S)$,
\begin{equation}\notag
{\cal L}_Z{\cal X}=\sum_{i=1}^m ({\cal L}_Z f^i){\cal X}_i + \sum_{i=1}^m
f^i({\cal L}_Z {\cal X}_i)
\end{equation}
where the last term belongs to $\Gamma(S)$ due to the projectability
of the ${\cal X}_i$s. Hence, for any ${\cal X} \in \Gamma(D^S)$ we have
shown that ${\cal L}_Z{\cal X}\in\Gamma(D^S),\ \forall Z\in\Gamma(S)$
and $S$ is a symmetry of $D^S$.

In this situation we can
project the Dirac structure onto $M/S$.  Given the projection $\tau:
M\rightarrow M/S$ the image of
$$ \tau_*\oplus (\tau^*)^{-1}:D^S\rightarrow T(M/S)\oplus T^*(M/S)$$
correctly defines a Dirac structure on $M/S$.

\vskip 5mm

{\bf Generalized Marsden-Ratiu reduction:} The third example is a
combination of the two previous ones and is inspired in the work of
Marsden and Ratiu \cite{MarRat}. It consists in reducing a Dirac
structure $D$ to a submanifold $N$ along a symmetry $S$ of $D^S$. $S$
is an integrable distribution as above and we view $N$ as a leaf of
another integrable distribution $\Phi\subset TM$.

As we have learnt $D^S$ induces a Dirac structure in
$M/S$.  The latter can be reduced to $N/(S\cap TN)$ and is the
generalization to any Dirac structure of the procedure of
\cite{MarRat} for Poisson manifolds.

One may wonder when the reduced Dirac structure, which we denote by
$\hat D$, is actually a Poisson structure on the orbit space $N/(S\cap
TN)$.  $\hat D$ is Poisson if and only if
\begin{eqnarray}\notag
\hat D \cap T (N/(S\cap TN))=\{0\}.
\end{eqnarray}

Equivalently, in terms of the original Dirac structure $D$, $\hat D$
is Poisson if and only if for any $p\in N$ we have that
\begin{eqnarray}\notag
(X_p,\xi_p)\in D_p \ {\rm s. t.}\ \xi_p\in (S^0)_p\cap
(TN^0)_p\Rightarrow X_p\in S.
\end{eqnarray}

This can be reformulated as
\begin{eqnarray}\notag
TM\oplus (S^0\cap TN^0)\subset D\cap (S\oplus T^*M)+TM
\end{eqnarray}
and taking the orthogonal with respect to the symmetric pairing
\begin{eqnarray}\label{MarRatcon}
TM\cap (D+S^0)\subset S+TN,
\end{eqnarray}
which can be phrased by saying that the (generalized) Poisson vector
fields of $S^0$ are sections of $S+TN$. This is precisely the same
condition that Marsden and Ratiu found for their reduction to yield a
Poisson structure in the reduced space.

We have then shown that Marsden and Ratiu reduction fits perfectly in
the framework of Dirac structures in the sense that the reduction of
an invariant Dirac structure always produces a Dirac structure
(assuming the existence of the different subbundles) in the reduced
space. In the particular case in which (\ref{MarRatcon}) holds the
final Dirac structure is actually Poisson.

\section{Deformation quantization of Poisson manifolds}

We end this chapter by introducing the problem of deformation
quantization of Poisson manifolds, which will be a central subject in
this dissertation due to its relation with the Poisson sigma model.

The program of deformation quantization and the notion of star product
was established by Bayen, Flato, Fronsdal, Lichnerowicz and
Sternheimer in the 1970s (\cite{BayFlaFroLicSte}). The idea is to
quantize the algebra of classical observables on a Poisson manifold
$M$ by deforming the commutative pointwise product of functions by a
non-commutative associative product, the star product, with the
Planck's constant controlling the non-commutativity.

More precisely, the problem of deformation quantization of a Poisson
 manifold $(M,\Pi)$ is to find an ${\mathbb R}[[\hbar]]$-bilinear
 product $\star$ (a {\it star product}) on $C^\infty(M)[[\hbar]]$
\begin{equation}
\begin{matrix}
&C^\infty(M)[[\hbar]] \times
C^\infty(M)[[\hbar]]&\rightarrow&C^\infty(M)[[\hbar]]\\
&(f,g)&\mapsto&f\star g\\ \notag
\end{matrix}
\end{equation}
such that $\forall f_1,f_2,f_3 \in C^\infty(M)$:
\begin{enumerate}
\item[(i)]$(f_1 \star f_2)\star f_3 =f_1 \star (f_2 \star f_3)$
\vspace{-2mm}
\item[(ii)]$f_1 \star f_2 = f_1f_2 + \{f_1,f_2\}\hbar + O(\hbar^2)$
\end{enumerate}
i.e. an associative deformation of the pointwise product on
$C^\infty(M)$ which at first order in $\hbar$ coincides with the
Poisson bracket.

The existence of star products for any symplectic manifold was shown
first by DeWilde and Lecomte (\cite{DewLec}). A more geometrical proof
was given later by Omori, Maeda and Yoshioka
(\cite{OmoMaeYos}). Finally, Fedosov provided a purely geometrical
construction (\cite{Fed85},\cite{Fed94}).

The case in which $M$ is a general Poisson manifold turned out to be
much harder and was solved by Kontsevich in 1997 (\cite{Kon}) as a
consequence of his formality theorem. He proved that every Poisson
manifold admits a star product and gave an explicit formula for
$M={\mathbb R}^n$ (see \cite{CatFelTom} for globalization aspects)
which can be expressed in terms of diagrams.

A diagram $\daleth$ of order $n$ consists of $n$ numbered vertices of
{\it first type} $\{1,\dots,n\}$ and two vertices of {\it second type}
labeled by $\{L,R\}$ (standing for Left and Right). Two ordered
oriented edges start at each of the vertices of the first type and end
at any vertex (first or second type), but an edge cannot start and end
at the same vertex. There are no edges starting at vertices of the
second type. The endpoints of the edges starting at vertex $i$ are
denoted $v_1(i),v_2(i)$.

One then associates to every diagram $\daleth$ a bidifferential
operator $B_\daleth$ as follows:

- Place a function $f$ at $L$ and a
function $g$ at $R$.

- Associate to each vertex $v$ of the first type the Poisson tensor
$\Pi^{j_1j_2}$.

- Associate a partial derivative $\partial_{j_1}$ ($\partial_{j_2}$)
  to the first (second) edge starting from $v$ and contract it with
  the first (second) index of $\Pi^{j_1j_2}$. The partial derivative
  acts on the function or Poisson tensor placed at the endpoint of the
  edge.

The Kontsevich's formula for the star product of $f,g \in C^\infty(M)$
reads\footnote{What Kontsevich calls $\hbar$ is here $i\hbar/2$. This
convention will be convenient in connection with the perturbative
quantization of the Poisson sigma model of Chapter
\ref{ch:quantumPSM}.}:
\begin{eqnarray}\label{Kontsevichformula}
f \star g = fg +\sum_{n \geq
1}\left(\frac{i\hbar}{2}\right)^n\frac{1}{n!}  \sum_{\daleth\text{
order n}}w_\daleth B_\daleth(f,g).
\end{eqnarray}

Consider the upper half plane $H_+$ endowed with the Poincar\'e metric
\begin{equation}\notag
\dd s^2 = \frac{\dd x^2+\dd y^2}{y^2}
\end{equation}
and define the configuration space
\begin{equation}\notag
H_n:=\{(u_1,\dots,u_n)\in H^n\vert \ u_k \in H,u_k\neq u_k' \
\mbox{for}\ k\neq k'\}.
\end{equation}

Now take a complex coordinate $z=x+iy$ and let $\phi(z,w)$ be the
angle (measured counterclockwise) between the geodesic passing through
$z$ and $i\infty$ and the geodesic passing through $z$ and $w$,
\begin{equation}\label{eq:Kontanglefunc}
\phi(z,w)=\frac{1}{2i}\log\frac{(z-w)(z-\bar w)}{(\bar z - \bar
w)(\bar z - w)}.
\end{equation}

 Define $\dd
\phi(z,w):=\partial_z\phi(z,w)\dd z + \partial_w\phi(z,w)\dd w$. The
weight $w_\daleth$ is given by:
\begin{eqnarray}\notag
w_\daleth=\frac{1}{(2\pi)^{2n}}\int_{H_n}\bigwedge_{i=1}^{n}
\dd\phi(u_i,u_{v_1(i)})\wedge\dd\phi(u_i,u_{v_2(i)}) 
\end{eqnarray}
setting $u_L=0,u_R=1$.

This combinatorial, diagrammatic construction of the Kontsevich's star
product resembles Feynman expansions in perturbative Quantum Field
Theory. It was Cattaneo and Felder who identified the theory from
which Kontsevich's formula can be recovered and it turned out to be
the Poisson sigma model, which we introduce in Chapter
\ref{ch:branesPSM}.

\newpage

\pagestyle{empty}
\chapter{Classical Poisson sigma model and branes}\label{ch:branesPSM}
\thispagestyle{empty}
\pagestyle{empty}
\cleardoublepage

\pagestyle{headings}
\renewcommand*{\chaptermark}[1]{\markboth{\small \scshape  \thechapter.\ #1}{}}
\renewcommand*{\sectionmark}[1]{\markright{\small \scshape \thesection. \ #1}}

In this chapter we introduce, at the classical level, the field theory
which is the main object of study of this dissertation: the Poisson
sigma model. In Section \ref{sec:IntroPSM} we discuss its original
motivation, some interesting theories which can be seen as particular
cases of the Poisson sigma model and a number of generalizations which
have been recently proposed.

In Section \ref{sec:branes} a thorough study of the classical Poisson
sigma model on a surface with boundary is carried out, identifying the
admissible branes.

In Section \ref{sec:HamPSM} it is shown how the phase space of the
model is related to the Poisson geometry of the brane.

\section{Introduction to the PSM}\label{sec:IntroPSM}

The Poisson sigma model (PSM) is a two-dimensional topological sigma
model defined on a surface $\Sigma$ and with a finite dimensional
Poisson manifold $(M,\Pi)$ as target. The fields are given by a bundle
map $(X,\eta): T\Sigma \rightarrow T^{*}M$, where $X:\Sigma
\rightarrow M$ is the base map and $\eta$ is a 1-form on $\Sigma$ with
values in $\Gamma(X^*T^*M)$. The action functional has the form
\begin{eqnarray} \label{PS}
S(X,\eta)=\int_\Sigma \langle \eta,\wedge \dd X \rangle + 
\frac{1}{2}\langle\Pi\circ X,\eta\wedge\eta\rangle,
\end{eqnarray}
where $\langle\cdot,\cdot\rangle$ denotes the pairing between vectors and
covectors of $M$.

In this section we shall not be concerned with the problem of defining
 consistent boundary conditions for the action (\ref{PS}), since it
 will be the central subject in the subsequent ones. Hence, in the
 remainder of this section we assume that $\Sigma$ is closed,
 i.e. $\partial\Sigma=\emptyset$.

If $X^{i}$ are local coordinates on $M$, $\sigma
^{\kappa},\ \kappa=1,2$ local coordinates on $\Sigma$, $\Pi^{ij}$ the
components of the Poisson structure in these coordinates and
$\eta_{i}=\eta_{i\kappa}\dd{\sigma}^{\kappa}$, the action reads
\begin{eqnarray} \label{PScoor}
S(X,\eta)=\int_\Sigma \eta_{i}\wedge \dd X^{i}+
\frac{1}{2}\Pi^{ij}(X)\eta_{i}\wedge \eta_{j}.
\end{eqnarray}

The equations of motion in the bulk are:
\begin{subequations}\label{eom}
\begin{align}
&\dd X^{i}+\Pi^{ij}(X)\eta_{j}=0 \label{eoma} \\ 
&\dd\eta_{i}+\frac{1}{2}\partial_{i}\Pi^{jk}(X)\eta_{j}
\wedge \eta_{k}=0. \label{eomb}
\end{align}
\end{subequations}

The necessity for the bivector field $\Pi$ to be a Poisson structure
is a consequence of the consistency of the equations of motion
(\ref{eom}). Differentiating (\ref{eoma}) we obtain:
\begin{equation}\notag
\partial_k\Pi^{ij}(X)\dd X^k\wedge\eta_j+\Pi^{ij}(X)\dd\eta_j=0
\end{equation}
which, using again (\ref{eoma}), yields
\begin{equation} \label{consisteom}
\Pi^{ij}(X)\dd\eta_j=\Pi^{kl}\partial_k\Pi^{ij}(X)\eta_l\wedge\eta_j.
\end{equation}
Contracting (\ref{eomb}) with $\Pi$ and applying
(\ref{consisteom}) gives
\begin{equation}
\frac{1}{2}\left(\Pi^{li}\partial_i\Pi^{jk}+\Pi^{ji}\partial_i\Pi^{kl}+
\Pi^{ki}\partial_i\Pi^{lj}\right)\eta_j\wedge\eta_k=0
\end{equation}
which holds for arbitrary $\eta$ only if $\Pi$ satisfies the
Jacobi identity (\ref{JacobiIdentity}).

The infinitesimal transformations
\begin{subequations}\label{symmetry}
\begin{align}
&\delta_{\epsilon}X^{i}=
\Pi^{ij}(X)\epsilon_{j}\label{symmetrya}\\
&\delta_{\epsilon}\eta_{i}=-\dd \epsilon_{i}-\partial_{i}\Pi^{jk}(X)\eta_{j}
\epsilon_{k}\label{symmetryb}
\end{align}
\end{subequations}
where $\epsilon(\sigma)=\epsilon_{i}(\sigma)\dd X^i$ is a section of
$X^*T^*M$, change the action (\ref{PScoor}) by a boundary term
\begin{eqnarray} \label{symmS}
\delta_{\epsilon}S=-\int_\Sigma \dd (\dd X^i \epsilon_i).
\end{eqnarray}
which vanishes if $\Sigma$ is closed. Up to trivial gauge
transformations (that is, symmetry transformations proportional to the
equations of motion) the expression (\ref{symmetry}) gives a complete
set of local symmetries for the PSM. In particular, the obvious
invariance of (\ref{PScoor}) under diffeomorphisms of $\Sigma$ should
be recovered by a suitable choice of the gauge parameter. If
$\sigma^\mu \mapsto \sigma^\mu + \xi^\mu(\sigma)$ is the infinitesimal
form of the diffeomorphism, we must take $\epsilon_i =
-\xi^\mu\eta_{i\mu}$ so that
\begin{subequations}\label{eq:diffsymmetry}
\begin{align}
&\delta_{\epsilon}X^{i}= {\cal L}_\xi X^i-\iota(\xi)(\dd
X^i+\Pi^{ij}(X)\epsilon_{j})\label{eq:diffsymmetrya}\\
&\delta_{\epsilon}\eta_{i}={\cal
L}_\xi\eta_i-\iota(\xi)(\dd\eta_i+\frac{1}{2}\partial_{i}\Pi^{jk}(X)\eta_{j}
\wedge\eta_{k})\label{eq:diffsymmetryb}
\end{align}
\end{subequations}
which on-shell gives the expected transformations of the fields.

It is worth pointing out that the form of the transformations
(\ref{symmetry}) is not invariant under a change of coordinates on
$M$. However, it is straightforward to check that they are
well-defined when evaluated on solutions of the equations of
motion. Let $X'^i(X^j)$ be a change of coordinates on $M$ and denote
by
\begin{equation}\notag
\epsilon'_i=\frac{\partial X^j}{\partial X'^i}\epsilon_j, \quad
\eta'_i=\frac{\partial X^j}{\partial X'^i}\eta_j, \quad
\Pi'^{ij}=\frac{\partial X'^i}{\partial X^k}\frac{\partial
X'^j}{\partial X^l}\Pi^{kl}
\end{equation}
the objects in the new coordinates $X'^i$. Using (\ref{symmetry}) we
can work out the transformation of the fields in these coordinates:
\begin{subequations}\notag
\begin{align}
&\delta_{\epsilon'}X'^{i}=
\Pi'^{ij}(X)\epsilon'_{j}\cr
&\delta_{\epsilon'}\eta'_{i}=-\dd
\epsilon'_{i}-\partial'_{i}\Pi'^{jk}(X)\eta'_{j}
\epsilon'_{k}-\frac{\partial X^j}{\partial X'^i}\frac{\partial^2
X'^r}{\partial X^j\partial X^l}(\dd
X^l+\Pi^{lk}\eta_k)\epsilon'_r
\end{align}
\end{subequations}
\addtocounter{equation}{-1}
\noindent which, on-shell, are of the form (\ref{symmetry}).

Also notice that
\begin{subequations}\label{OpenAlgebra}
\begin{align}
 &[\delta_\epsilon,\delta_{\epsilon'}]X^i=\delta_{[\epsilon,\epsilon']^*}
X^i \label{commga} \\
&[\delta_\epsilon,\delta_{\epsilon'}]\eta_i=
\delta_{[\epsilon,\epsilon']^*} \eta_i
-\epsilon_k\epsilon_{l}'\partial_i\partial_j
\Pi^{kl}(\dd X^{j}+\Pi^{js}(X)\eta_{s})\label{commgb}
\end{align}
\end{subequations}
where $[\epsilon,\epsilon']^{*}_k := -\partial_k\Pi^{ij}(X)
\epsilon_i\epsilon'_j$.  The term in parenthesis in \eqref{commgb} is
the equation of motion \eqref{eoma}. Hence, the commutator of two
transformations of type (\ref{symmetry}) is a transformation of the
same type only on-shell and the gauge transformations (\ref{symmetry})
form an open algebra. As we shall see in Chapter \ref{ch:quantumPSM}
this makes the quantization of the model quite involved, since the
standard BRST techniques do not work and the more sophisticated
Batalin-Vilkovisky procedure is needed.

\vskip 0.5cm

The PSM was introduced in \cite{SchStr94},\cite{Ike} as a
generalization of the underlying mathematical structure of
two-dimensional gravity models and Yang-Mills theories. In relation to
the latter, it is worth noticing that given a volume form
$\varepsilon$ on $\Sigma$ and a Casimir function $C$ on $M$ we can add
a non-topological term
\begin{equation} \label{nontopterm}
S_C(X)=\lambda\int_\Sigma\varepsilon C(X(\sigma)),\ \lambda\in{\mathbb
R}
\end{equation}
to the action (\ref{PScoor}) without spoiling the gauge symmetry
(\ref{symmetry}). The equation of motion (\ref{eoma}) does not change
whereas (\ref{eomb}) now reads:
\begin{equation}\notag
\dd\eta_{i}+\frac{1}{2}\partial_{i}\Pi^{jk}(X)\eta_{j}
\wedge \eta_{k}+\lambda\partial_iC(X)=0.
\end{equation}

2D Yang-Mills theories are related in this framework to linear Poisson
structures (see Example \ref{exLinPoi}). Let $M$ be a linear space and
$\Pi^{ij}(X)=f^{ij}_kX^k$. The Jacobi identity implies that $M$ is the
dual of a Lie algebra $\gl g$ of which $f^{ij}_k$ are the structure
constants. After integrating by parts, the topological action
(\ref{PScoor}) takes the form
\begin{equation} \label{chPS:BFtheory}
S_{{\rm lin}}=\int_\Sigma X^i F_i
\end{equation}
with $F_i:=\dd\eta_i+\frac{1}{2}f^{jk}_i\eta_j\wedge\eta_k$. The field
$\eta\in \Omega(\Sigma)\otimes {\gl g}$ can be viewed as a 1-form
connection and $F$ as the curvature two-form. The action
(\ref{chPS:BFtheory}) actually corresponds to BF-theory
(\cite{Hor}). A first order formulation of Yang-Mills theory is
obtained by adding to $S_{\rm lin}$ a term (\ref{nontopterm}) with
$C(X)=\sum_i X^iX^i$, the quadratic Casimir of ${\gl g}$. Namely,
\begin{equation}\notag
S^{fo}_{\rm YM}= \int_\Sigma X^i
\left(\dd\eta_i+\frac{1}{2}f^{jk}_i\eta_j\wedge\eta_k\right)+\lambda\varepsilon\sum_i
X^iX^i
\end{equation}
which can be seen to be equivalent to the ordinary second order
formulation of Yang-Mills theory
\begin{equation}
S_{\rm YM}=-\frac{1}{4\lambda}\int_\Sigma \tr (F\wedge *F)
\end{equation}
by using the field equations to get rid of $X$.

\vskip 5mm

It is remarkable that most two-dimensional models of pure gravity can
be formulated as a PSM (see \cite{Str00} for a thorough
discussion). As an example we work out the case of $R^2$-gravity,
whose action is given by
\begin{equation} \label{Rsquared}
S_{R^2}=\frac{1}{4}\int_\Sigma\sqrt{
{\rm det}(g)}\left(\frac{1}{4}R^2+1\right)d^2\sigma
\end{equation}
where $g$ is a (Riemannian) metric on $\Sigma$ and $R$ is the scalar
curvature of the torsionless connection compatible with $g$. The
action (\ref{Rsquared}) can be written as a PSM with $\mbox{dim}(M)=3$
via Einstein-Cartan variables. Let $e^a,a=1,2$ denote the zweibein and
$\omega$ the connection 1-form. The torsion two-form in these
variables reads\footnote{$\varepsilon=e^1\wedge e^2$ is the volume
form of the metric $g$, which in Einstein-Cartan variables is just
$g=e^1\wedge e^1+e^2\wedge e^2$.}  $De^a:=\dd
e^a+\varepsilon^a_b\omega e^b$. Thus one can see that (\ref{Rsquared})
is equivalent to the first-order action
\begin{equation} \label{foRsquared}
S^{\rm fo}_{R^2}=\int_\Sigma X^a De^a+ X^3\dd \omega
+\left(\frac{1}{4}-(X^3)^2\right)\varepsilon.
\end{equation}
With the identification $(\eta_1,\eta_2,\eta_3)\equiv(e^1,e^2,\omega)$ and
 after an integration by parts, the action (\ref{foRsquared}) is a PSM
 with Poisson structure given by (\ref{exnonlinPoisson}).

\vskip 5mm

Interestingly, the PSM can also be viewed as a limit of the action of
the bosonic string in a generic background of massless fields
(\cite{BauLosNek}). The latter reads (setting the dilaton field to
zero for simplicity)
\begin{equation} \label{eq:Sstring}
S_{\mbox{str}}=\frac{1}{4\pi\alpha'}\int_\Sigma\dd^2\sigma
\sqrt{\mbox{det}(h)}
\big[g_{ij}(X)h^{\mu\nu}+\epsilon^{\mu\nu}B_{ij}(X)\big] \partial_\mu
X^i\partial_\nu X^j
\end{equation}
where $h$ is the metric on $\Sigma$, $g$ the metric on the target
space $M$ and $B$ the two-form on $M$. Let us write a first order
action equivalent to (\ref{eq:Sstring}). To this end, introduce a
one-form $\eta$ on $\Sigma$ with values in $X^*(T^*M)$ and take
\begin{equation}\notag
S_{\mbox{str}}^{fo}=\int_\Sigma \eta_i\wedge\dd
X^i+\pi\alpha'G^{ij}(X)\eta_i\wedge
*\eta_j+\frac{1}{2}\Pi^{ij}(X)\eta_i\wedge\eta_j
\end{equation}
where
\begin{equation}\notag
(g+B)^{-1}=G+\frac{1}{2\pi\alpha'}\Pi.
\end{equation}

The Seiberg-Witten limit (\cite{SeiWit}) consists in taking the limit
$\alpha'\to 0$ keeping $G$ and $\Pi$ fixed. This implies
\begin{eqnarray}\notag
&&g\sim (2\pi\alpha')^2\Pi^{-1}G(\Pi^{t})^{-1},\cr\cr
&&B\sim 2\pi\alpha'\Pi^{-1}.
\end{eqnarray}

One can show (see \cite{BauLosNek}) that the conditions for the theory
to describe a conformal sigma model impose that in the Seiberg-Witten
limit $\dd B=0$ and then $\dd \Pi^{-1}=0$. Hence, in this limit we
obtain
\begin{equation}\notag
S_{\mbox{string}}^{fo}=\int_\Sigma  \eta_i\wedge\dd X^i +
\frac{1}{2}\Pi^{ij}\eta_i\wedge\eta_j
\end{equation}
which is a PSM with (symplectic) Poisson structure $\Pi$.

\subsection{Generalizations of the PSM}\label{sec:generalizationsPSM}

Several generalizations of the action (\ref{PScoor}) and the equations
of motion (\ref{eom}) have been proposed by Strobl {\it et al.}. In
this section we briefly discuss some of them.

\vskip 5mm

{\it -- WZ-Poisson sigma model.}

\vskip 2mm

In \cite{KliStr} a modification of the action (\ref{PScoor}) by adding
a WZ term
\begin{equation}\notag
S_{WZ}=\int_\Sigma \dd^{-1}H
\end{equation}
was considered. Here $H$ is a closed 3-form on $M$. The consistency of
the model leads to the interesting notion of twisted Poisson manifolds
(\cite{SevWei}), in which the Jacobi identity is violated by a term
proportional to $H$. We shall consider the WZ-Poisson sigma model and
twisted Poisson manifolds in Chapter \ref{ch:SusyWZP}.

\vskip 5mm

{\it -- Lie algebroid sigma models.}

\vskip 2mm

The idea behind this generalization of the PSM (due to Strobl) is the
observation that the equations of motion of the PSM (\ref{eom}) can be
interpreted as a morphism of Lie algebroids (in the sense described in
\cite{BojKotStr}) from $T\Sigma$ to $T^*M$. Hence, the aim is to
generalize the PSM so that the new theories, associated to general Lie
algebroids, yield equations of motion which have an analogous
geometrical interpretation. In this framework, Strobl has found new
gravity models (\cite{Str03}) and generalizations of Yang-Mills
theories (\cite{Str04}). Let us sketch what the mathematical concepts
underlying these constructions are.

Let us take a Lie algebroid $E$ of dimension $n$ over an
$m$-dimensional manifold $M$. If $\{X^i\}_{i=1}^m$ are local
coordinates on $M$ and $\{b_I\}_{I=1}^n$ denotes a local basis of $E$,
the bracket and anchor give rise to structure functions $c_{IJ}^K(X)$ and
$\rho_I^i(X)$,
\begin{equation}\notag
[b_I,b_J]=c_{IJ}^K b_K, \quad\quad \rho(b_I)=\rho_I^i\partial_i.
\end{equation}

The compatibility conditions in the definition of a Lie algebroid (see
Section \ref{sec:Diracstructures}) are translated into differential
equations for the structure functions:
\begin{subequations}
\begin{align}
&c_{IJ}^Sc_{KS}^L+c_{IJ,i}^L\rho_K^i+\mbox{cycl.}(IJK)=0
\label{eq:strucfuncLAa}\\
&c_{IJ}^K\rho_K^i-\rho_I^j\rho_{J,j}^i+\rho_J^j\rho_{I,j}^i=0.
\label{eq:strucfuncLAb}
\end{align}
\end{subequations}
where a comma denotes a partial derivative.

For the particular case of a Poisson manifold $M$, $E=T^*M$, $b_I\sim
\dd X^i$, $\rho_J^i\sim \Pi^{ji}$ and $c_{IJ}^K\sim
\partial_k\Pi^{ij}$. It is straightforward to check that in this case
(\ref{eq:strucfuncLAb}) gives the Jacobi identity and
(\ref{eq:strucfuncLAa}) its derivative, so that it is a consequence of
the former.

Although we shall not use it, let us mention that differential
calculus on $T^*M$ can be generalized to any Lie algebroid $E$. Take
the local basis of $E^*$, $\{b^I\}_{I=1}^n$, dual to
$\{b_I\}_{I=1}^n$. The differential
$\dd_E:\Omega^k_E(M)\rightarrow\Omega^{k+1}_E(M)$, with
$\Omega^k_E(M):=\Gamma(\wedge^kE^*)$, $\Omega^0_E(M):=C^\infty(M)$, is
defined by
\begin{eqnarray}\notag
&&\dd_E f = f_{,i}\rho_I^ib^I \cr\cr
&&\dd_E b^I=-\frac{1}{2}c_{JK}^Ib^J\wedge b^K
\end{eqnarray}

\noindent and extended by a graded Leibniz rule. One can show that, in
particular, $\dd_E^2=0$.

\vskip 5mm

The generalization of the equations of motion of the PSM is:
\begin{eqnarray}\label{eq:LAmorphisms}
&&\dd X^i-\rho_I^i\eta^I=0\cr
&&\dd\eta^I+\frac{1}{2}c_{JK}^I\eta^J\wedge\eta^K=0
\end{eqnarray}
and the generalization of the gauge symmetries
(\ref{symmetry}) (which also have a geometrical meaning in terms of
homotopies) is:
\begin{eqnarray}\notag
&&\delta_\epsilon X^i=\rho^i_I\epsilon^I\cr\cr
&&\delta_\epsilon\eta^I=\dd\epsilon^I+c_{JK}^I\eta^J\epsilon^K\quad
\mbox{(on-shell)}.
\end{eqnarray}

In \cite{Str03} an interpretation of these equations in terms of
gravity theories was discussed. The paper \cite{Str04} is devoted, as
said above, to a generalization of Yang-Mills theories in the context
of Lie algebroids. Actions whose equations of motion are
(\ref{eq:LAmorphisms}) can be constructed for arbitrary
$d=\mbox{dim}(\Sigma)$, but only through the introduction of Lagrange
multipliers in the general case. Namely,
\begin{equation}\notag
S=\int_\Sigma B_i\wedge(\dd
X^i-\rho_I^i\eta^I)+B_I\wedge(\dd\eta^I+\frac{1}{2}c_{JK}^I\eta^J\wedge\eta^K)
\end{equation}
where $B_i,B_I$ are respectively $(d-2)$- and $(d-1)$- forms. Only for
very special choices of $E$ and $\mbox{dim}(\Sigma)$ one can construct
topological actions yielding Lie algebroid morphisms up to
homotopies. For $\mbox{dim}(\Sigma)=2$ and $E=T^*M$ one has the PSM,
whereas for $\mbox{dim}(\Sigma)=3$ and $E$ a quadratic Lie algebra we
get the Chern-Simons theory.

\vskip 5mm

{\it -- Dirac sigma models.}

\vskip 2mm

The Dirac sigma model, introduced in \cite{KotSchStr}, is a
generalization of the $G/G$ WZW model and the WZ-Poisson sigma model
at the same time. The fields of the model are given by a bundle map
from the tangent bundle of a surface $\Sigma$ to a Dirac structure
$D$. In order to ensure that the equations of motion produce Lie
algebroid morphisms from $T\Sigma$ to $D$, metric tensors both on
$\Sigma$ and $M$ are needed, acting as a sort of regulators.

These models are mathematically appealing, and in principle might be
related to the quantization of Dirac structures in the same way as the
PSM is related to the quantization of Poisson manifolds. However, at
the present time the applications of the Dirac sigma models are still
to be investigated.

\section{Classically admissible branes} \label{sec:branes}

In this section we carry out the classical study of the PSM defined on
a surface with boundary and search for the boundary conditions (BC)
which make the theory consistent.

In order to preserve the topological character of the theory one must
choose the BC independent of the point of the boundary, as far as we
move along one of its connected components.  For the sake of clarity
we shall restrict ourselves in this section to one connected
component. In the next section we shall discuss the relation between
the BC in the possible different connected components of the boundary.

In surfaces with boundary a new term appears in the variation of the
action (\ref{PScoor}) under a change of $X$ when performing the
integration by parts:
\begin{eqnarray}\notag
\delta_{X} S=-\int_{\partial\Sigma}\delta X^i\eta_{i}+
\int_\Sigma\delta X^i(\dd\eta_{i}+
\frac{1}{2}\partial_{i}\Pi^{jk}(X)\eta_{j}\wedge \eta_{k})
\end{eqnarray}

The BC must be such that the boundary term vanishes. Let us take the
field
\begin{eqnarray}\label{BCX}
X|_{\partial \Sigma}:\partial \Sigma\rightarrow N
\end{eqnarray}
for an arbitrary (for the moment) closed submanifold $N$ of $M$
(brane, in a more stringy language). Cattaneo and Felder concluded in
\cite{CatFel03} that maximally symmetric branes are given by the
coisotropic submanifolds of $M$. In the sequel we show that much more
general branes are admissible, however.

Denote by $\eta_{\bf t} = \eta_{i\bf t}\dd X^i$ the contraction of
$\eta$ with vector fields tangent to $\partial\Sigma$. The BC
(\ref{BCX}) implies that $\delta X\in T_{_X} N$ at every point of the
boundary, and consequently $\eta_{\bf t}$ must belong to $T_{_X}N^0$
(the fiber over $X$ of the conormal bundle of $N$).

On the other hand, by continuity, the equations of motion in the bulk
must be satisfied also at the boundary. In particular,
$$\partial_{\bf t} X=\Pi^\sharp\eta_{\bf t}$$ where by $\partial_{\bf
t}$ we denote the derivative in the direction of a vector on $\Sigma$
tangent to the boundary.  As $\partial_{\bf t} X$ belongs to $T_{_X}N$
it follows that $\eta_{\bf t}\in\Pi^{\sharp-1}_{_X}(T_{_X}N)$.

Both conditions for $\eta_t$ imply that
\begin{eqnarray}\label{BCpsi}
\eta_{\bf t}(u)\in\Pi^{\sharp-1}_{_{X(u)}}(T^{}_{_{X(u)}}N)\cap T_{_{X(u)}}N^0,
\mbox{ for any }u\in \partial\Sigma 
\end{eqnarray}
which is the boundary condition we shall take for $\eta_{\bf t}$. We
should check now that the BC are consistent with the gauge
transformations (\ref{symmetry}).

In order to cancel the boundary term (\ref{symmS})
$\epsilon\vert_{\partial\Sigma}$ must be a smooth section of
$TN^0$ and if (\ref{symmetry}) is to preserve the boundary
condition of $X$, $\epsilon\vert_{\partial\Sigma}$ must belong to
$\Pi^{\sharp-1}(TN)$. Hence,
\begin{eqnarray}\label{BCepsilon}
\epsilon(u)=\Pi^{\sharp-1}_{_{X(u)}}(T^{}_{_{X(u)}}N)\cap T_{_{X(u)}}N^0,
\mbox{ for any }u\in \partial\Sigma.
\end{eqnarray}

Next, we shall show that the the gauge transformations
(\ref{symmetry}) with (\ref{BCepsilon}) also preserve (\ref{BCpsi})
provided that the brane satisfies some (mild) regularity
conditions. Let us restrict ourselves to {\it weakly regular branes},
i.e. branes verifying the weak regularity condition
(\ref{weakregularity}). This allows to choose adapted coordinates
$(X^a,X^\mu,X^A)$ satisfying the properties (i), (ii), (iii)$'$ of
Section \ref{reduction}. Recall that in these coordinates the Poisson
tensor satisfies:
\begin{eqnarray}\notag
&&\Pi^{\mu\nu}\vert_N = 0,\quad \Pi^{\mu A}\vert_N = 0\cr
&&{}\cr
&&{\rm det}(\Pi^{AB})\vert_N \neq 0 \quad \text{in an open dense set}.
\end{eqnarray}

The boundary condition (\ref{BCpsi}) translates in these coordinates
into $\eta_{\bf t} = \eta_{\mu {\bf t}}\dd X^\mu$. Hence, we must show
that $\delta\eta_{a {\bf t}}=\delta\eta_{A{\bf t}} = 0$. Recalling
(\ref{BCepsilon}) we also may write $\epsilon\vert_N = \epsilon_\mu
\dd X^\mu$ and therefore,
$$\delta\eta_{a {\bf t}}=
\partial_a \Pi^{\mu\nu}\vert_N\eta_{\mu {\bf t}}\epsilon_\nu\vert_N$$
which vanishes because $\Pi^{\mu\nu}\vert_N = 0 \Rightarrow
\partial_a\Pi^{\mu\nu}\vert_N = 0$.

Showing that
$$\delta\eta_{A{\bf t}}=\partial_A\Pi^{\mu\nu}\vert_N\eta_{\mu
{\bf t}}\epsilon_\nu\vert_N$$
also vanishes on $N$ is more tricky, but it
does, as a consequence of the Jacobi identity:
\newpage
\begin{eqnarray}\notag
\Pi^{AB} 
\partial_A\Pi^{\mu\nu}
+
\Pi^{a B} 
\partial_a\Pi^{\mu\nu}
+
\Pi^{\rho B} 
\partial_\rho\Pi^{\mu\nu}
&&\cr
+ \Pi^{A\nu} 
\partial_A\Pi^{B\mu}
+
\Pi^{a B} 
\partial_a\Pi^{B\mu}
+
\Pi^{\rho\nu} 
\partial_\rho\Pi^{B\mu}
&&\cr
+ \Pi^{A\mu} 
\partial_A\Pi^{\nu B}
+
\Pi^{a \mu} 
\partial_a\Pi^{\nu B}
+
\Pi^{\rho\mu} 
\partial_\rho\Pi^{\nu B}
&=& 0.
\end{eqnarray}

Evaluating on $N$ and using $\Pi^{\mu\nu}\vert_N = 
\Pi^{\mu A}\vert_N=0$ and $\partial_a\Pi^{\mu\nu}\vert_N = 
\partial_a\Pi^{\mu A}\vert_N=0$,
one may check that all terms except the first one vanish. Then,
$$\Pi^{AB}\vert_N \partial_A\Pi^{\mu\nu}\vert_N = 0.$$

Using now that $\Pi^{AB}\vert_N$ is invertible in an open dense subset
and by continuity, we conclude that $\partial_A\Pi^{\mu\nu}\vert_N =
0$ and then $\delta\eta_{A{\bf t}} = 0$.  A similar derivation proves
that the gauge transformations close on-shell at the boundary (see
(\ref{commgb})).

\vskip 0.2cm

Summarizing, the {\it weakly regular submanifolds} of $M$ are
classically admissible branes for the PSM. Notice that for a weakly
regular submanifold $N$ the differentials of the $X^\mu$ coordinate
functions span at every point $X\in N$ a subspace of $T_X^*M$ which
coincides with $\{(\dd f)_X | f\in{\cal F}\cap{\cal I}\}$. Hence, we can
write the boundary conditions imposed on the fields as:
\vskip 0.2cm

(XBC)\qquad $X(u) \in N,\forall u\in \partial\Sigma$

\vskip 0.2cm

($\eta$BC)\qquad $\eta_{\bf t}(u)\in\{(\dd f)_{X(u)} | f\in{\cal
F}\cap{\cal I}\}, \ \forall u\in \partial\Sigma$

\vskip 0.2cm

\noindent and in order to have gauge transformations compatible at the
boundary with the BC one must have

\vskip 0.2cm

($\epsilon$BC)\qquad $\epsilon(u)\in\{(\dd f)_{X(u)} | f\in{\cal
F}\cap{\cal I}\}, \ \forall u\in \partial\Sigma$.

\section{Hamiltonian study of the PSM} \label{sec:HamPSM}

We proceed now to the Hamiltonian study of the model with the BC of
the previous section (in each connected component of the boundary)
when $\Sigma = {\mathbb R}\times [0,1]$ (open string). The fields in
the Hamiltonian formalism are a smooth map $X:[0,1]\rightarrow M$ and
a 1-form $\eta$ on $[0,1]$ with values in the pull-back $X^*T^*M$;
in coordinates, $\eta=\eta_{i} \dd X^i \dd\sigma$.

The space of maps $(X,\eta)$ can be given the structure of a Banach
manifold by requiring $X$ to be differentiable and $\eta$ to be
continuous. This infinite dimensional manifold is equipped with a
canonical symplectic structure $\omega$. The action of $\omega$ on two
vector fields (denoted for shortness $\delta,\delta'$) reads
\begin{eqnarray} \label{Omega}
\omega(\delta,\delta') = \int_{0}^{1}(\delta X^i \delta' 
\eta_{i} - \delta' X^i \delta \eta_{i})\dd\sigma.
\end{eqnarray}

The phase space ${\cal P}(M;N_0,N_{1})$ of the theory is defined by
the constraints:
\begin{eqnarray} \label{constraint}
\partial X^i + \Pi^{ij}(X)\eta_{j} = 0,\ i=1,\dots,{\rm dim}(M).
\end{eqnarray}
and the BC: $X(0)\in N_0$ and $X(1)\in N_1$ for two closed
submanifolds $N_{u}\subset M$, ${u}=0,1$. We use the notation
$\partial\equiv\partial_\sigma.$

This geometry, with a boundary consisting of two connected components,
raises the question of the relation between the BC at both ends. Note
that due to (\ref{constraint}) the field $X$ varies within a
symplectic leaf of $M$. This implies that in order to have solutions
the symplectic leaf must have non-empty intersection both with $N_0$
and $N_{1}$.  In other words, only points of $N_0$ and $N_{1}$ that
belong to the same symplectic leaf lead to points of ${\cal
P}(M;N_0,N_{1})$. In the following we shall assume that this condition
is met for every point of $N_0$ and $N_1$ and correspondingly for the
tangent spaces. That is, if we denote by $J_0, J_1$ the maps
\begin{eqnarray}\notag
J_0:{\cal P}(M;N_0,N_1)&\longrightarrow & N_0\cr
(X,\eta)&\longmapsto& X(0)
\end{eqnarray}
\begin{eqnarray}\notag
J_1:{\cal P}(M;N_0,N_1)&\longrightarrow & N_1\cr
(X,\eta)&\longmapsto& X(1)
\end{eqnarray}
we are assuming that both maps are surjective submersions.

Vector fields tangent to the phase space satisfy the linearization of
(\ref{constraint}), i.e. $\delta\eta_{j}$ and $\delta X^i$ 
are such that
\begin{eqnarray} \label{linearization}
\partial\delta X^i = \partial_j\Pi^{ki}(X)
\eta_{k}\delta X^j + \Pi^{ji}\delta \eta_{j}
\end{eqnarray}
with $\delta X(u)\in T_{_{X(u)}}N_u$, $u=0,1$.

The solution to the differential equation (\ref{linearization}) is
\begin{eqnarray} \label{solution}
\delta X^{i}(\sigma)= R^{i}_{j}(\sigma,0)\delta X^{j}(0) - 
\int_{0}^{\sigma}R^{i}_{j}(\sigma,\sigma')\Pi^{jk}(X(\sigma'))
\delta \eta_{k}(\sigma')d\sigma'
\end{eqnarray}
where $R$ is given by the path-ordered integral
$$
R(\sigma,\sigma')
= \overleftarrow{P {\rm exp}
[\int_{\sigma'}^{\sigma}A(z)dz]},
\qquad A^{i}_{j}(z)=(\partial_j\Pi^{ki})(X(z))\eta_{k}(z)$$
and the boundary conditions are such that $\forall \xi_0\in T_{X(0)}N_0^0,\ \forall \xi_1\in T_{X(1)}N_1^0$,
\begin{eqnarray} \label{BCtangentvectors}
&&\xi_{0i}\delta X^i(0) = 0,\cr
&&{}\cr
&&\xi_{1 i}\left(R^{i}_{j}(1,0)\delta X^{j}(0) -
\int_{0}^{1}R^{i}_{j}(1,\sigma')\Pi^{jk}(X(\sigma')) \delta
\eta_{k}(\sigma')d\sigma'\right)=0.\qquad\quad
\end{eqnarray}

The canonical symplectic 2-form restricted to ${\cal P}(M;N_0,N_{1})$,
$\omega_P$, is only presymplectic. It is instructive to give a
detailed calculation of its kernel. Assume $(\delta X(\sigma), \delta
\eta(\sigma))\in {\rm Ker}(\omega_P)$. Then,
$\omega(\delta,\delta')=0$ for any $(\delta' X(\sigma), \delta'
\eta(\sigma))$ tangent to ${\cal P}(M;N_0,N_{1})$. Using
(\ref{solution}) and with a change in the order of integration we get:
\begin{eqnarray}\notag
&&0= -\delta'X^j(0)\int_0^1
 R^i_j(\sigma,0)\delta\eta_i(\sigma)\dd\sigma+ \cr
 &&+\int_0^1\left[\delta X^i(\sigma)+\int_\sigma^1
 R^k_j(\sigma',\sigma)\Pi^{ji}(X(\sigma))\delta\eta_k(\sigma')\dd\sigma'\right]\delta'\eta_i(\sigma)\dd\sigma.
\end{eqnarray}

{}From (\ref{BCtangentvectors}) we deduce that there exist covectors
$\xi_0\in T_{X(0)}N_0^0$ and $\xi_1\in T_{X(1)}N_1^0$ such that
\begin{eqnarray}
&&-\int_0^1 R^i_j(\sigma,0)\delta\eta_i(\sigma)\dd\sigma =
 \xi_{0j}+\xi_{1 i}R^{i}_{j}(1,0),\cr
&&{}\cr
&&\delta X^i(\sigma)+\int_\sigma^1
 R^k_j(\sigma',\sigma)\Pi^{ji}(X(\sigma))\delta\eta_k(\sigma')\dd\sigma'=
 -\xi_{1 k}R^{k}_{j}(1,\sigma)\Pi^{ji}(X(\sigma)).\notag
\end{eqnarray}

Now, defining
\begin{equation} \label{defepsilon}
\epsilon_j(\sigma) = \int_\sigma^1
R^i_j(\sigma',\sigma)\delta\eta_i(\sigma')\dd\sigma'+\xi_{1
i}R^{i}_{j}(1,\sigma)
\end{equation}
we can write
\begin{equation}\notag
\delta X^i(\sigma)=\Pi^{ij}(X)\epsilon_j(\sigma).
\end{equation}

Finally, differentiating (\ref{defepsilon}) we find
\begin{equation}\notag
\delta\eta_i(\sigma)=-\partial\epsilon_i(\sigma)-
\partial_i\Pi^{jk}(X(\sigma))\eta_j(\sigma)\epsilon_k(\sigma).
\end{equation}

Hence, the kernel is given by:
\begin{eqnarray} \label{Hamsymm}
&&\delta_{\epsilon}X^{i}=\Pi^{ij}(X)\epsilon_{j}\cr
&&\delta_{\epsilon}\eta_{i}=-\partial\epsilon_{i}-
\partial_{i}\Pi^{jk}(X)\eta_{j} \epsilon_{k}
\end{eqnarray}
where $\epsilon$ is subject to the BC
$$\epsilon({u})\in \Pi_{_{X({u})}}^{\sharp -1}(T^{}_{_{X({u})}} N_{u})
\cap (T_{_{X({u})}}N_{u})^0,\quad {\rm for}\ {u}=0,1$$
because (\ref{Hamsymm}) must be tangent to ${\cal P}(M;N_0,N_1)$.

As discussed in Example \ref{expresymp} the presymplectic structure
induces a Poisson algebra ${\cal A}_\omega$ on the phase space ${\cal
P}(M;N_0,N_1)$. On the other hand, the canonical reduction of $\Pi$
explained in Section \ref{reduction} defines Poisson algebras in $N_0$
and $N_1$. We turn now to study the relation between them. First, we
wish to figure out when a function on the phase space of the form
$F(X,\eta)=f(X(0))$, $f\in{C}^\infty(M)$ belongs to ${\cal A}_\omega$,
i.e. when it has a Hamiltonian vector field $\delta_F$.  Solving the
corresponding equation
\begin{equation}\notag
\int_0^1(\delta_FX^i\delta'\eta_i-\delta'X^i\delta_F\eta_i)\dd\sigma
= \partial_i f(X(0))\delta'X^i(0)
\end{equation}
along the same lines as above one can show that
the general solution is of the form (\ref{Hamsymm}) with
\begin{equation}\label{cond0}
\epsilon(0)-\dd f_{_{X(0)}}\in
T_{_{X(0)}}N_0^0,
\qquad\epsilon(0)\in \Pi_{_{X(0)}}^{\sharp -1}(T^{}_{_{X(0)}} N_0)
\end{equation}
and  
$$\epsilon(1)\in \Pi_{_{X(1)}}^{\sharp -1}(T^{}_{_{X(1)}} N_1)
\cap T_{_{X(1)}}N_1^0.$$

As we have already learnt (see proof of Theorem \ref{imagegaugeinv}),
equation (\ref{cond0}) can be solved in $\epsilon(0)$ if and only if
$F$ is a gauge invariant function (i.e. it is invariant under
(\ref{Hamsymm})).  This is equivalent to saying that $f+{\cal I}_0$
belongs to the Poisson algebra ${\cal C}(\Pi,M,N_0)$.  (Here ${\cal
I}_0$ is the ideal of functions that vanish on $N_0$).

Given two such functions $F_1$ and $F_2$ associated to the classes
$f_1+{\cal I}_0$ and $f_2+{\cal I}_0 \in{\cal C}(\Pi,M,N_0)$ and with
gauge parameter $\epsilon_1$ and $\epsilon_2$ respectively, one
immediately computes the Poisson bracket
$\{F_1,F_2\}_P=\omega(\delta_{F_1},\delta_{F_2})$ to give
\begin{equation}\notag
\{F_1,F_2\}_P=\Pi^{ij}\epsilon_{1i}(0)\epsilon_{2j}(0)
\end{equation}

This coincides with the restriction to $N_0$ of $\{f_1+{\cal
I}_0,f_2+{\cal I}_0\}_{N_0}$ and defines a Poisson homomorphism
between ${\cal C}(\Pi,M,N_0)$ and the Poisson algebra of ${\cal
P}(M,N_0,N_1)$. This homomorphism is $J_0^*$, the pull-back defined by
$J_0$, and the latter turns out to be a Poisson map. In an analogous
way we can show that $J_1$ is an anti-Poisson map and besides
$$\{f_0\circ J_0, f_1\circ J_1\}=0\quad {\rm for\ any}\  
f_{u}\in{\cal C}(\Pi,M,N_{u}),\ {u}=0,1.$$

The previous considerations 
can be summarized in the following diagram
\begin{equation}\notag
\begin{matrix}
&\displaystyle
J^*_0
&\displaystyle
&\displaystyle
J^*_1
&\displaystyle
\cr
{\cal C}(\Pi,M,N_0)
&\displaystyle
\longrightarrow
&\displaystyle
{\cal A}_\omega
&\displaystyle
\longleftarrow
&\displaystyle
{\cal C}(\Pi,M,N_1)
\cr
\end{matrix}
\end{equation}
in which $J_0^*$ is a Poisson homomorphism, $J_1^*$ antihomomorphism
and the image of each map is the commutant (with respect to the
Poisson bracket) of the other. In particular it implies that the
reduced phase space is finite-dimensional. This can be considered as a
generalization of the symplectic dual pair to the context of Poisson
algebras.

\subsection{Symplectic groupoid structure of the phase space}

As we already know, a Poisson bracket on a manifold $M$ defines a Lie
bracket on the space of 1-forms on $M$. This bracket is determined by:
\begin{enumerate}
\item[(i)] $[\dd f,\dd g]=\dd \{f,g\},\ f,g\in C^\infty(M)$
\vspace{-0.1cm}
\item[(ii)] $[\alpha,f\beta]=f[\alpha,\beta]+\Pi^\sharp\beta(f)\alpha,\
\alpha,\beta\in\Omega^1(M),f\in C^\infty(M)$
\end{enumerate}
which makes $T^*M$ into a Lie algebroid over $M$ with anchor
$\rho = \Pi^\sharp$.

\vskip 2mm

As mentioned in the introductory chapter, Cattaneo and Felder proved
in \cite{CatFel00} that the reduced phase space of the Poisson sigma
model defined on $\Sigma = [0,1]\times {\mathbb R}$ and with $X$ free
at the boundary can be viewed (in the integrable case) as the
symplectic groupoid for the Poisson manifold $M$. In this section we
recall the basic concepts and constructions and study what happens
when a second-class brane is introduced.

The global version of Lie algebroids are Lie groupoids
(\cite{CanWei}). A {\it Lie groupoid} over a manifold $M$ is a
manifold ${\cal G}$ together with the following structure maps (all in
the smooth category) satisfying a set of axioms. An injection
$j:M\hookrightarrow {\cal G}$, two surjections $l,r:{\cal
G}\rightarrow M$ and a product on pairs $g,h\in\cal G$ defined only if
$r(g)=l(h)$. Denote ${\cal G}_{x,y}=l^{-1}(x)\cap r^{-1}(y)$. These
maps must verify:
\begin{enumerate}
\item[(i)] $l \circ j = r \circ j = {\rm id}_M$.
\vspace{-0.2cm}
\item[(ii)] If $g\in{\cal G}_{x,y}$ and $h\in{\cal G}_{y,z}$, then
$gh\in{\cal G}_{x,z}$.
\vspace{-0.2cm}
\item[(iii)] $j(x)g=gj(y)=g$, if $g\in{\cal G}_{x,y}$.
\vspace{-0.2cm}
\item[(iv)] For every $g\in{\cal G}_{x,y}$ there exists
$g^{-1}\in{\cal G}_{y,x}$ such that $gg^{-1}=j(x)$.
\vspace{-0.2cm}
\item[(v)] The product on $\cal G$ is associative whenever it is defined.
\end{enumerate}

Some examples of Lie groupoids are:
\begin{example}
$M$ a point and $\cal G$ a Lie group. $j$ maps to the identity of the
group and $l,r$ are trivial.
\end{example}
\begin{example}
${\cal G}$ a vector bundle over $M$, $l=r$ the projection, $j$
the zero section, the multiplication given by $(p,v_1)(p,v_2)=(p,v_1+v_2)$
and the inverse $(p,v)^{-1}=(p,-v)$.
\end{example}
\begin{example}
${\cal G}=M\times M$. Then, $l,r$ are projections onto the left and
right components respectively and $j$ the diagonal map. The
multiplication is defined by $(p_1,p_2)(p_2,p_3)=(p_1,p_3)$ and the
inverse by $(p_1,p_2)^{-1}=(p_2,p_1)$.
\end{example}

Assume that $\cal G$ is equipped with a symplectic structure
$\omega$. $\cal G$ is called the symplectic groupoid of the Poisson
manifold $M$ if the following axioms are also satisfied:
\begin{enumerate}
\item[(vi)] $j(M)$ is a Lagrangian (i.e. maximally isotropic)
submanifold.
\vspace{-0.2cm}
\item[(vii)] $l$ is a Poisson map and $r$ is anti-Poisson.
\vspace{-0.2cm}
\item[(viii)] Define ${\cal G}_0=\{(g,h)\in {\cal G}\vert r(g)=l(h)\}$
and let $\tau_1,\tau_2$ the projections onto the first and second
factor. If $\mu:{\cal G}_0\rightarrow {\cal G}$ denotes the product on
${\cal G}_0$, then $\mu^*\omega =\tau^*_1\omega+\tau^*_2\omega$.
\vspace{-0.2cm}
\item[(ix)] The inverse map is anti-Poisson.
\end{enumerate}

The basic example of a symplectic Lie groupoid is:
\begin{example}
$M={\mathfrak g}^*$ the dual of a Lie algebra with the
Kostant-Kirillov Poisson structure. Define $j:{\mathfrak
g}^*\rightarrow T^*G$ as the inclusion for any Lie group integrating
the Lie algebra ${\mathfrak g}$. The maps
$l,r:T^*G\rightarrow{\mathfrak g}^*$ are naturally defined by left
(right) translations to the cotangent space at the identity of
$G$. Take two elements $(g_1,\xi_1)$ and $(g_2,\xi_2)$ such that
$r(g_1,\xi_1)=l(g_2,\xi_2)$. The multiplication law is given by
$(g,\xi)=(g_1,\xi_1)(g_1,\xi_1)$ with $g=g_1g_2$ and $\xi=(\dd
R_h(g)^*)^{-1}\xi_1=(\dd L_g(h)^*)^{-1}\xi_2$.
\end{example}

\vskip 0.5cm

The question of integrability for general Lie algebroids was solved by
Crainic and Fernandes in \cite{CraFer2}, much inspired by the work of
Cattaneo and Felder \cite{CatFel00} on the particular case of Poisson
manifolds. They proved that the reduced phase space of the PSM, ${\cal
G}:={\cal P}(M;M,M)/\sim$ (where the quotient is taken with respect to
the gauge transformations (\ref{Hamsymm}) with $\epsilon$ vanishing at
the boundary) is the symplectic groupoid of the Poisson manifold $M$
(in the integrable case, or equivalently, when ${\cal G}$ is a
manifold).

The algebraic groupoid structure is defined by composition of
paths. One can show (\cite{CatFel00}) that in each equivalence class
$[(X,\eta)]$ in $\cal G$ there exists a representative with
$\eta(0)=\eta(1)=0$. Then, the composition law
$[(X,\eta)]=[(X_1,\eta_1)][(X_2,\eta_2)]$ is given by choosing such
representatives and
\begin{equation}\notag
X(\sigma)=
\begin{cases}
X_1(2\sigma),\quad 0\leq\sigma\leq\frac{1}{2}\cr
X_2(2\sigma-1),\quad \frac{1}{2}\leq\sigma\leq 1
\end{cases}
\end{equation}
\begin{equation}\notag
\eta(\sigma)=
\begin{cases}
2\eta_1(2\sigma),\quad 0\leq\sigma\leq\frac{1}{2}\cr
2\eta_2(2\sigma-1),\quad \frac{1}{2}\leq\sigma\leq 1
\end{cases}
\end{equation}
as long as $X_1(1)=X_2(0)$.

In this case the remaining objects entering the definition of a
symplectic groupoid are as follows. The map $j$ sends a point $x\in M$
to the class of the constant solution $X(\sigma)=x,\eta(\sigma)=0$ and
the maps $r,l$ give the values of $X$ at the endpoints. Namely,
$l(X,\eta)=X(0),r(X,\eta)=X(1)$. The symplectic structure is provided
by (\ref{Omega}), which is obviously symplectic on $\cal G$. We refer
the reader to \cite{CatFel00} for the explicit proof that ${\cal G}$
satisfies (i)-(ix).

\vskip 0.5cm

It would be interesting to investigate what happens when we set a
second-class brane at both endpoints of the string. Hence, let us take
$N$ a second-class brane. ${\cal P}(M;N,N)$ is given by paths
\begin{eqnarray} \label{constraintSCB}
\partial X^i + \Pi^{ij}(X)\eta_{j} = 0,\ i=1,\dots,{\rm dim}(M).
\end{eqnarray}
satisfying $X^A(0)=X^A(1)$ in adapted coordinates. The characteristic
foliation of $\omega$ restricted to ${\cal P}(M;N,N)$ is of the form
(\ref{Hamsymm}) with $\epsilon(0)=\epsilon(1)=0$. We want to study the
structure of the reduced phase space ${\cal P}(M;N,N)/\sim$.

Locally, one can choose $\epsilon_A(\sigma)$ so that
$X^A(\sigma)=0,\forall\sigma\in [0,1]$. This is easily seen by
writing
\begin{equation}
\delta_\epsilon X^A=\Pi^{Aa}\epsilon_a+\Pi^{AB}\epsilon_B
\end{equation}
and recalling that $\Pi^{AB}$ is invertible in a neighborhood of $N$.

The characteristic foliation is tangent to ${\cal P}(M;N,N)$ and a
representative satisfying $X^A(\sigma)=0$ still must obey
(\ref{constraintSCB}), which implies
\begin{equation}
\eta_A=-\omega_{AB}\Pi^{Ba}\eta_a
\end{equation}
which gives, for lower-case indices:
\begin{equation} \label{branegroupoid}
\partial X^a = -\Pi_{\cal D}^{ab}\eta_b
\end{equation}
where $\Pi_{\cal D}$ is the Dirac bracket (\ref{DiracComponents}).

The remaining freedom in the choice of representative is given by
gauge transformations which preserve $X^A(\sigma)=0$. This corresponds
to those $\epsilon$ verifying
$\epsilon_A=-\omega_{AB}\Pi^{Ba}\epsilon_a$ and leaves the freedom for
lower-case indices:
\begin{equation} \label{branegroupoid2}
\delta_\epsilon X^a=\Pi_{\cal D}^{ab}\epsilon_b
\end{equation}

Analogously, one can check that the remaining freedom for $\eta_a$ is
\begin{equation} \label{branegroupoid3}
\delta_\epsilon\eta_a = -\partial\epsilon_a-\partial_a\Pi^{bc}_{\cal
D}\eta_b\epsilon_c
\end{equation}

Hence, from (\ref{branegroupoid}), (\ref{branegroupoid2}),
(\ref{branegroupoid3}) we deduce that ${\cal P}(M;N,N)/\sim$ can be
viewed (if smooth), up to topological obstructions, as the symplectic
groupoid integrating the brane equipped with the Dirac bracket. Not
only it might happen that $\Pi^{AB}$ be degenerate for paths far
enough from the brane, but the obstructions might appear in a
different, more essential way. Notice that if the intersection of $N$
with a symplectic leaf is disconnected, there exist paths with can not
be taken to lie on $N$ by means of gauge transformations. Below we
give an example of a Poisson manifold and a second-class brane where
this obstacle shows up.

\begin{example}
Consider $M={\mathbb R}^3$ with coordinates $(x,y,z)$ and Poisson
structure defined by
\begin{equation}\notag
\{x,z\}=x,\quad \{y,z\}=y,\quad \{x,y\}=0
\end{equation}
\end{example}

The symplectic leaves of this Poisson structure are the level sets of
the Casimir function $\theta =\mbox{arctan}(y/x)$. Take $N$ a closed
curve (without self-intersections) winding twice around the $z$
axis. It is clear that $N$ can be chosen so that it is second-class
and has disconnected interesection with some symplectic leaves.

\newpage
\pagestyle{empty}
\chapter{Branes in the quantum Poisson sigma model}
\label{ch:quantumPSM}
\thispagestyle{empty}
\pagestyle{empty}
\cleardoublepage

\pagestyle{headings}
\renewcommand*{\chaptermark}[1]{\markboth{\small \scshape  \thechapter.\ #1}{}}
\renewcommand*{\sectionmark}[1]{\markright{\small \scshape \thesection. \ #1}}

In \cite{CatFel99} Cattaneo and Felder gave a field theoretical
interpretation of Kontsevich's formula (\cite{Kon}) for the
deformation quantization of a Poisson manifold. They showed that
Kontsevich's formula can be obtained from Feynman expansion of certain
Green's functions of the PSM when $\Sigma$ is the unit disk $D$ and
the base map $X:\Sigma \rightarrow M$ has free boundary conditions.

We have proven in Section \ref{sec:branes} that classically the field
$X$ can be consistently restricted at the boundary $\partial\Sigma$ to
an almost arbitrary submanifold $N$. Then, we have shown in Section
\ref{sec:HamPSM} that the symplectic structure on the reduced phase
space of the model is related to the Poisson bracket canonically
induced on (a subset of) $C^\infty(N)$.

On the light of these results it is natural to conjecture that the
perturbative quantization of the model with general $N$ be related to
the deformation quantization of the induced Poisson bracket on
(certain functions on) $N$. We shall work out in detail the case in
which $N$ can be defined by a set of {\it second-class constraints}
which is, in some sense, opposite to the coisotropic one. The
quantization of the coisotropic case (\cite{CatFel03}) presents some
intricacies due to the fact that gauge transformations do not vanish
at the boundary (see Section \ref{sec:quantumcoisotropic}). If $N$ is
defined by second-class constraints ({\it second-class brane}) they do
vanish and one would expect to have a clean quantization recovering
Kontsevich's formula, this time not for $\Pi$ but for the Dirac
bracket on $N$. We show that this expected result holds and that it
emerges in quite a different way from the coisotropic case. Finally,
we shall give the quantization of the PSM with a general brane defined
by a mixture of first and second class constraints.

\section{Quantization of gauge theories}

The content of this section is now standard material in modern Quantum
Field Theory. However, the Batalin-Vilkovisky scheme, needed for the
quantization of gauge theories with open algebras like the PSM, is
much less known than its simplified version for closed algebras (the
standard BRST method). That is why we find it useful to give a brief
introduction to this topic. The exposition is based on the excellent
reference (\cite{Jon}).

\subsection{Gauge algebras}

Let $\varphi^i$ be the fields of our theory and $S[\varphi]$ the action
functional. With the notation
\begin{equation} \notag
y_i(\varphi)=\frac{\overleftarrow{\delta}S}{\delta\varphi^i}
\end{equation}
the stationary points of $S$ are given by the field equations
$y_i=0,\forall i$.

A symmetry transformation of the action $S$ is given by a set of
independent operators $R^i_\alpha(\varphi)$ verifying
\begin{equation} \label{defofsymm}
y_i(\varphi)R^i_\alpha(\varphi)\epsilon^\alpha = 0
\end{equation}
where we adopt the DeWitt notation, understanding an integration in
space-time whenever there is a summation over $i$.

If (\ref{defofsymm}) holds only for constant $\epsilon^\alpha$ we have
a {\it global symmetry}. If it holds for space-time dependent
$\epsilon^\alpha$ we talk about a {\it gauge symmetry} and the
summation over $\alpha$ includes an integration in space-time. The
problems when quantizing a theory with gauge symmetries manifest in
several ways. For instance, the quadratic part of the action is not
invertible and the perturbative expansion is not well-defined. From a
path integral point of view, the source of the problems is that when
naively integrating over the fields, many physically equivalent
configurations are added in a redundant way and one cannot make sense
of the resulting expressions. Hence, some prescriptions for the
quantization of gauge theories must be developed.

We shall assume that our set of gauge generators $R^i_\alpha(\varphi)$ is
complete in the following sense:
\begin{equation} \label{completeness}
y_i(\varphi)f^i(\varphi)=0 \Rightarrow
f^i(\varphi)=R^i_\alpha(\varphi)\epsilon^\alpha+y_j(\varphi)M^{ij}(\varphi)
\end{equation}
for some $M^{ij}(\varphi)$ such that
$M^{ij}(\varphi)=(-1)^{\varepsilon_i\varepsilon_j+1}M^{ji}(\varphi)$, where
$\varepsilon_i$ is the Grassmann parity of $\varphi^i$.

Denote by $\varepsilon_\alpha$ the Grassmann parity of
$\epsilon^\alpha$. Due to (\ref{completeness}) the most general form
of the graded commutator of two gauge transformations is given by
\begin{equation} \label{generalcomm}
\frac{\overleftarrow{\delta}R^i_\alpha}{\delta\varphi^j}R^j_\beta-
(-1)^{\varepsilon_\alpha
\varepsilon_\beta}\frac{\overleftarrow{\delta}R^i_\beta}
{\delta\varphi^j}R^j_\alpha = 2(-1)^{\varepsilon_\alpha}R^i_\gamma
T^\gamma_{\alpha\beta} -
4(-1)^{\varepsilon_i+\varepsilon_\alpha}y_jE^{ji}_{\alpha\beta}
\end{equation}
for some $T^\gamma_{\alpha\beta}(\varphi)$ and
$E^{ji}_{\alpha\beta}(\varphi)$. The signs and numerical factors have
been chosen for later convenience. If $E^{ji}_{\alpha\beta}(\varphi)=0$
one calls the gauge algebra {\it closed}. If
$E^{ji}_{\alpha\beta}(\varphi)\neq0$ the gauge algebra is called {\it
open}.

\subsection{BRST quantization}

The idea of the BRST method (\cite{BecRouSto},\cite{Tyu}) is to
enlarge the space of fields so that the resulting theory has no gauge
symmetries and the integration over field configurations is
well-defined. At the end of the day, one should be able to eliminate
the states associated to the fields which were not present in the
original theory. Let us formulate these ideas in a precise way.

Besides the ${\mathbb Z}_2$ gradation corresponding to the Grassmann
parity we introduce a $\mathbb Z$ gradation called {\it ghost
number}. The action must have even Grassmann parity and ghost number
zero.

For every gauge generator we introduce a {\it ghost field} $c^\alpha$,
an {\it antighost field} $b_\alpha$ and a Lagrange multiplier
$\lambda_\alpha$. The Grassmann parities are
$\varepsilon_{c^\alpha}=\varepsilon_{b_\alpha}=\varepsilon_\alpha +
1$, $\varepsilon_{\lambda_\alpha}=\varepsilon_\alpha$. Finally, we
assign ghost numbers $\gh(\varphi^i)=\gh(\lambda_\alpha)=0,\
\gh(c^\alpha)=1,\ \gh(b_\alpha)=-1$.

The essential object in the BRST formalism is an odd (right)
derivation of ghost number one $\delta$. It is defined by its action
on the basic fields:
\begin{eqnarray}
&&\delta\varphi^i = R^i_\alpha(\varphi)c^\alpha \notag\\
&&\delta c^\alpha = T^\alpha_{\beta\gamma}(\varphi)c^\gamma c^\beta  \notag\\
&&\delta b_\alpha = \lambda_\alpha  \notag\\
&&\delta\lambda_\alpha = 0\label{BRSTtrans}
\end{eqnarray}
and extendend to any functional of the fields through the Leibniz
rule, $\delta(F_1F_2)=F_1\delta F_2 + (-1)^{\varepsilon_{F_2}}\delta
F_1 F_2$. It is straightforward to check that $\delta^2=0$ for closed
gauge algebras.

Now, given a set of admissible gauge-fixing conditions $F^\alpha
(\varphi)$ we define the gauge-fixing fermion $\Psi= b_\alpha
F^\alpha$. The gauge fixed action is defined as
\begin{equation} \label{BRSTgfaction}
S_{\rm gf}= S + \delta\Psi.
\end{equation}
$S_{\rm gf}$ has no gauge symmetries and can be used for path integral
quantization of the theory. Notice that the original local symmetry
has been replaced by the global BRST symmetry given by\footnote{More
precisely, by ${\tilde \delta}_\epsilon :=\epsilon \delta$.}
$\delta$. Recall, however, that this has been done at the expense of
introducing spurious fields (which would translate into spurious
states in a Hilbert space formulation) and we should give some
prescription to get rid of them after quantizing. The key point is the
nilpotency of $\delta$, which allows to define a cohomology on the
space of fields. A gauge invariant observable translates into a
non-trivial cohomology class of $\delta$ at ghost number zero.

Notice that this construction fails when dealing with open algebras
because $\delta$ is not nilpotent anymore. Instead, $\delta^2 \varphi^i$
is proportional to the field equations. The appropriate procedure in
this case is given by the Batalin-Vilkovisky quantization method. In
the next section we discuss in detail how the quantization of gauge
theories with {\it closed } algebras fits into the BV scheme. Then, we
shall be ready to give the suitable prescriptions for open algebras,
which is our main objective.

\subsection{BV formalism} \label{BVformalism}

Assume we have a gauge theory with closed gauge algebra given by a
classical action $S$. Let us denote by $\varphi^A$ the original fields
$\varphi^i$, the ghosts $c^\alpha$, the antighosts $b_\alpha$ and the
Lagrange multipliers $\lambda_\alpha$. Their BRST transformation rules
(\ref{BRSTtrans}) are summarized as
\begin{equation}\notag
\delta\varphi^A={\cal R}^A(\varphi^B).
\end{equation}

In the BV formalism
(\cite{BatVil81},\cite{BatVil83},\cite{BatVil832},\cite{BatVil84},\cite{BatVil85}) we
double the number of fields by introducing an {\it antifield}
$\varphi^+_A$ for each field $\varphi^A$ such that:
\begin{enumerate}
\item[(i)]$\varphi^+_A$ has Grassmann parity opposite to that of $\varphi^A$
\vspace{-0.25cm}
\item[(ii)]$\text{gh}(\varphi^+_A)=-1-\text{gh}(\varphi^A)$.
\end{enumerate}

We define the BV action by
\begin{equation}\notag
S_{BV}^0[\varphi,\varphi^+]= S[\varphi^i]+\varphi^+_A{\cal R}^A(\varphi).
\end{equation}

Finally, given a gauge-fixing fermion $\Psi$ the partition function is
taken to be
\begin{eqnarray}\label{BVZ}
Z=\int e^{\frac{i}{\hbar}S_{BV}^0[\varphi,\varphi^+]}\delta\left(\varphi_A^+ -
\frac{\overleftarrow{\delta}\Psi}{\delta
\varphi^A}\right)\cal{D}\varphi\cal{D}\varphi^+.
\end{eqnarray}

It is easy to show that after integrating over the antifields one is
left with the partition function defined by the action
(\ref{BRSTgfaction}), hence recovering the BRST quantization
formalism.

The essential object in the BV formalism, which allows its
generalization for the quantization of theories with open gauge
algebras, is the {\it Batalin-Vilkovisky antibracket} for functionals
of the fields and antifields:
\begin{eqnarray}\notag
(F,G)=\frac{\overleftarrow{\delta}F}{\delta \varphi^A}
\frac{\overrightarrow{\delta}G}{\delta
\varphi^+_A}-\frac{\overleftarrow{\delta}F}{\delta \varphi^+_A}
\frac{\overrightarrow{\delta}G}{\delta
\varphi^A}.
\end{eqnarray}

The construction we have given for closed gauge algebras ensures that
the {\it classical master equation}
\begin{equation}\notag
(S_{BV}^0,S_{BV}^0)=0
\end{equation}
is satisfied. It might be necessary to add quantum corrections to
$S_{BV}^0$ in order to keep the BRST symmetry at the quantum level. It
is possible to show that this is equivalent to impose on
$S_{BV}=S_{BV}^0 + O(\hbar)$ that it be a solution of the {\it quantum
master equation}
\begin{equation} \label{qmasterequation}
(S_{BV},S_{BV})-2i\hbar \Delta S_{BV}=0
\end{equation}
where
\begin{equation}\notag
\Delta F =
(-1)^{\varepsilon_A+1}\frac{{\overleftarrow\delta}}{\delta\varphi^+_A}
\frac{{\overleftarrow\delta}}{\delta\varphi^A}F.
\end{equation}

For open algebras the prescription is as follows. Take the BRST
transformations (\ref{BRSTtrans}). The starting point is
\begin{equation}
S_{BV}^0=S+ \varphi^+_A\delta\varphi^A \notag.
\end{equation}
The problem is that now $\delta$ is not nilpotent and $S_{BV}^0$ does
not satisfy the classical master equation. Hence, one first must add
terms so that it is satisfied and then the quantum corrections needed
to obtain a solution of the quantum master equation
(\ref{qmasterequation}).

\section{Perturbative quantization of the PSM on the disk}

Take $\Sigma$ the unit disk $D=\{\sigma\in \mathbb{R}^2, |\sigma| \leq
1\}$. Cattaneo and Felder showed in \cite{CatFel99} that the
perturbative expansion of certain Green's functions of the
PSM defined on $D$ with $N=M$ (free BC) yields the
Kontsevich $\star$-product corresponding to the Poisson manifold
$M$, namely
\begin{eqnarray}\label{GreenFunctions}
\left<f(X(p_1))g(X(p_2))\delta(X(p_3)-x)\right>=f\star g(x)
\end{eqnarray}
where $p_1$, $p_2$ and $p_3$ are three cyclically ordered points at
the boundary of $D$, $x\in M$, and the expectation value is calculated
using (\ref{BVZ}). In a later work (\cite{CatFel03}) the same authors
studied the quantization with a general coisotropic brane $N$, which
turned out to be related with the deformation quantization of the
submanifold $N \hookrightarrow M$.

In both papers \cite{CatFel99}, \cite{CatFel03} the Green's functions
(\ref{GreenFunctions}) are worked out in the same fashion: first, one
takes the {\it Lorentz gauge} $\dd{*\eta}=0$, where the Hodge operator
acting on 1-forms requires the introduction of a complex structure on
$D$. The Feynman expansion in powers of $\hbar$ is then performed
around the constant classical solution where $X(\sigma)=x\in N$ and
the rest of the fields vanish. In this case expanding in powers of
$\hbar$ amounts to expanding in powers of $\Pi$ or, equivalently, to
expanding around zero Poisson structure.

The next section is a review of the seminal work
\cite{CatFel99}. Then, we shall try to work out (\ref{GreenFunctions})
following the steps enumerated in the last paragraph when $N$ is
non-coisotropic. We shall see that when second-class constraints are
present the propagator does not exist, but a natural redefinition of
the unperturbed (or quadratic) part of the action will yield a
well-defined perturbative expansion, showing that the non-coisotropic
branes also make sense at the quantum level. However, this leads to a
messy expression whose interpretation is far from clear. In Section
\ref{cleversolution} we shall see that a change of the gauge fixing is
illuminating in order to unravel the relation between the quantization
of the PSM with a non-coisotropic brane and the Kontsevich formula. In
Section \ref{sec:quantumgeneralbrane} we deal with the quantization of
a general strongly regular brane. Section \ref{sec:quantumcoisotropic}
is a summary of the results of \cite{CatFel03} on the quantization of
coisotropic branes.

\subsection{Quantization with free boundary conditions. The Kontsevich formula}

As we already know, the gauge algebra of the PSM is open (see
(\ref{OpenAlgebra})) and we must use the BV techniques learnt in
Section \ref{BVformalism} to formulate the quantum theory. The BRST
transformations ({\ref{BRSTtrans}) in the case of the PSM model read
\begin{subequations}\notag
\begin{align}
&\delta X^{i}=
\Pi^{ij}(X)\beta_{j}\\
&\delta \eta_{i}=-\dd \beta_{i}-\partial_{i}\Pi^{jk}(X)\eta_{j}
\beta_{k}\\
&\delta \beta_i=\frac{1}{2}\partial_i\Pi^{jk}(X)\beta_j\beta_k\\
&\delta \gamma^i=\lambda^i\\
&\delta \lambda^i=0
\end{align}
\end{subequations}
where $\beta_i$ are the ghost fields, $\gamma^i$ the antighost fields
and $\lambda^i$ the Lagrange multipliers. Applying the machinery of
Section \ref{BVformalism} one gets the BV action for the PSM
(\cite{CatFel99})\footnote{The path integral quantization of the
PSM in the particular case of 2D gravity was first
carried out in \cite{KumLieVas}.}
\begin{eqnarray} \label{SBV}
S_{BV}&=&\int_D \eta_i\wedge\dd X^i + \frac{1}{2}\Pi^{ij}(X)\eta_i\wedge\eta_j+
X_i^+\Pi^{ij}(X)\beta_j-\notag \\
& -&\eta^{+i}\wedge(\dd \beta_i+\partial_i\Pi^{kl}(X)
\eta_k\beta_l)-\frac{1}{2}\beta^{+i}\partial_i\Pi^{jk}(X)\beta_j\beta_k-
\notag\\&-&\frac{1}{4}
\eta^{+i}\wedge\eta^{+j}\partial_i\partial_j\Pi^{kl}(X)\beta_k\beta_l-
\lambda^i\gamma^+_i.
\end{eqnarray}

Let us take the {\it Lorentz gauge} $\dd{*\eta}=0$. The gauge fixing
fermion is then
$$\Psi= \int_D \gamma^i\dd{*\eta_i}$$
and the gauge fixed action with the antifields integrated out becomes
\begin{eqnarray}\label{gfaccion}
S_{gf}&=&\int_D \eta_i\wedge\dd X^i + \frac{1}{2}\Pi^{ij}(X)\eta_i\wedge\eta_j
- *{\dd\gamma}^i
\wedge(\dd \beta_i+\partial_i\Pi^{kl}(X)
\eta_k\beta_l)- \cr
&-&\frac{1}{4}
*{\dd\gamma}^i\wedge
*{\dd\gamma}^j\wedge
\partial_i\partial_j\Pi^{kl}(X)\beta_k\beta_l-
\lambda^i\dd{*\eta}_i.
\end{eqnarray}

Now write $X^i(\sigma)=x^i+\xi^i(\sigma)$ and choose
\begin{eqnarray}\label{unpgfaccion}
S_0&=&\int_D \eta_i\wedge\dd{\xi^i} 
- *\dd\gamma^i 
\wedge\dd \beta_i
-\dd{*\eta_i}\lambda^i
\end{eqnarray}
as the quadratic part which defines the propagators. This is the
perturbative expansion defined in \cite{CatFel99} in order to make
contact with Kontsevich's formula. Therein the field $X$ is taken to
be free at the boundary, i.e. the brane is the whole manifold $M$ and
consequently the contraction of $\eta$ with vectors tangent to
$\partial D$ must vanish. The fields $\beta_i,\gamma_i$ and
$\lambda^i$ must also vanish at the boundary.

It is convenient to map conformally the disk onto the upper complex
half plane $H_+$ (recall the conformal invariance of $S_{gf}$) and use
a complex coordinate $z\in H_+$. The boundary of the disk is mapped
onto the (compactified) real line.

The equations defining the propagators are obtained as Schwinger-Dyson
equations of the form
\begin{equation}\label{SchDys}
\frac{\delta}{\delta\varphi^j(w,{\bar w})}\int \varphi^i(z,{\bar z})
e^{\frac{i}{\hbar}S_0}{\cal D}\varphi = 0
\end{equation}
where $\varphi^i$ stands for any field entering the gauge-fixed action
$S_{gf}$. For example, taking $\varphi^i(z,{\bar z})=\beta_i(z,{\bar
z})$, $\varphi^j(w,{\bar w})=\beta_j(w,{\bar w})$, we get
\begin{equation}\notag
\partial_w\partial_{{\bar w}}\langle\gamma^j(w,{\bar
w})\beta_i(z,{\bar z})\rangle_0 = \frac{i\hbar}{4}\delta^j_i\delta^2(w-z).
\end{equation}

Hence, the propagators for the $\beta_i$ and $\gamma^i$ fields are
given by the Green's function of the Laplacian with Dirichlet boundary
conditions. Namely,
\begin{eqnarray}\notag
\langle \gamma^i(w,\bar w)\beta_j(z,\bar z)\rangle_0&=&
\frac{i\hbar}{2\pi}\delta_j^i\log\frac{|w-z|}{|w-\bar z|}
\end{eqnarray}
with $\partial_{\bar z}\frac{1}{z}=\pi\delta^2(z)$.

The other non vanishing components of the propagator are conveniently
expressed in terms of the complex fields $\zeta^j=\xi^j+i\lambda^j$
and $\bar\zeta^j=\xi^j-i\lambda^j$. The relevant Schwinger-Dyson
equations (\ref{SchDys}) yield
\vskip 3mm
\begin{eqnarray}\notag
&&\partial_z\langle{\bar\zeta}(w,\bar w)\eta_{\bar z}(z,\bar
z)\rangle_0-\partial_{\bar z}\langle{\zeta}(w,\bar w)\eta_{z}(z,\bar
z)\rangle_0=-\hbar\delta^2(w-z)\cr {}\cr
&&\partial_z\langle{\zeta}(w,\bar w)\eta_{\bar z}(z,\bar
z)\rangle_0-\partial_{\bar z}\langle{\bar\zeta}(w,\bar w)\eta_{z}(z,\bar
z)\rangle_0=0\cr {}\cr
&&\partial_{\bar w}\langle{\zeta}(w,\bar w)\eta_{z}(z,\bar
z)\rangle_0=-\frac{\hbar}{2}\delta^2(w-z)\cr {}\cr
&&\partial_{w}\langle{\bar\zeta}(w,\bar w)\eta_{\bar z}(z,\bar
z)\rangle_0=\frac{\hbar}{2}\delta^2(w-z)\cr {}\cr
&&\partial_z\langle{\zeta}(w,\bar w)\eta_{\bar z}(z,\bar
z)\rangle_0+\partial_{\bar z}\langle{\bar\zeta}(w,\bar w)\eta_{z}(z,\bar
z)\rangle_0=0\cr {}\cr
&&\partial_z\langle{\bar\zeta}(w,\bar w)\eta_{\bar z}(z,\bar
z)\rangle_0+\partial_{\bar z}\langle{\zeta}(w,\bar w)\eta_{z}(z,\bar
z)\rangle_0=0\cr {}\cr
&&\partial_{w}\langle{\bar\zeta}(w,\bar w)\eta_{z}(z,\bar
z)\rangle_0=0\cr {}\cr
&&\partial_{\bar w}\langle{\zeta}(w,\bar w)\eta_{\bar z}(z,\bar
z)\rangle_0=0.
\end{eqnarray}

It is not difficult to show that the general solution (i.e. before
imposing the boundary conditions) is:
\begin{eqnarray}\label{eq:propanoBC}
\langle\zeta^i(w,\bar w)\eta_j(z,\bar z)\rangle_0 &=&
\frac{\hbar}{2\pi}\delta_j^i\left (-\frac{\dd z}{w-z}+ f_j(w,z)\dd z
+g_j(w,\bar z)\dd\bar z\right )\cr\cr
\langle\bar\zeta^i(w,\bar w)\eta_j(z,\bar z)\rangle_0 &=&
\frac{\hbar}{2\pi}\delta_j^i\left (\frac{\dd \bar z}{\bar w-\bar z}+
\bar f_j(\bar w,\bar z)\dd \bar z +\bar g_j(\bar w,z)\dd z\right)\qquad\quad
\end{eqnarray}
where no sum in $j$ is assumed and $f_j, \bar f_j, g_j, \bar g_j$ are
holomorphic in their arguments with domains given by $w, z\in
H_+$. The boundary conditions imply $f_i=\bar f_i=0$ and
\begin{eqnarray}\notag
&&g_i(w,\bar z)=\frac{1}{w- \bar z},\qquad \bar g_i(\bar w,
z)=\frac{-1}{\bar w- z}.
\end{eqnarray}

\vskip 5mm

Define $\dd_z:=\dd z\frac{\partial}{\partial z}+\dd {\bar
z}\frac{\partial}{\partial {\bar z}}$. We can rewrite the propagators
in terms of the Green's function of the Laplacian\footnote{When no
confusion is possible we shall simplify the notation related to the
variables on which the complex functions depend. Hence, we shall write
$\beta_j(z)$ instead of $\beta_j(z,{\bar z})$, and so on.}
\begin{equation}\notag
\psi(z,w)=\log\frac{|z-w|}{|z-\bar w|}
\end{equation}
and the Kontsevich's angle function (recall (\ref{eq:Kontanglefunc}))
\begin{equation}\notag
\phi(z,w)=\frac{1}{2i}\log\frac{(z-w)(z-\bar w)}{(\bar z - \bar
w)(\bar z - w)}.
\end{equation}
This will be convenient to identify the different ingredients of the
Kontsevich formula in the perturbative expansion.

Thus we have
\begin{eqnarray}\notag
&&\langle \gamma^k(w)\beta_j(z)
\rangle_0=\frac{i\hbar}{2\pi}\delta^k_j\psi(z,w)\cr
&&\langle
\xi^k(w)\eta_j(z)
\rangle_0=\frac{i\hbar}{2\pi}\delta^k_j\dd_z\phi(z,w)\cr
&&\langle\lambda^k(w)\eta_j(z)\rangle_0=-\frac{i\hbar}{2\pi}\delta^k_j\dd_z\psi(z,w).
\end{eqnarray}

Observing that $*\dd_w\psi(z,w)=\dd_w\phi(z,w)$, so that
$$\langle *\dd\gamma^k(w)\beta_j(z)\rangle_0 =
\delta^k_j\frac{i\hbar}{2\pi}\dd_w\phi(z,w)$$
it follows that the propagators can be combined into a superpropagator
\begin{equation}\notag
\langle\xi^k(w)\eta_j(z)\rangle_0+\langle
*\dd\gamma^k(w)\beta_j(z)\rangle_0=\frac{i\hbar}{2\pi}\delta^k_j\dd\phi(z,w)
\end{equation}
where $\dd := \dd_z + \dd_w$. In terms of superfields
${\tilde\eta}_j(z,\theta)=\beta_j(z)+\theta^\mu\eta_{j\mu}(z)$ and
${\tilde\xi}^k(w,\rho)=\xi^k(w)+\rho^mu\eta_\mu^+(w)$ with
$\eta^{+j}=*\dd\gamma^j$, the superpropagator reads
\begin{equation}\notag
\langle {\tilde\xi}^k(w,\rho){\tilde\eta}_j(z,\theta)\rangle_0
=\frac{i\hbar}{2\pi}\delta^k_jD\phi(z,w)
\end{equation}
with $D:=\theta^\mu\frac{\partial}{\partial
z^\mu}+\rho^mu\frac{\partial}{\partial w^\mu}$.

The perturbative series is defined by taking $S_{gf}=S_0+S_1$ and
expanding:
\begin{equation}\notag
\langle {\cal O}\rangle = \int e^{\frac{i}{\hbar}S_{gf}}{\cal
O}=\sum_{n=0}^{\infty}\frac{i^n}{\hbar^n n!}\int
e^{\frac{i}{\hbar}S_0}(S_1)^n{\cal O}.
\end{equation}

The Feynman expansion is obtained by expanding $S_1$ and the
observable $\cal O$ in powers of ${\tilde\xi},{\tilde\eta}$, giving
the vertices
\begin{equation}\notag
S_1=\frac{1}{2}\int_D\int\dd^2\theta\sum_{k=0}^\infty\frac{1}{k!}
\partial_{j_1}\dots\partial_{j_k}\Pi^{lr}(x){\tilde\xi}^{j_1}\dots
{\tilde\xi}^{j_k}{\tilde\eta}_l{\tilde\eta}_r.
\end{equation}

Take three cyclically ordered points at $\partial D$, $0,1$ and
$\infty$. The perturbative expansion of the expectation value of
\begin{equation}\notag
{\cal O}=f({\tilde X}(0))g({\tilde X}(1))\delta(X(\infty)-x)
\end{equation}
reproduces Kontsevich formula for $\Pi$ except for the presence of
tadpoles, i.e. diagrams containing at least an edge which starts and
ends at the same vertex. The divergence of the Poisson tensor,
$\partial_i\Pi^{ij}$, and the ill-defined expression $\dd\phi(z,z)$
enter into the tadpole diagrams. However, it is easy to
regularize these terms by a point-splitting procedure defining
\begin{equation}\notag
\dd_z\phi(z,z)=\kappa(z;\tau):=\lim_{\epsilon\to
0}\dd_z\phi(z,z+\epsilon\tau(z))
\end{equation}
where $\tau(z)$ is a vector field which does not vanish in the
interior of the disk. The limit exists but depends on
$\tau(z)$. Writing $\tau(z)=r(z)e^{i\alpha(z)}$ in polar coordinates,
then $\kappa(z;\tau)=\dd\alpha(z)$. Choosing $\alpha$ constant we can
get rid of the tadpole diagrams\footnote{Equivalently, one can add a
counterterm proportional to $\partial_i\Pi^{ij}$ an renormalize the
tadpoles to zero.} and the Kontsevich formula is obtained.

\vskip 3mm

\subsection{Perturbation expansion for non-coisotropic branes}
\label{sec:pertexpnoncois}

Now, our aim is to show that the perturbative expansion defined in the
previous section makes sense only for coisotropic branes (recall that
$N=M$ is coisotropic) and breaks down when second-class constraints
are present. In other words, we want to show that if the brane $N$ is
not coisotropic the propagator defined by (\ref{unpgfaccion}) cannot
fulfill the appropriate boundary conditions.

In the adapted coordinates of Section 2.1 the index $i$ splits into
$a,\mu,A$ where $\xi^a$ are coordinates along the brane (free at the
boundary) and $\xi^\mu$ and $\xi^A$ are respectively first-class and
second-class coordinates transverse to the brane and must vanish at
the boundary. For the rest of the fields we have the following
boundary conditions,
\vskip 1mm
-- Dirichlet: $\lambda^a$, $\lambda^A$, $\eta_{a{\bf t}}$,
$(*\eta_{\mu})_{\bf t}$, $\eta_{A{\bf t}}$, $\beta_a$, $\beta_A$,
$\gamma^a$ and $\gamma^A$.

-- Neumann: $\beta_\mu$, $\lambda^\mu$ and $\gamma^\mu$.
\vskip 1mm
The propagators for the $\beta$ and $\gamma$ fields are given by the
Green's function of the Laplacian with appropriate boundary
conditions:
\begin{eqnarray}\label{betagamma}
\langle \gamma^a(w,\bar w)\beta_b(z,\bar z)\rangle_0&=&
\frac{i\hbar}{2\pi}\delta_b^a\log\frac{|w-z|}{|w-\bar z|}\cr\cr
\langle \gamma^\mu(w,\bar w)\beta_\nu(z,\bar z)\rangle_0&=&
\frac{i\hbar}{2\pi}\delta_\nu^\mu\log(|w-z||w-{\bar z}|)\cr\cr
\langle \gamma^A(w,\bar w)\beta_B(z,\bar z)\rangle_0&=&
\frac{i\hbar}{2\pi}\delta_B^A\log\frac{|w-z|}{|w-\bar z|}.
\end{eqnarray}

The problem comes when we try to impose the boundary conditions to the
remaining components of the propagator (\ref{eq:propanoBC}). It is
straightforward to prove that the boundary conditions imply
$f_a=f_\mu=\bar f_a=\bar f_\mu=0$ and
$$g_a(w,\bar z)=-g_\mu(w,\bar z)=\frac{1}{w- \bar z},
\qquad \bar g_a(\bar w, z)=
-\bar g_{\mu}(\bar w,z)=\frac{-1}{\bar w- z}.$$

However, if we try to fulfill the boundary conditions for the
components corresponding to the second-class constraints ($A,B,...$
indices) we find a contradiction. Let us show this in detail.

We shall omit the second-class subscripts for simplicity. The BC for
$\eta_{\bf t}$ implies
\begin{eqnarray}\notag
&&\langle\zeta(w,\bar w)(\eta_z(z,\bar z)+\eta_{\bar z}(z,\bar
z))\rangle_0\vert_{z={\bar z}}=0\cr\cr
&&\langle{\bar\zeta}(w,\bar w)(\eta_z(z,\bar z)+\eta_{\bar z}(z,\bar
z))\rangle_0\vert_{z={\bar z}}=0
\end{eqnarray}
which translates into
\begin{eqnarray}\label{eq:contradiction1}
&&g(w,{\bar z})=\frac{1}{w-{\bar z}}-h(w,{\bar z}),\ \mbox{s.t. }
h(w,u)=f(w,u),u\in{\mathbb R},\cr
&&{\bar g}({\bar w},z)=-\frac{1}{{\bar w}-z}-{\bar h}({\bar w},z),\
\mbox{s.t. } {\bar h}(w,u)={\bar f}(w,u),u\in{\mathbb R}.\qquad
\end{eqnarray}

Dirichlet BC on second-class components of the field $X$ impose
\begin{eqnarray}\notag
&&\langle(\zeta(w,\bar w)+{\bar\zeta}(w,\bar w))\eta_z(z,\bar
z)\rangle_0\vert_{w={\bar w}}=0\cr\cr
&&\langle(\zeta(w,\bar w)+{\bar\zeta}(w,\bar w))\eta_{\bar z}(z,\bar
z)\rangle_0\vert_{w={\bar w}}=0
\end{eqnarray}
so that
\begin{eqnarray}\label{eq:contradiction2}
&&{\bar g}({\bar w},z)=\frac{1}{{\bar w}-z}-h'({\bar w},z),\
\mbox{s.t. } h'(u,z)=f(u,z),u\in{\mathbb R},\cr
&&g(w,{\bar z})=-\frac{1}{w-{\bar z}}-{\bar h}'(w,{\bar z}),\
\mbox{s.t. } {\bar h}'(u,{\bar z})={\bar f}(u,{\bar z}),u\in{\mathbb
R}.\qquad
\end{eqnarray}

{}From (\ref{eq:contradiction1}) and (\ref{eq:contradiction2}) we
deduce that $f$ and $\bar f$ extend to entire functions in $z$ and $w$
and consequently they are constant. In addition, also from
(\ref{eq:contradiction1}) and (\ref{eq:contradiction2}) we derive the
relation
\begin{equation}\notag
\bar f(w,\bar z) - f(w,\bar z) = -\frac{2}{w-\bar z}
\end{equation}
which is obviously impossible if $f$ and $\bar f$ are entire. Hence,
the propagator does not exist.

At this point one might be tempted to conclude that only coisotropic
branes make sense at the quantum level. But this is a too sloppy
conclusion since we have only shown that the perturbative expansion
defined by the choice of (\ref{unpgfaccion}) as the unperturbed part
ceases to exist when second-class constraints appear. The question is
whether there is a different definition of the perturbative expansion
leading to a well-defined result in this case.  The situation is
reminiscent to the contradiction found by Dirac in imposing the
second-class constraints on the states of the physical Hilbert space;
he proposed to circumvent this difficulty with the help of the Dirac
bracket \cite{Dir}.

{}From now on we shall restrict to branes which satisfy the {\it
strong regularity condition} (\ref{regularity}) and, in order to make
the presentation simpler, we shall assume that there are no
first-class constraints, i.e. we restrict to {\it second-class
branes}.
 
The strategy to solve the problem is the well-known technique of using
our opponent's strength against him. The origin of the non-existence
of the propagator for the second-class coordinates is that
$\det(\Pi^{AB})\not=0$ implies that if $X^A=0$ at the boundary then
$\eta_{A{\bf t}}$ must also vanish. And the propagator cannot satisfy
these two conditions simultaneously. But precisely due to the fact
that $\det(\Pi^{AB})\not=0$ and given that the $\eta_{A}$ fields
appear at most quadratically in the gauge fixed action
(\ref{gfaccion}) we can perform the Gaussian integration over them in
order to get an effective action $S^{eff}$. This action can be used to
compute the correlation functions of observables that do not involve
$\eta_A$ fields as it is our case.

Once the integration has been performed 
there is a splitting of $S^{eff}$
which defines a consistent perturbative expansion. Take $S^{eff}=
S^{eff}_0 +S^{eff}_{pert}$ with
\begin{eqnarray}\notag
S^{eff}_0=\int_D \eta_a\wedge\dd \xi^a 
-\dd{*\eta}_a \lambda^a
+ \omega_{AB}(x)\dd\xi^A\wedge *\dd\lambda^B
- *\dd\gamma^i 
\wedge\dd \beta_i.
\end{eqnarray}

The $\beta$, $\gamma$ propagators are as before (see eq.
(\ref{betagamma})), as well as those for $\zeta^a$ and $\eta_a$. In
addition, $S^{eff}_0$ yields well-defined propagators
for the other fields, the only non-zero components being
\begin{equation}\notag
\langle\lambda^A(w,\bar w)\xi^B(z,\bar z)\rangle_0^{eff} =
\frac{i\hbar}{2\pi}\Pi^{AB}(x)\log\frac{|w-z|}{|w-\bar z|}.
\end{equation} 

Now one can expand $S^{eff}_{pert}$ into vertices and define a
perturbative expansion for Green's functions of the form
(\ref{GreenFunctions}). However, from the resulting perturbative
series it seems very hard to find out whether the formula
(\ref{GreenFunctions}) defines an associative product.  A simpler
derivation providing a positive answer is given in the next section.

\subsection{Second-class branes and the Kontsevich formula for the Dirac bracket}
\label{cleversolution}

Let us take advantage of our opponent's strength in a more profound
sense. Using that $\Pi^{AB}$ is invertible at every point of $N$, and
consequently in a tubular neighborhood of $N$, we can show that the
gauge fixing
\begin{eqnarray}\label{GaugeFixing}
\dd{*\eta_a}=0,\qquad \xi^A=0
\end{eqnarray}
is reachable, at least locally: $\dd{*\eta_a}=0$ can be obtained by
choosing suitably $\epsilon_a$ in (\ref{symmetryb}). Now, write
(\ref{symmetrya}) for upper-case Latin indices
$$\delta_{\epsilon}\xi^{A}=\Pi^{A{a}}(x+\xi)\epsilon_{{a}}+
\Pi^{AB}(x+\xi)\epsilon_{B}.$$
Since $\Pi^{AB}$ is invertible one can solve for $\epsilon_B$ and get
$\xi^A=0$.
\vskip .2cm
We want to stress that the analog of (\ref{GaugeFixing}) is not an
admissible gauge-fixing in the coisotropic case. For second-class
branes both the Lorentz gauge and (\ref{GaugeFixing}) are admissible
but, as we shall see, the latter makes the perturbative quantization
transparent and is the appropriate approach to the problem.

Let us go back to the BV action (\ref{SBV}), set the indices of the
antighosts $\gamma$ and Lagrange multipliers $\lambda$ upstairs or
downstairs as demanded by (\ref{GaugeFixing}) and take
\begin{eqnarray}\notag
\Psi= \int_D \gamma^{a}\dd{*\eta}_{a} + \int_D\gamma_A X^A
\end{eqnarray}
where $\gamma_A$ are anticommuting 2-form fields on $\Sigma$. 

On the submanifold $\varphi_i^+ =
\frac{\overrightarrow{\delta}\Psi}{\delta \varphi^i}$ we have
\begin{subequations}\notag
\begin{align}
&\beta^{+{a}}=\beta^{+A}=0 \cr
&\eta^{+{a}}=*\dd \gamma^{a},\eta^{+A}=0 \cr
&X^+_{a}=0,\ X^+_A=-\gamma_A \cr
&\gamma^+_{a}=\dd {*\eta}_{a},\ \gamma^{+A}=X^A.
\end{align}
\end{subequations}
\addtocounter{equation}{-1}

And the gauge fixed action with the antifields integrated out reads
now
\begin{eqnarray}
\tilde S_{gf}&=&\int_D \eta_{i}\wedge\dd X^{i} +
\frac{1}{2}\Pi^{ij}(X)\eta_i\wedge\eta_j-*\dd \gamma^{a}\wedge(\dd
\beta_{a}+\partial_{a}\Pi^{kl}(X)\eta_k\beta_l)- \nonumber\\
&-&\frac{1}{4}{*\dd\gamma}^{a}\wedge *{\dd
\gamma}^{b}\partial_{a}\partial_{b}\Pi^{kl}(X)\beta_k\beta_l-
\lambda^{a}\dd{*\eta_{a}}-\gamma_A\Pi^{Ai}(X)\beta_i
- \lambda_AX^A. \nonumber
\end{eqnarray}

Recall that we are interested in calculating the expectation value of 
functionals depending only on $X$. Hence, integration over 
$\lambda_A$ sets $X^A=0$ and we can
write:
\begin{eqnarray}
\tilde S_{gf}'&=&\int_D \eta_{a}\wedge\dd X^{a} + 
\frac{1}{2}\Pi^{ij}(X)\eta_i\wedge\eta_j-{*\dd\gamma}^{a}\wedge(\dd \beta_{a}+
\partial_{a}\Pi^{kl}(X)\eta_k\beta_l)- \nonumber\\
&-&\frac{1}{4}{*\dd\gamma}^{a}\wedge {*\dd\gamma}^{b}\partial_{a}\partial_{b}
\Pi^{kl}(X)\beta_k\beta_l-\lambda^{a}\dd{*\eta}_{a}-
*\gamma_A\Pi^{Ai}(X)\beta_i \nonumber
\end{eqnarray}
where $\Pi$ is evaluated on $X^A=0$.

Now, integrating over $\gamma_A$ forces
\begin{eqnarray} \label{condition}
&&\Pi^{Ai}\beta_i=0 \Leftrightarrow \beta_A =-
\omega_{AB}\Pi^{B{a}}\beta_{a}
\end{eqnarray}
which is a crucial relation which can be used to get rid of the
components of the fields with upper-case indices and get an effective
action depending only on the lower-case components. Notice that
writing
$$*\dd\gamma^{a}\wedge\partial_{a}\Pi^{kl}(X)\eta_k\beta_l=
*\dd\gamma^{a}\wedge\partial_{a}(\Pi^{{b}{c}}(X)\eta_{b}\beta_{c}
+ \Pi^{Ai}(X)\eta_A\beta_i+\Pi^{{b} A}(X)\eta_{b}\beta_A)$$ and
applying (\ref{condition}) to the second and third terms in
parentheses we obtain the Dirac Poisson structure
(\ref{DiracComponents}) in a beautiful way:
$$*\dd\gamma^{a}\wedge\partial_{a}\Pi^{kl}\eta_k\beta_l=
*\dd\gamma^{a}\wedge\partial_{a}\Pi^{{b}{c}}_{\cal D}\eta_{b}\beta_{c}.$$

Doing the same for the term quadratic in $\dd\gamma$ in $S_{BV}^{eff}$ we get:
\begin{eqnarray} \label{SBVeff2}
\tilde S_{gf}''&=&\int_D \eta_{a}\wedge\dd X^{a} + \frac{1}{2}\Pi^{ij}(X)
\eta_i\wedge\eta_j-*\dd \gamma^{a}\wedge(\dd \beta_{a}+\partial_{a}
\Pi_{\cal D}^{bc}(X)\eta_b\beta_c)- \notag\\
&-&\frac{1}{4}{*\dd\gamma}^{a}\wedge *\dd \gamma^{b}\partial_{a}
\partial_{b}\Pi_{\cal D}^{cd}(X)\beta_c\beta_d-\lambda^{a}
\dd{*\eta_{a}}-i\hbar\log\det(\Pi^{AB}(X)) \notag
\end{eqnarray}
where the last term in the action comes from the Jacobian corresponding 
to the delta distribution
$$\delta(\Pi^{Ai}\beta_i)=\delta(\beta_A +
\omega_{AB}\Pi^{B{a}}\beta_{a})\det (\Pi^{BC}(X)).$$

The final step is to integrate out the $\eta_A$ fields.  The integral
is Gaussian (due again to the non-degeneracy of $\Pi^{AB}$) and the
determinant coming from it cancels the contribution from the $\delta$
function. Finally,
\begin{eqnarray}\notag
\tilde S_{gf}^{eff}&=&\int_D \eta_{a}\wedge\dd X^{a} +
\frac{1}{2}\Pi_{\cal D}^{ab}(X)\eta_a\wedge\eta_b-*\dd \gamma^{a}\wedge(\dd
\beta_{a}+\partial_{a}\Pi_{\cal D}^{cd}(X)\eta_c\beta_d)\cr
&-&\frac{1}{4}*{\dd\gamma}^{a}\wedge *\dd
\gamma^{b}\partial_{a}\partial_{b}\Pi_{\cal D}^{cd}(X)
\beta_c\beta_d-\lambda^{a}\dd{*\eta_{a}}
\end{eqnarray}
which is Cattaneo and Felder's gauge-fixed BV action for a
PSM defined on $D$, with target $(N,\Pi_{\cal D})$
(recall that we set $X^A=0$) and boundary conditions such that $X^{a}$
is free and $\eta_{a}$ vanishes on vectors tangent to
$\partial D$. In other words, we have ended up with the situation
studied in \cite{CatFel99}. Invoking the results therein we can deduce
our announced relation, namely that the perturbative expansion of
$$\left<f(X(0))g(X(1))\delta(X(\infty)-x)\right>,\ f,g\in C^\infty(M)$$
yields the Kontsevich's formula for $\Pi_{\cal D}$ applied to the restrictions 
to $N$ of $f$ and $g$. 

In the derivation of this result the second-class boundary conditions
seem to play no role, as they are not used to compute any
propagator. Notice, however, that the gauge fixing (\ref{GaugeFixing})
makes sense only if the fields $\xi^A$ vanish at the boundary before
fixing the gauge. As stressed above the fact that
$\det(\Pi^{AB})\not=0$ is also essential. This ties inextricably the
present result and the use of second-class branes together.

We would like to stress the interesting cancellation of the
determinant coming from the integration of the $\gamma_A$ and
$\beta_A$ fields with that coming from the integration of the $\eta_A$
fields. It would be worth finding out whether there is some underlying
symmetry behind it.

\subsection{Quantization with a general brane}
\label{sec:quantumgeneralbrane}

Once the quantization of the PSM with a second-class brane has been
understood, it is straightforward to describe the procedure for the
quantization of the model with an arbitrary strongly regular brane
defined by both first and second-class constraints. The appropriate
gauge fixing fermion in the general case is
\begin{eqnarray}\label{generalPsi}
\Psi= \int_D \gamma^{a}\dd{*\eta}_{a} + \gamma^{\mu}\dd{*\eta}_{\mu} +
\gamma_A X^A.
\end{eqnarray}

Then, we integrate out the second-class components of the fields
exactly as above and we are left with
\begin{eqnarray}\notag
\tilde S_{gf}^{eff}&=&\int_D \eta_{p}\wedge\dd X^{p} +
\frac{1}{2}\Pi_{\cal D}^{pq}(X)\eta_p\wedge\eta_q-*\dd \gamma^{p}\wedge(\dd
\beta_{p}+\partial_{p}\Pi_{\cal D}^{qr}(X)\eta_q\beta_r)\cr
&-&\frac{1}{4}*{\dd\gamma}^{p}\wedge *\dd
\gamma^{q}\partial_{p}\partial_{q}\Pi_{\cal D}^{rs}(X)
\beta_r\beta_s-\lambda^{p}\dd{*\eta_{p}}
\end{eqnarray}
where the indices now run over $a$ and $\mu$ values and $\Pi_{\cal D}$
is the Dirac bracket of (\ref{DiracCompFirst}). This is Cattaneo and Felder's
gauge-fixed BV action for the PSM with target given by
local coordinates $(X^a,X^\mu)$, Poisson structure $\Pi_{\cal D}$ and
a coisotropic brane defined by $X^\mu=0$. At this point we can apply
the results of \cite{CatFel03}.

An interesting question is how the choice of a set of second-class
constraints affects the final result. In our case the choice of second
class constraints was made through the gauge fixing fermion,
i.e. a different choice amounts to a change of
gauge-fixing. Since the expectation values of gauge-invariant
observables do not depend on this particular choice, we conclude that
the final result is independent of the choice of second-class
constraints.

The whole derivation parallels that of Dirac's quantization of
constrained systems (\cite{Dir}): one gets rid of the second-class
constraints by defining the appropriate Dirac bracket that can be
quantized with the first class constraints imposed on the states.  It
is nice that this result is obtained in a quantum-field theoretical
context by tuning the boundary conditions of the fields.

\vskip 0.2 cm

{\it Remark:} The need for strong regularity in the quantum case can
be seen, for example, from the fact that we need $\det(\Pi^{AB})\neq
0$ at every point of a tubular neighborhood of $N$ in order to perform
the Gaussian integration over the $\eta_A$ fields.

\subsection{Comments on the coisotropic case}\label{sec:quantumcoisotropic}

The classical and quantum study of the PSM with coisotropic boundary
conditions was performed in detail by Cattaneo and Felder in
\cite{CatFel99}. As we have already mentioned, the results on the
perturbative quantization in this case have some subtleties which we
would like to discuss in a more detailed way.

We proved in Section \ref{sec:pertexpnoncois} that the perturbative
expansion valid in the free case ($N=M$), i.e. around $\Pi=0$, is also
valid for any coisotropic brane $N$. This expansion was used in
\cite{CatFel99} and its relation with deformation quantization was
investigated. Concretely, it was found that the perturbative expansion
on the disk $D$ of
\begin{equation}\notag
f\star g(x):=\int_{X(p_3)=x}e^{\frac{i}{\hbar}S}f(X(p_1))g(X(p_2))
\end{equation}
with $p_1,p_2,p_3$ cyclically ordered at the boundary of $D$ and $x\in
N$, leads to a modification of the Kontsevich's formula
(\ref{Kontsevichformula}). It can be described as follows:
\begin{equation}\label{eq:starproductcois}
f\star g = fg + \sum_{n\geq1}\frac{\varepsilon^n}{n!}\sum_{\daleth\in
G_{n,2}}w_\daleth B_\daleth(f,g),\qquad f,g\in C^\infty(N)
\end{equation}
where $\varepsilon=i\hbar/2$. A graph in $G_{n,2}$ has $n$ vertices of
the first type and 2 vertices of the second type. The edges are
oriented and they also come in two types, called in \cite{CatFel99}
straight and wavy. Exactly as described for the Kontsevich's formula
(\ref{Kontsevichformula}), two edges emerge from each vertex of the
first type and no vertex emerges from a vertex of the second type. An
edge cannot start and end at the same vertex.

Take local adapted coordinates on $M$, $(x^a,x^\mu),
 a=1,\dots,\mbox{dim}(N), \mu=\mbox{dim}(N)+1, \dots,
 \mbox{dim}(M)$. Recall that functions on $N$ which inherit a Poisson
 bracket from $\Pi$ are those invariant under the flow of the vector
 fields
\begin{equation}\notag
Z^\mu = \Pi^{\mu i}\partial_i,\qquad \mu=\mbox{dim}(N)+1, \dots,
 \mbox{dim}(M)
\end{equation}
on $N$, i.e. gauge-invariant functions.

The bidifferential operator $B_\daleth$ associated to the graph
$\daleth$ is constructed as the one entering
(\ref{Kontsevichformula}). The only difference is that if an edge is
straight, the sum is performed only over $a$ indices whereas if the
edge is wavy, it is performed only over $\mu$ indices.

The weight $w_\daleth,\daleth\in G_{n,2}$ is
\begin{equation}\notag
w_\daleth =
\frac{1}{(2\pi)^{2n}}\int_{H_n}\prod_{\mbox{\footnotesize{edges
}}e}\dd\phi_e.
\end{equation}
Let $e$ be an edge going from $k$ to $k'$. $\dd\phi_e$ is
$\dd\phi(z_k,z_{k'})$ if $e$ is straight and $\dd\phi(z_{k'},z_k)$ if
it is wavy, where $\phi(z,w)$ is the Kontsevich's angle function
(\ref{eq:Kontanglefunc}).

Of course, the product defined by (\ref{eq:starproductcois}) coincides
with Kontsevich's formula (\ref{Kontsevichformula}) when
$N=M$. However, for general coisotropic $N$ the formula
(\ref{eq:starproductcois}) does not define an associative product. The
description of the obstructions requires some new objects which we
proceed to introduce.

Let $Y,Y'$ be two vector fields on $M$ and define a differential
operator $A(Y)$ on $C^\infty(N)[[\varepsilon]]$ and a function
$F(Y,Y')\in C^\infty(N)[[\varepsilon]]$ by
\begin{subequations}
\begin{align}
&A(Y)f=Yf+\sum_{n\geq1}\frac{\varepsilon^n}{n!}\sum_{\daleth\in
G_{n+1,1}}w_\daleth B_\daleth(Y)f,\\
&F(Y,Y')=\sum_{n\geq0}\frac{\varepsilon^n}{n!}\sum_{\daleth\in
G_{n+2,0}}w_\daleth B_\daleth(Y,Y').
\end{align}
\end{subequations}
The bidifferential operators $B_\daleth$ are defined as for $G_{n,2}$
with the following exceptions: graphs in $G_{n+1,1}$ have one
additional vertex of the first type corresponding to $Y$ and only one
vertex of the second type. Exactly one edge starts from the vertex of
the first type associated to $Y$. Graphs in $G_{n+2,0}$ have two new
vertices of the first type corresponding to $Y$ and $Y'$ and there are
no vertices of the second type.

The claim in \cite{CatFel99} is that if
\begin{equation}\notag
F(Z^\mu,Z^\nu)=0, \qquad \mbox{dim}(N)+1\leq \mu,\nu \leq \mbox{dim}(M)
\end{equation}
then, formula (\ref{eq:starproductcois}) defines an associative
product on
\begin{equation}\notag
C^\infty_{inv}(N):=\{f\in C^\infty(N)[[\varepsilon]]\ \vert \
A(Z^\mu)f=0, \ \mbox{dim}(N)+1\leq \mu \leq \mbox{dim}(M)\}
\end{equation}
which in general is not a deformation of the set of gauge-invariant
functions on $N$ as discussed in \cite{CatFel99}.

\section{The cyclicity of the star product}
\label{sec:cyclicity}

In this section we would like to draw the attention to an interesting
application of the field theoretical interpretation of Kontsevich's
formula given by Cattaneo and Felder in \cite{CatFel99}. Let us set
the issue of branes aside and take $M={\mathbb R}^n$. As we already
know, Kontsevich's formula for the Poisson structure $\Pi$ can be
obtained as the perturbative expansion of the following Green's
functions of the PSM defined on the disk with free boundary conditions
for $X$:
\begin{eqnarray}\notag
f\star g(x)=\left<f(X(p_1))g(X(p_2))\delta(X(p_3)-x)\right>
\end{eqnarray}
with $p_1,p_2,p_3$ three cyclically ordered points at the boundary of
the disk. More generally, one can consider the expectation value
\begin{equation}\notag
\left<f(X(p_1))g(X(p_2))h(X(p_3))\right>
\end{equation}
which seems invariant under cyclic permutations of $f,g,h$ because the
action of the PSM is invariant under orientation preserving
diffeomorphisms of the disk. Hence, one might be tempted to conclude
that the Kontsevich's star product satisfies
\begin{equation} \label{cyclicity}
\int_{{\mathbb R}^n}(f\star g)h \ \dd^n x =\int_{{\mathbb R}^n}(g\star
h)f \  \dd^n x.
\end{equation}

However, as proved in \cite{FelSho}, this claim is not generally
true. The correct statement is given by the following

\begin{theorem}
Take a constant volume form $\Omega$ and a Poisson tensor $\Pi$ such
that ${\rm div}_\Omega \Pi =0$. Then, the Kontsevich's star product
satisfies (\ref{cyclicity}).
\end{theorem}

The pitfall of the heuristic argument above resides in the fact that
the regularization of tadpole diagrams in the perturbative expansion
of the PSM involves a counterterm proportional to
$\partial_i\Pi^{ij}$, which breaks topological invariance. However, if
the Poisson tensor is divergence-free the heuristic argument applies.

We would like to point out that the cyclicity property
(\ref{cyclicity}) implies, taking $h=1$,
\begin{equation}\label{closedsp}
\int_{{\mathbb R}^n}f\star g \ \dd^n x =\int_{{\mathbb R}^n}fg \ \dd^n
x
\end{equation}
and the star product is called {\it closed}. It is interesting to
notice that the implication also works in the opposite
direction. Assume that (\ref{closedsp}) holds. Take $g=g'\star h$,
\begin{equation}\notag
\int_{{\mathbb R}^n}f\star (g'\star h) \ \dd^n x =\int_{{\mathbb
R}^n}f(g'\star h) \ \dd^n x.
\end{equation}

Now, using associativity
\begin{equation}\notag
\int_{{\mathbb R}^n}(f\star g')\star h \ \dd^n x =\int_{{\mathbb
R}^n}f(g'\star h) \ \dd^n x.
\end{equation}

Finally, applying the closedness property to the left-hand side of the
last equality gives the desired result,
\begin{equation}\notag
\int_{{\mathbb R}^n}(f\star g') h \ \dd^n x =\int_{{\mathbb
R}^n}(g'\star h)f \ \dd^n x.
\end{equation}

\newpage
\pagestyle{empty}
\chapter{Lie bialgebras and Poisson-Lie groups} \label{ch:Liebial}
\thispagestyle{empty}
\pagestyle{empty}
\cleardoublepage

\pagestyle{headings}
\renewcommand*{\chaptermark}[1]{\markboth{\small \scshape  \thechapter.\ #1}{}}
\renewcommand*{\sectionmark}[1]{\markright{\small \scshape \thesection. \ #1}}

{\it Poisson-Lie groups} are Lie groups endowed with a Poisson
structure satisfying a compatibility condition with the group
multiplication. The tangent space at the identity $e$ of the group,
$\mathfrak g$, is a Lie algebra. In addition, the Poisson structure
induces in a natural way a Lie algebra structure on the cotangent
space at $e$, ${\mathfrak g}^*$, which is the dual space of
${\mathfrak g}$. The compatibility condition between the group
multiplication and the Poisson structure yields a compatibility
condition between both Lie brackets, defining a {\it Lie bialgebra}
structure.

The theory of Lie bialgebras and Poisson-Lie groups was developed
mainly by Drinfeld and Semenov-Tian-Shansky. These objects, which play
a fundamental role in the theory of classical integrable systems, are
classical limits of the so-called {\it quantum groups}, a notion which
emerged from the quantum inverse scattering method, due to Faddeev and
collaborators. The classical counterpart plays a fundamental role in
the theory of classical integrable systems.

In this chapter we introduce the basic facts on Lie bialgebras and
Poisson-Lie groups which will be needed in the study of the PSM when
the target is a group manifold.

\section{Poisson-Lie groups and Lie bialgebras}
Consider a Lie group $G$ equipped with a Poisson structure $\{\cdot
,\cdot \}_{G}$.  It is natural to demand the Poisson structure to be
compatible with the group multiplication. To this end, endow $G\times
G$ with the product Poisson structure
\begin{equation}\notag
\{f_1,f_2\}_{G\times G}(g_{1},g_{2})= \{f_1(\cdot, g_{2}),f_2(\cdot,
g_{2})\}_{G}(g_{1}) +
\{f_1(g_{1},\cdot),f_2(g_{1},\cdot)\}_{G}(g_{2})
\end{equation}
for $f_1,f_2\in C^\infty(G\times G)$. $G$ is called a {\it Poisson-Lie
group} if the multiplication $\mu:G\times G\rightarrow G$ is a Poisson
map, i.e. if
\begin{eqnarray}\label{poissoncomp}
\{f,h\}_{G}(g_{1}g_{2})=\{f(\cdot g_{2}),h(\cdot g_{2})\}_{G}(g_{1}) +
\{f(g_{1}\cdot),h(g_{1}\cdot)\}_{G}(g_{2})
\end{eqnarray}
for $f,h \in C^{\infty}(G)$. It is evident from (\ref{poissoncomp})
that a Poisson-Lie structure always vanishes at the unit $e$ of
$G$. Therefore, the linearization of the Poisson structure at $e$
provides a Lie algebra structure on ${\mathfrak g}^*=T^*_{e}(G)$ by
the formula
\begin{eqnarray}\notag
[\dd f(e),\dd h(e)]_{{\mathfrak g}^*}=\dd \{f,h\}_G(e),\
f,h \in C^\infty(G).
\end{eqnarray}

The infinitesimal form of equation (\ref{poissoncomp}) yields a
compatibility condition between the Lie brackets of ${\mathfrak g}$
and ${\mathfrak g}^*$. Taking $X,Y \in {\mathfrak g},\ \xi,\zeta \in
{\mathfrak g}^*$ and denoting by $\left<\cdot,\cdot\right>$ the
natural pairing between elements of a vector space and its dual, it
reads\footnote{$\ad ^*$ denotes the coadjoint
representation of a Lie algebra on its dual vector space. Hence, $\xi
\in {\mathfrak g}^* \mapsto \ad ^*_\xi$ is the coadjoint
representation of $({\mathfrak g}^*,[,]_{{\mathfrak g}^*})$ on
${\mathfrak g}$.}
\newpage
\begin{eqnarray}\label{comp} 
\left<[\xi,\zeta]_{{\mathfrak g}^*},[X,Y]\right>&+&\left<\ad_Y^*\zeta,
\ad_\xi^*X\right>-\left<\ad_Y^*\xi,\ad_\zeta^*X\right>-\cr
&-&\left<\ad_X^*\zeta,\ad_\xi^*Y\right>+\left<\ad_X^*\xi,\ad_\zeta^*Y\right>=0
\end{eqnarray}
This is the condition defining a {\it Lie bialgebra} structure for
${\mathfrak g}$ (or, by symmetry, for ${\mathfrak g}^*$).

Now take ${\mathfrak g}\oplus{\mathfrak g}^*$ equipped with the
non-degenerate symmetric bilinear form
\begin{eqnarray} \label{scalarproduct}
&&(X+\xi \vert Y+\zeta)=\left<\zeta,X\right>+\left<\xi,Y\right>,\ X,Y \in
{\mathfrak g},\ \xi,\zeta \in {\mathfrak g}^*.
\end{eqnarray}

There exists a unique Lie algebra structure on ${\mathfrak
g}\oplus{\mathfrak g}^*$ such that ${\mathfrak g}$ and ${\mathfrak
g}^*$ are Lie subalgebras and that (\ref{scalarproduct}) is invariant:
\begin{eqnarray} \label{Liesuma}
\hskip -0.6cm&&[X+\xi,Y+\zeta]=[X,Y]+[\xi,\zeta]_{{\mathfrak
g}^*}-ad^{*}_{X}\zeta + ad^{*}_{Y}\xi + ad^{*}_{\zeta}X -
ad^{*}_{\xi}Y.\qquad\quad
\end{eqnarray}

The vector space ${\mathfrak g}\oplus {\mathfrak g}^*$ with the Lie bracket
(\ref{Liesuma}) is called the double of ${\mathfrak g}$ and is denoted
by ${\mathfrak g}\bowtie{\mathfrak g}^*$ or ${\mathfrak d}$.

If $G$ is connected and simply connected, (\ref{comp}) is enough to
integrate $[\cdot,\cdot]_{{\mathfrak g}^*}$ to a Poisson structure on
$G$ that makes it Poisson-Lie and the Poisson structure is
unique. Hence, there is a one-to-one correspondence between
Poisson-Lie structures on $G$ and Lie bialgebra structures on
${\mathfrak g}$.  The symmetry between ${\mathfrak g}$ and ${\mathfrak
g}^*$ in (\ref{comp}) implies that one has also a Poisson-Lie group
$G^*$ with Lie algebra $({\mathfrak g}^*,[\cdot,\cdot]_{{\mathfrak g}^*})$
and a Poisson structure $\{\cdot,\cdot\}_{G^*}$ whose linearization at $e$
gives the Lie bracket of ${\mathfrak g}$. $G^*$ is the {\it dual
Poisson-Lie group} of $G$. The connected and simply connected Lie
group with Lie algebra ${\mathfrak g}\bowtie{\mathfrak g}^*$ is known
as the {\it double group} of $G$ and denoted by ${\bf D}$.

$G$ and $G^*$ are subgroups of ${\bf D}$ and there exists a
 neighborhood ${{\bf D}_0}$ of the identity of ${\bf D}$ such that
 every element $\nu \in {{\bf D}_0}$ can be factorized as $\nu = ug =
 {\tilde g}{\tilde u},\ g,{\tilde g}\in G, \ u,{\tilde u} \in G^{*}$
 and both factorizations are unique (notice that $G_0 := G \cap G^*
 \subset {\bf D}$ is a discrete subgroup). These factorizations define
 a local left action of $G^{*}$ on $G$ and a local right action of $G$
 on $G^{*}$ by
\begin{eqnarray}\notag
&&{}^{u}g = {\tilde g}\cr
&&u^{g} = {\tilde u}.
\end{eqnarray}

Starting with the element $gu \in {\bf D}$ we can define in an
analogous way a left action of $G$ on $G^{*}$ and a right action of
$G^{*}$ on $G$. These are known as {\it dressing transformations} or
{\it dressing actions}. The symplectic leaves of $G$
(resp. $G^{*}$) are the connected components of the orbits of the
right or left dressing action of $G^{*}$ (resp. $G$).

There is a natural Poisson structure on ${\bf D}$ which will be
important for us since it will show up in the analysis of the reduced
phase space of the Poisson-Lie sigma models. Its main symplectic leaf
is ${\bf D}_0=GG^*\cap G^*G$. We write its inverse in ${\bf D}_0$,
which is a symplectic form defined at a point $ug = {\tilde g}{\tilde
u} \in {\bf D}_0$ as
\begin{equation} \label{HeisenbergDouble}
\Omega (ug) = \left<\dd\tilde{g}\tilde{g}^{-1}\stackrel{\wedge}{,}
\dd{u}{u}^{-1}\right> + \left<g^{-1}\dd
g\stackrel{\wedge}{,}\tilde{u}^{-1}\dd\tilde{u}\right>
\end{equation}
where $\langle\cdot,\cdot\rangle$ acts on the values of the
Maurer-Cartan one-forms. ${\bf D}$ endowed with the Poisson structure
yielding $\Omega$ is known as the {\it Heisenberg double}
(\cite{FalGaw91},\cite{AleMal}).

A {\it Poisson-Lie subgroup} $H \subset G$ is a subgroup which is
Poisson-Lie and such that the inclusion $i:H \hookrightarrow G$ is a
Poisson map. In particular $H$ is a coisotropic submanifold of
$G$. Let us call ${\mathfrak h}\subset {\mathfrak g}$ the Lie algebra
of $H$ and ${\mathfrak h}^0\subset {\mathfrak g}^*$ its
annihilator. $H$ is a Poisson-Lie subgroup if and only if ${\mathfrak
h}^0$ is an ideal of ${\mathfrak g}^*$, i.e. $[\xi,\zeta]_{{\mathfrak
g}^*}\in {\mathfrak h}^0,\ \forall \xi\in {\mathfrak
g}^*,\forall\zeta\in {\mathfrak h}^0$. This property permits to
restrict the bialgebra structure to ${\mathfrak h}$, which is then
called a {\it Lie subbialgebra} of ${\mathfrak g}$. The Poisson-Lie
group $H$ associated to ${\mathfrak h}$ is a subgroup of $G$. However,
in general there is no natural way to realize the dual Poisson-Lie
group $H^*$ as a subgroup of $G^*$.

\subsection{Poisson-Lie structures on simple Lie groups}
\label{sec:PLstructsimpLG}

Let us take G a complex, simple, connected and simply connected Lie
group and give the above construction explicitly. The (essentially
unique) non-degenerate, invariant, bilinear form ${\rm tr}(\ )$ on
${\mathfrak g}$ establishes an isomorphism between ${\mathfrak g}$ and
${\mathfrak g}^*$. The Poisson structure $\Pi$ contracted with the
right-invariant forms on $G$, $\theta_R (X)=\tr(\dd gg^{-1}X),\ X \in
{\mathfrak g}$, will be denoted by $\pi$,
\begin{equation} \label{PoissonContracted}
\pi_{g}(X,Y)=\iota(\Pi_{g})\theta_R(X)\wedge\theta_R(Y).
\end{equation}

For a general Poisson-Lie structure on G (\cite{LuWei}),
\begin{eqnarray} \label{PoissonLie}
\pi_{g}^{r}(X,Y)={\frac{1}{2}}\ \tr(XrY-X\Ad_{g}r\Ad_{g}^{-1}Y)
\end{eqnarray}
where $r:{\mathfrak g}\rightarrow {\mathfrak g}$ is an antisymmetric
endomorphism such that
\begin{eqnarray} \label{YangBaxter}
r[rX,Y]+r[X,rY]-[rX,rY]=\alpha[X,Y],\ \alpha \in \mathbb{C}
\end{eqnarray}
which is sometimes called {\it modified Yang-Baxter identity}. Such an
operator is what we shall understand by an {\it
$r$-matrix}\footnote{In the literature what we call $r$ is often
denoted by $R$, keeping $r$ for elements of ${\mathfrak
g}\otimes{\mathfrak g}$.}.

It is possible to show that $\Ad_{g_{0}}r = r\Ad_{g_{0}},\ g_{0} \in
G_{0}$.

The matrix $r$ allows to define a second Lie bracket on ${\mathfrak
g}$,
\begin{eqnarray} \label{rproduct}
[X,Y]_{r}={\frac{1}{2}}([X,rY]+[rX,Y])
\end{eqnarray}
which is nothing but the linearization of (\ref{PoissonLie}) at the unit
of $G$. Denoting by ${\mathfrak g}_r$ the vector space ${\mathfrak g}$
equipped with the Lie bracket $[\cdot,\cdot]_{r}$, we have that ${\mathfrak
g}_r$ is isomorphic to $({\mathfrak g}^{*},[\cdot,\cdot]_{{\mathfrak g}^*})$.

In fact, every Lie bialgebra structure on a simple Lie algebra is
given by an $r$-matrix as defined above. The pair $({\mathfrak g},r)$
is called a {\it factorizable} (resp. {\it non-factorizable} or {\it
triangular}) Lie bialgebra if $\alpha\neq 0$ (resp. $\alpha=0$). The
corresponding Poisson-Lie groups will be called either factorizable or
triangular accordingly.

Using ${\rm tr}(\ )$ it is easy to show that ${\mathfrak
g}\bowtie{\mathfrak g}^*\cong({\mathfrak g}\oplus{\mathfrak
g},[\cdot,\cdot]_{\mathfrak d})$ as Lie algebras, where
\vskip -1mm
{\footnotesize\begin{subequations}
\begin{align}
&[(X,Y),(X',Y')]_{\mathfrak d} = \notag \\
&([X,X']+{\frac{1}{2}}([X,rY']+[rY,X']+r[Y',X]+ \notag
r[X',Y]),[X,Y']+[Y,X']+[Y,Y']_{r}).
\end{align}
\end{subequations}}
\addtocounter{equation}{-1}
\vskip -3mm

We would like to point out that $r$-matrices are used not only to
construct Poisson-Lie structures but also to construct more general Poisson
structures on Lie groups. Following Semenov-Tian-Shansky (\cite{Sem85}),
if $r$ and $r'$ are antisymmetric and satisfy (\ref{YangBaxter}) with
the same value of $\alpha$, we can define the Poisson structure
(contracted with the right-invariant forms),
\begin{eqnarray} \label{generalPoisson}
\pi_{g}^{r,r'}(X,Y)=\frac{1}{2}\tr(XrY+XAd_{g}r'Ad_{g}^{-1}Y).
\end{eqnarray}
$G$ equipped with (\ref{generalPoisson}) is denoted by
$G_{r,r'}$. Given three $r$-matrices $r,r'$ and $r''$ with the same
value of $\alpha$, the map $G_{r,-r'}\times G_{r',r''}\rightarrow
G_{r,r''}$ is Poisson. When $r'=r$ and $r''=-r$ we get
(\ref{PoissonLie}) and the Poisson structure is, indeed, Poisson-Lie.

\section{The double of a Lie bialgebra}

 It turns out that the double ${\mathfrak d}$ has a very different
aspect for $\alpha \neq 0$ and $\alpha=0$. In this section we rederive
an approach (appeared already in \cite{Sto}) that allows us to
understand the cases $\alpha \neq 0$ and $\alpha = 0$ in a unified
way.

Consider the Lie algebra ${\mathfrak G}={\mathfrak g}[[\varepsilon]]$
of polynomials on a variable $\varepsilon$ with coefficients in
${\mathfrak g}$ (always a simple complex Lie algebra from now on)
where
\begin{eqnarray}\notag
[\sum_{m=0}^{M} X_m \varepsilon^m,\sum_{n=0}^{N} X'_n
\varepsilon^n]=\sum_{m=0}^{M}\sum_{n=0}^{N}\varepsilon^{m+n}[X_m,X'_n].
\end{eqnarray}
${\mathfrak G}_{\alpha}=(\varepsilon^2 - \alpha){\mathfrak G}$ is an
ideal of ${\mathfrak G}$, hence ${\mathfrak G}/{\mathfrak G}_{\alpha}$
inherits in a canonical way a Lie algebra structure from ${\mathfrak
G}$. In practice, it is more useful to think of ${\mathfrak
G}/{\mathfrak G}_{\alpha}$ as the set $\{X+Y\varepsilon\vert \ X,Y \in
{\mathfrak g},\ {\varepsilon}^2 = \alpha\}$. Then, the Lie bracket of
two elements of ${\mathfrak G}/{\mathfrak G}_{\alpha}$ can be written
as
\begin{eqnarray}\notag
[X+Y\varepsilon,X'+Y'\varepsilon]= [X,X']+\alpha [Y,Y'] +
([X,Y']+[Y,X'])\varepsilon.
\end{eqnarray}

There exists an isomorphism of Lie algebras between $({\mathfrak
g}\oplus{\mathfrak g},[\cdot,\cdot]_{\mathfrak d})$ and ${\mathfrak G}/{\mathfrak
G}_{\alpha}$ given by $(X,Y)\mapsto X +
{\frac{1}{2}}rY+{\frac{1}{2}}Y\varepsilon$ and, consequently, ${\mathfrak
d}\cong {\mathfrak G}/{\mathfrak G}_{\alpha}$. Furthermore,
$${\mathfrak g}\cong \{X \mid X \in {\mathfrak g}\}, \ {\mathfrak g}_r
\cong \{rX + X\varepsilon \mid X\in {\mathfrak g}\} \subset {\mathfrak
G}/{\mathfrak G}_{\alpha}.$$

\subsection{Factorizable Lie bialgebras}

It is clear that if the Lie bialgebra is factorizable (i.e. $\alpha
\neq 0$), the subalgebras $\{(1+\frac{\varepsilon}{\sqrt \alpha})X\mid
X\in {\mathfrak g}\}$, $\{(1-\frac{\varepsilon}{\sqrt \alpha})X\mid
X\in {\mathfrak g}\}$ commute with one another and both are isomorphic
to ${\mathfrak g}$. In fact, ${\mathfrak G}/{\mathfrak G}_{\alpha}
\cong {\mathfrak g}\oplus {\mathfrak g}$ with the natural Lie bracket:
$$[(X_{1},X_{2}),(Y_{1},Y_{2})] = ([X_{1},Y_{1}],[X_{2},Y_{2}])$$ with
the isomorphism given by $X + Y\varepsilon \mapsto (X + Y, X - Y)$.
%In the sequel, if not stated otherwise, ${\mathfrak g}\oplus
%{\mathfrak g}$ will denote the direct sum of Lie algebras. 
Hence, the
double of a factorizable Lie bialgebra is isomorphic to ${\mathfrak
g}\oplus {\mathfrak g}$ and ${\bf D}=G\times G$.

As deduced from (\ref{YangBaxter}) $r_{\pm}={\frac{1}{2}}(r \pm
\sqrt{\alpha})$ are Lie algebra morphisms from ${\mathfrak g}_{r}$ to
${\mathfrak g}$, i.e.
\begin{equation} \label{Liealghomomorphisms}
r_{\pm}[X,Y]_{r} = [r_{\pm}X,r_{\pm}Y]
\end{equation}
and we have the following embeddings of ${\mathfrak g}$ and
${\mathfrak g}_{r}$ in ${\mathfrak g}\oplus {\mathfrak g}$:
$${\mathfrak g}_d = \{(X,X)\mid X\in {\mathfrak g}\},\quad {\mathfrak
g}_r = \{(r_{+}X,r_{-}X) \mid X \in {\mathfrak g}\}.$$

We shall use the same notation ${\mathfrak g}_r$ for $({\mathfrak
g},[\cdot,\cdot]_r)$ and for its embedding in ${\mathfrak g}\oplus {\mathfrak
g}$. This should not lead to any confusion.

The map $X \mapsto (r_{+}X,r_{-}X)$ is non-degenerate as long as
$\alpha \neq 0$. Thus we can recover $X$ through the formula
$X={\alpha}^{-{\frac{1}{2}}}(r_{+}X-r_{-}X)$. We shall often use the
notation $X_{\pm}=r_{\pm}X$.

Notice that ${\mathfrak g}_{\pm}:= r_{\pm}{\mathfrak g}$ are Lie
subalgebras of ${\mathfrak g}$ and denote by $G_{\pm}$ the subgroups
of $G\times G$ integrating ${\mathfrak g}_{\pm}$. We have the
following embeddings of $G$ and $G^*$:
$$G_d = \{(g,g)\in {\bf D} \ \vert \ g\in G\},\quad G_r =
\{(g_+,g_-)\in {\bf D} \ \vert \ g_+\in G_+,g_-\in G_-\}.$$

The dressing transformations are given by the solutions of the
factorization problem
\begin{equation}\notag
(h_+,h_-)(g,g)=({\tilde g},{\tilde g})({\tilde h}_+,{\tilde h}_-).
\end{equation}

Consider the map
\begin{eqnarray}\label{mappi}
 \begin{matrix}\nu:&G_r&\longrightarrow&G \cr
&(g_+,g_-)&\longmapsto&g_-g_+^{-1}.
\end{matrix}
\end{eqnarray}
It is a submersion so that $\nu(G_r)$ is an open connected subset of
$G_r$ containing the unit. Clearly, $\nu(g_+,g_-)=\nu({\tilde
g}_+,{\tilde g}_-)$ if and only if ${\tilde g}_\pm = g_\pm g_0$ with
$g_0\in G_0$.

\vskip 3mm

We can write now explicitly the Poisson-Lie structure on $G_{r}$ dual
to (\ref{PoissonLie}). After contraction with the right-invariant
forms on the group $G_{r}$, $\theta^{r}_R(X)=\tr[(dg_{+}g_{+}^{-1}-
dg_{-}g_{-}^{-1})X]$ for $X \in {\mathfrak g}$, it takes the form
\begin{eqnarray} \label{dualPoissonLie}
\pi_{(g_{+},g_{-})}^{r}(X,Y)=\tr\left(X(Ad_{g_{+}}-Ad_{g_{-}})
(r_{-}Ad_{g_{+}}^{-1}-r_{+}Ad_{g_{-}}^{-1})Y\right)
\end{eqnarray}
which verifies, in particular, that its linearization at the unit of
$G_r$ gives the Lie bracket of $\mathfrak g$. The symplectic leaves of
(\ref{dualPoissonLie}) are the connected components of the preimages
by the map $\nu$ (\ref{mappi}) of the conjugacy classes in $G$ (see
\cite{Sem85}).

\vskip 3mm

Given an $r$-matrix in $G$ it is possible to define an $r$-matrix in
${\bf D}=G \times G$ (which we shall denote by $R$) in a natural way:
$R:= P_{d}-P_{r}$ where $P_{d}$ and $P_{r}$ are the projectors on ${\gl
g}$ and ${\gl g}_{r}$ respectively, parallel to the complementary
subalgebra. Hence, $R_{+}=P_{d},\ R_{-}=-P_{r}$.

Using the explicit description of the double group we can write the
symplectic structure (\ref{HeisenbergDouble}) at a point
$(h_{+}g,h_{-}g)=({\tilde g}{\tilde h}_{+}, {\tilde g}{\tilde
h}_{-})\in {\bf D}_{0}$ as
\begin{eqnarray}\label{HeiDoubleFact}
\Omega ((h_{+}g,h_{-}g)) &=&{\rm tr}(\dd\tilde{g}\tilde{g}^{-1}\wedge 
(\dd h_{+}h_{+}^{-1}-
\dd h_{-}h_{-}^{-1}))+\cr
&+&\tr(g^{-1}\dd g\wedge (\tilde{h}_{+}^{-1}\dd\tilde{h}_{+}-
\tilde{h}_{-}^{-1}\dd\tilde{h}_{-}))
\end{eqnarray}
which corresponds to ${\bf D}_{R,R}$ with the notation introduced in Section
\ref{sec:PLstructsimpLG}.

\vskip 0.5 cm

The $r$-matrix allows to write the closed 3-form $\chi=
\frac{1}{3}\tr[(g^{-1}\dd g)^{\wedge3}]$ on $G$ as the differential of
a 2-form on $\nu(G_r)\subset G$. Let us show this in detail because it
will be important in Chapter \ref{PLSM} in relation with the $G/G$ WZW
model. The pullback of $\chi$ by $\nu$ has the form
\begin{eqnarray}\label{pichi}
\nu^*\chi&=&\frac{1}{3}\tr[(g_{-}^{-1} \dd\gm -\gpinv \dd\gp)^{\wedge3}]\cr\cr
           &=&\dd\ \tr[g_{-}^{-1} \dd\gm \wedge\gpinv \dd\gp]+\cr
&&+{1\over 3}\tr[(g_{-}^{-1} \dd\gm)^{\wedge3}]
  -{1\over 3}\tr[(\gpinv \dd\gp)^{\wedge3}].
\end{eqnarray}
Now define $\gamma=\gpinv \dd\gp -g_{-}^{-1}
\dd\gm\in\Omega^1(G_r)\otimes{\gl g}$. One has
$r_{\pm}\gamma=g_{_\pm}^{-1}\dd g_{_\pm}$ and
\begin{eqnarray}
\tr[(\gpinv \dd\gp)^{\wedge3}]&=&\tr[(r_+\gamma)^{\wedge3}]\cr
                    &=&{1\over 2}\tr[r_+(\gamma\wedge r\gamma+r\gamma\wedge\gamma)
                           \wedge r_+\gamma]\cr
                    &=&-{1\over 2}\tr[(\gamma\wedge r\gamma + r\gamma\wedge\gamma)
                             \wedge r_-r_+\gamma]
\end{eqnarray}
where we have used (\ref{Liealghomomorphisms}) and the antisymmetry of
$r$ with respect to the bilinear form $\tr( \ )$.  The same result is
obtained for $\tr[(g_{-}^{-1} \dd\gm)^{\wedge3}]$, so that the last
two terms in (\ref{pichi}) cancel each other. Hence, $\chi= \dd\rho$
on $\nu(G_r)$ where $\rho$ is a 2-form on $\nu(G_r)$ such that
 \begin{eqnarray}\label{rho}
\nu^*\rho=\tr[g_{-}^{-1} d\gm \wedge\gpinv d\gp]\,.
\end{eqnarray}

A straightforward computation leads to an equivalent expression 
for the symplectic form $\Omega$ on the leaf
${\bf D}_0$ in the Heisenberg double:
\begin{eqnarray}\label{Omega2}
\Omega(hg_+,hg_-)&=&\tr[(g^{-1}\dd g+\dd gg^{-1}+g^{-1}h^{-1}\dd hg)h^{-1}\dd h]\cr&&
\hskip -2mm+\,\rho(g)-\rho(hgh^{-1}).
\end{eqnarray}               
The right hand side may be viewed as the 2-form on the space of pairs
$(h,g)$ such that both $g=g_-g_+^{-1}$ and $hgh^{-1} =\tilde
g_-{\tilde g}_+^{-1}$ are in $\nu(G_r)$. We shall denote it then as
$\Omega(h,g)$.

%%Then, if ${\mathfrak g}$ is simple (as we will always assume),
%%${\mathfrak d}$ is semisimple and the Manin triple can be constructed
%%as $({\mathfrak g}\oplus {\mathfrak g},{\mathfrak g}_{d},{\mathfrak
%%g}_{r})$ with the scalar product $<(X,Y),(X',Y')> =\\tr(XX'-YY')$.

\vskip 0.4 cm

\begin{example}
Let ${\mathfrak g}$ be a simple Lie algebra over ${\mathbb C}$ and
$\Delta$ its set of roots. The decomposition in root spaces reads
\begin{eqnarray} \label{decomposition}
{\mathfrak g} = {\mathfrak t} \oplus \bigg(\bigoplus_{\beta \in \Delta}
{\mathfrak g}_{\beta}\bigg)
\end{eqnarray}
where ${\mathfrak t}$ is a Cartan subalgebra of ${\mathfrak g}$ and
$${\mathfrak g}_{\beta}=\{{\mathbb C}X_{\beta}\mid
[T,X_{\beta}]=\beta(T)X_{\beta},\forall T\in {\mathfrak t}\}.$$ Given
a splitting into positive and negative roots, $\Delta = \Delta_+ \cup
\Delta_-$, any element $X \in {\mathfrak g}$ can be written in a
unique way as $X = X^{(+)} + X^{(-)} + T$, where $X^{(\pm)} \in {\rm
span}(X_\beta,\ \beta \in \Delta_{\pm})$ and $T\in{\mathfrak t}$.

The {\it standard $r$-matrix} is defined by $r=r_+ + r_-$ with
\begin{eqnarray}\label{standardr}
&&r_+ X = X^{(+)}+\frac{1}{2}T \notag\\
&&r_- X = -X^{(-)}-\frac{1}{2}T
\end{eqnarray}
which is a factorizable $r$-matrix with $\alpha=1$.

Take as a particular case ${\mathfrak g}={\mathfrak sl}(n,{\mathbb
C})$ with the standard $r$-matrix. Then, ${\mathfrak sl}(n,{\mathbb
C})_r \subset {\mathfrak sl}(n,{\mathbb C})\oplus {\mathfrak
sl}(n,{\mathbb C})$ consists of pairs $(X_+,X_-)$ where $X_+$ (resp.$
X_-$) is upper (resp. lower) triangular and ${\rm diag}(X_+)=-{\rm
diag}(X_-)$.

At the group level, $SL(n,{\mathbb C})_r \subset SL(n,{\mathbb
C})\times SL(n,{\mathbb C})$ is the set of pairs $(g_+,g_-)$ such that
$g_+$ (resp. $g_-$) is upper (resp. lower) triangular and ${\rm
diag}(g_+)={\rm diag}(g_-)^{-1}$.
\end{example}

\subsection{Triangular Lie bialgebras}

In this subsection we describe the double of a triangular Lie
bialgebra and the double of the associated Poisson-Lie groups. We
shall use these results for writing explicitly the Poisson structure
dual to (\ref{PoissonLie}).

If $\alpha = 0$, $r_{\pm}$ degenerate to ${\frac{1}{2}}r$, the map $X
\mapsto ({\frac{1}{2}}rX,{\frac{1}{2}}rX)$ is not invertible and ${\mathfrak
G}/{\mathfrak G}_{0}$ is no longer isomorphic to ${\mathfrak g}\oplus
{\mathfrak g}$. Indeed, ${\mathfrak G}/{\mathfrak G}_{0}=\{X +
Y\varepsilon \mid {\varepsilon}^2 = 0\}$ is not semisimple, for the
elements $X\varepsilon$ form an abelian ideal as seen from the Lie
bracket
\begin{eqnarray} \label{bracket}
[X+Y\varepsilon,X'+Y'\varepsilon]=[X,X']+([X,Y']+[Y,X'])\varepsilon.
\end{eqnarray}

This is the Lie algebra of the tangent bundle of $G$, $TG\cong G\times
{\mathfrak g}$, with the natural group structure given by the semidirect
product
\begin{eqnarray} \label{semidirect}
(g,X)(g',X')=(gg',\Ad_{g}X'+X).
\end{eqnarray}
Hence, the double of a triangular Lie bialgebra is isomorphic to the
tangent bundle of $G$ with the product given by (\ref{semidirect}).

We can represent the elements of the double as
\begin{eqnarray}\notag
{\bf D}= \left\{ \begin{pmatrix}e&0 \\ X&e\end{pmatrix} g \mid g\in G,\ X\in {\mathfrak g}\right\}
\end{eqnarray}
where the product is now the formal product of matrices, resulting the
semidirect product mentioned above. Its Lie algebra with this notation is
\begin{eqnarray}\notag
{\mathfrak d}=\left\{ \begin{pmatrix}X&0 \cr Y&X \end{pmatrix} \mid  X,Y \in {\mathfrak g}\right\}
\end{eqnarray}
and the Lie bracket (\ref{bracket}) is given by the formal commutator
of matrices. The embeddings of ${\mathfrak g}$ and ${\mathfrak g}_r$
in ${\mathfrak d}$ are given by
\begin{eqnarray}\notag
{\mathfrak g}_d = \left\{ \begin{pmatrix}X&0 \cr 0&X\end{pmatrix} \mid  X \in {\mathfrak g}\right\},\quad {\mathfrak g}_r = \left\{ \begin{pmatrix}rX&0 \cr X&rX \end{pmatrix} \mid  X \in {\mathfrak g}\right\}.
\end{eqnarray}

%The Manin triple in this case may be seen as $({\mathfrak
%d},{\mathfrak g}_{d},{\mathfrak g}_{r})$ with the scalar product
%$$<\begin{pmatrix}X&0\cr Y&X \end{pmatrix},\begin{pmatrix}X'&0\cr
%Y'&X' \end{pmatrix}>=\\tr(XY'+YX')$$

Both subalgebras exponentiate to subgroups $G_d,G_{r} \subset {\bf D}$. 
%(we
%shall use the same notation for $G$ and its embedding as a subgroup of
%the double). 
Clearly,
\begin{eqnarray}\notag
G_d=\left\{ \begin{pmatrix}g&0 \cr 0&g \end{pmatrix}\mid g\in G \right\}
\end{eqnarray}
whereas for $G_{r}$ the description is less explicit. It is the subgroup of ${\bf D}$ generated by elements of the form:
\begin{eqnarray} \label{Gr}
{\begin{pmatrix}e&0 \cr Y&e \end{pmatrix}}e^{rX} \quad {\rm with}\quad
Y=\int^{1}_{0}\Ad_{e^{srX}}X \dd s,\ X\in {\mathfrak g}.
\end{eqnarray}
%i.e. every element of $G_r$ can be written as a product of a finite
%number of elements of the form (\ref{Gr}). 
We will denote a general element of $G_r$ by
\begin{eqnarray}
{\bar Y}={\begin{pmatrix}e&0 \cr Y&e \end{pmatrix}}h_Y
\end{eqnarray}
where, in the general case, $Y$ belongs to a dense subset of ${\mathfrak g}$ and determines
$h_{Y}$ up to multiplication by an element of $G_0$. This means that
$${\rm if} \quad  {\begin{pmatrix}e&0 \cr Y&e \end{pmatrix}}h_Y \in G_{r},\ {\rm then}$$
$${\begin{pmatrix}e&0 \cr Y&e\end{pmatrix}}h'_Y \in G_{r} \Leftrightarrow h'_{Y} = h_{Y}g_{0}, \ g_{0} \in G_{0}.$$
As a consequence, the notation ${\bar Y}$ has a small ambiguity, but
we shall use of it for brevity wherever it does not lead to
confusion.

The right-invariant forms on $G_{r}$ are
$$\theta^r_R(Y)=\tr\left(\left(\dd X+[X,\dd
h_{X}h_{X}^{-1}]\right)Y\right)=\tr\left(\dd
\left(\Ad_{h_X}^{-1}X\right)\Ad_{h_X}^{-1}Y\right)$$
for $Y\in{\mathfrak g}$ and $\begin{pmatrix}e&0\cr X&e\end{pmatrix}h_X\in G_r$.
Whereas the left invariant forms read
$$\theta^r_L(Y)=\tr\left(Y\Ad^{-1}_{h_X}\dd X\right).$$

The concrete realization of $G^*$ described above allows us to write
the Poisson structure dual to (\ref{PoissonLie}) in explicit terms. It
can be checked after a straightforward (although lengthy) calculation
that the dual Poisson-Lie structure contracted with the
right-invariant forms is
\begin{eqnarray} \label{dualPoissonLieTriangular}
\pi_{\bar X}^{r}(Y,Z)=
\tr\left(Y[X,Z]-[X,Y]\Ad_{h_X}r\Ad_{h_X}^{-1}[X,Z]\right).
\end{eqnarray}

In the triangluar case, the symplectic structure of the Heisenberg
double (\ref{HeisenbergDouble}) at a point ${\bar X}g = {\tilde
g}{\tilde{\bar X}} \in {\bf D}_0$ can be written as
\begin{equation}
\Omega ({\bar X}g) = \tr \left(\dd {\tilde g} {\tilde
g}^{-1}\wedge(\dd X+[X,\dd h_X h_X^{-1}]) + g ^{-1}\dd g\wedge
\Ad_{h_{\tilde X}}^{-1}\dd {\tilde X} \right).
\end{equation}

Let us finish this chapter with two examples of triangular Lie
bialgebras:

\begin{example}\label{ex:zeroPL}
$r=0$ is an $r$-matrix with $\alpha = 0$. It endows $G$ with trivial
Poisson bracket $\{\cdot,\cdot\}_G = 0$ and ${\mathfrak g}^*$ with trivial
Lie bracket $[\cdot,\cdot]_{{\mathfrak g}^*}=0$. The dual Poisson Lie group
$G^*$ is ${\mathfrak g}^*$ viewed as an abelian group and equipped
with the Kostant-Kirillov Poisson bracket defined in Example
\ref{KostantKirillov}.
\end{example}

\vskip 0.5 cm

\begin{example}
Take ${\mathfrak g}$ a complex simple Lie algebra. If $\tau_{\mathfrak
t}:{\mathfrak g}\rightarrow {\mathfrak t}$ is the projector onto the
Cartan subalgebra ${\mathfrak t}$ with respect to the decomposition
(\ref{decomposition}) and ${\cal O}:{\mathfrak t}\rightarrow
{\mathfrak t}$ is an antisymmetric endomorphism of ${\mathfrak t}$
with respect to ${\rm tr}(\ )$, then $r={\cal
O}\tau_{\mathfrak t}$ is an $r$-matrix with $\alpha = 0$.

It is worth studying in detail the structure of $G_{r}$ for this
$r$. As we know, $G_{r}$ is generated by elements of the form
\begin{eqnarray}\label{dualgroup}
\begin{pmatrix}e&0\cr Y&e \end{pmatrix}e^{rX}, 
\ Y=\int^{1}_{0}\Ad_{e^{srX}}X\dd s,\ X\in {\mathfrak g}.
\end{eqnarray}

The elements $\{rX\mid X\in{\mathfrak g}\}$ span a subalgebra of
${\mathfrak t}$ (therefore abelian) and its exponentiation will be a
subgroup of the Cartan subgroup of $G$, so we concentrate on $Y$.

Take $$X=T + \sum_{\beta \in \Delta} a_{\beta}X_{\beta}, \ T \in
{\mathfrak t}.$$

Observing that
$$\Ad_{e^{srX}}=e^{\mbox{{\scriptsize ad}}_{srX}}$$
we straightforwardly obtain
\begin{eqnarray}\notag
\Ad_{e^{srX}}X = T + \sum_{\beta \in \Delta} e^{s\beta
(rT)}a_{\beta}X_{\beta}
\end{eqnarray}
and, therefore,
\begin{eqnarray}\notag
Y = T + \sum_{\beta \in \Delta}{1\over{\beta(rT)}}\left(e^{\beta
(rT)}-1\right)a_{\beta}X_{\beta}
\end{eqnarray}
where, if some $\beta(rT)=0$, the limit $\beta(rT)\rightarrow 0$
must be understood in the last expression.

The product of $n$ elements of the form (\ref{dualgroup})
$$\begin{pmatrix}e&0\cr
Y_{1}&e \end{pmatrix}e^{rX_{1}}\dots \begin{pmatrix}e&0\cr Y_{n}&e
\end{pmatrix}e^{rX_{n}}:=\begin{pmatrix}e&0\cr Y&e \end{pmatrix}T_{Y}$$
can be computed explicitly:

\begin{eqnarray}\notag
%\label{generaldualgroup}
T_{Y} &=& e^{rT_{1} + \dots + rT_{n}}\cr\cr
Y &=& T_{1}+ \dots +T_{n} +\cr\cr
&+&\sum_{\beta \in
\Delta}\bigg[{1\over{\beta(rT_{1})}}\left(e^{\beta (rT_{1}+ \dots
+rT_{n})}-e^{\beta (rT_{2}+ \dots +rT_{n})}\right)a_{\beta,1} +\cr
&+& {1\over{\beta(rT_{2})}}\left(e^{\beta (rT_{2}+ \dots
+rT_{n})}-e^{\beta (rT_{3} + \dots +rT_{n})}\right)a_{\beta,2} + \cr
&+& \dots + {1\over{\beta(rT_{n})}}\left(e^{\beta
(rT_{n})}-1\right)a_{\beta,n}\bigg]X_{\beta}
\end{eqnarray}
where we have used the notation
$$X_i=T_i + \sum_{\beta \in \Delta} a_{\beta ,i}X_{\beta}, \ T_i \in
{\mathfrak t}.$$

It is clear that, in this case, $G_{0}=e$ and that $Y$ fills
${\mathfrak g}$. As a consequence, the dressing actions are globally
defined, ${\bf D}=GG_{r}=G_{r}G$ and the factorizations are unique.
\end{example}

\newpage
\pagestyle{empty}
\chapter{The Poisson sigma model over Lie groups}\label{PLSM}
\thispagestyle{empty}
\pagestyle{empty}
\cleardoublepage

\pagestyle{headings}
\renewcommand*{\chaptermark}[1]{\markboth{\small \scshape  \thechapter.\ #1}{}}
\renewcommand*{\sectionmark}[1]{\markright{\small \scshape \thesection. \ #1}}

In this chapter we shall be interested in the Poisson sigma model when
the target is a Lie group. In this case it is natural to demand the
compatibility between its Poisson and group structures and this leads
to the concept of Poisson-Lie group introduced in Chapter
\ref{ch:Liebial}. Examples of Poisson-Lie sigma models have been
studied in \cite{FalGaw02},\cite{AleSchStr} in connection to $G/G$
Wess-Zumino-Witten theories. Here we pursue a systematic study of the
matter.

One of the simplest examples of Poisson-Lie sigma model is the linear
one: $M$ is a vector space (abelian group) and its Poisson structure
is linear. This model is related to $BF$ and Yang-Mills theories and
can be considered as dual of the trivial Poisson-Lie sigma model in an
(in general) non-abelian group with vanishing Poisson bracket.  Our
models can be regarded as the simplest generalizations of the linear
ones with which they share some properties that will be stressed in
the sequel.

Another aspect that will deserve our attention is that of duality,
i.e. we shall try to relate the two dual models, which is not obvious
at the Lagrangian level. However, in the Hamiltonian approach, in the
open geometry, the duality becomes evident: it consists in the
exchange of bulk and boundary degrees of freedom. It would be
interesting to relate this fact with the non abelian T-duality
discussed in
\cite{AleKliTse},\cite{KliSev951},\cite{Klisev952},\cite{KliSev96},
but this point is not clear to us at the present time.

\vskip 1 cm

 When the target manifold is a Lie group the action of the Poisson
 sigma model can be recasted in terms of a set of fields adapted to
 the group structure. $T^{*}G$ can be identified, by right
 translations, with $G\times{\mathfrak g}^*$ and, using ${\rm tr}(\
 )$, with $G \times {\mathfrak g}$. Then, in (\ref{PS}) we can take
 $A\in\Omega^1(\Sigma)\otimes{\mathfrak g}$ and $g:\Sigma\rightarrow
 G$ as fields (equivalently, $\eta\equiv\tr(\dd gg^{-1}\wedge A)$) and
 use the Poisson structure contracted with the right-invariant forms
 on $G$ (\ref{PoissonContracted}). Denoting by
 $\pi_g^\sharp:{\mathfrak g}\rightarrow {\mathfrak g}$ the
 endomorphism induced by $\pi_g$ using ${\rm tr}(\ )$ we can write the
 action of the Poisson sigma model as
 \begin{equation} \label{PSgroup}
S(g,A)=\int_{\Sigma}{\tr(\dd gg^{-1}\wedge A)-{\frac{1}{2}}\tr(A\wedge
\pi_g^\sharp A)}.
\end{equation}

In particular, for the Poisson-Lie structure (\ref{PoissonLie}) we have
\begin{eqnarray}\label{PLS}
S_{PL}(g,A)=\int_{\Sigma}{\tr(\dd gg^{-1}\wedge A)-{1\over4}\tr(A\wedge
(r-\Ad_{g}r\Ad_{g}^{-1})A)}
\end{eqnarray}
which is the action of what we shall call {\it Poisson-Lie sigma
model} with target $G$.

The equations of motion are
\begin{subequations} \label{eomG}
\begin{align}
&\dd gg^{-1}+{\frac{1}{2}}(r-\Ad_{g}r\Ad_{g}^{-1})A=0\label{eomGa} \\
&\dd {\tilde A}+[{\tilde A},{\tilde A}]_{r}=0,\ {\tilde A}:=\Ad_{g}^{-1}A
\label{eomGb}
\end{align}
\end{subequations}
from which a zero curvature equation can be also derived for $A$,
\begin{eqnarray}\label{Aflat}
&&\dd A+[{A},{A}]_{r}=0.
\end{eqnarray}

The infinitesimal gauge symmetry of the action, for
$\beta:\Sigma\rightarrow {\mathfrak g}$ is
\begin{subequations} \label{gaugesymmetryG}
\begin{align}
&\delta_{\beta}g g^{-1}={\frac{1}{2}}(\Ad_{g}r\Ad_{g}^{-1}-r)\beta 
\label{gaugesymmetryGa} \\
&\delta_{\beta}A=\dd \beta +
[A,\beta]_{r}-{\frac{1}{2}}[\dd gg^{-1}+{\frac{1}{2}}(r-\Ad_{g}r\Ad_{g}^{-1})A,\beta]
\label{gaugesymmetryGb}
\end{align}
\end{subequations}
which corresponds to the right dressing vector fields of \cite{LuWei}
translated to the origin by right multiplication in $G$. Its
integration (local as in general the vector field is not complete)
gives rise to the dressing transformation of $g$. On-shell,
$[\delta_{\beta_{1}},\delta_{\beta_{2}}]=\delta_{[\beta_{1},\beta_{2}]_{r}}$,
so that the `structure functions' of the commutation relations are
field-independent, unlike in the general case (see equation
(\ref{OpenAlgebra})). Thus we can talk properly about a gauge
group. In fact, the gauge group of the Poisson sigma model over $G$ is
its dual $G^*$.

Up to here we have not needed to distinguish between $\alpha=0$ and
$\alpha\neq 0$. In order to study further (\ref{PLS}) and its dual
model we need to make use of the embeddings of $G$ and $G^{*}$ in the
double ${\bf D}$. As we have learnt, ${\bf D}$ is very different in the
factorizable and triangular cases and we must analyse them
separately.

\section{Factorizable Poisson-Lie sigma models} \label{FactPLSM}

Without loss of generality we take $\alpha=1$ in equation
(\ref{YangBaxter}). Since the equation of motion for $A$ is
independent of $g$, we first solve the equations (\ref{eomGb}) and
(\ref{Aflat}). This is in contrast with the general case of
\cite{BojStr}. Locally,
\begin{subequations}\notag
\begin{align}
&A=h_{+}\dd h_{+}^{-1}-h_{-}\dd h_{-}^{-1} \label{solutionA} \\
&{\tilde A}= {\tilde h}_{+}\dd{\tilde h}_{+}^{-1}-{\tilde h}_{-}\dd{\tilde
h}_{-}^{-1}. \label{solutionAtilde}
\end{align}
\end{subequations}

We need to work a bit more to get explicit solutions for $g$. 
The equation of motion (\ref{eomGa}) can be equivalently written as
$$\dd gg^{-1}+(r_{+}-Ad_{g}r_{+}Ad_{g}^{-1})A=0$$
and recalling that ${\tilde A}=Ad_{g}^{-1}A$,
$$\dd gg^{-1}+r_{+}A-Ad_{g}r_{+}{\tilde A}=0.$$

The operator $r_{+}$ projects on to the $+$ component of the elements
of ${\gl g}$ and yields
$$\dd gg^{-1}+h_{+}\dd h_{+}^{-1}-Ad_{g}{\tilde h}_{+}\dd{\tilde
h}_{+}^{-1}=0$$
which is equivalent to
\begin{eqnarray}\notag
{\tilde h}_{+}^{-1}g^{-1}h_{+}d(h_{+}^{-1}g{\tilde h}_{+})=0 \Rightarrow h_{+}^{-1}g{\tilde h}_{+}={\hat g},\quad {\hat g}=g(\sigma_0) \in G.
\end{eqnarray}

The same procedure can be carried out by using $r_{-}$,
\begin{eqnarray}\notag
h_{-}^{-1}g{\tilde h}_{-}={\hat g}.
\end{eqnarray}

The solutions of the equations of motion can then be expressed by the
single relation in ${\bf D}$:
\begin{eqnarray} \label{solFact}
(h_+(\sigma)\hat g,h_-(\sigma)\hat g)=
(g(\sigma){\tilde h}_+(\sigma),g(\sigma)\tilde h_-(\sigma))
\end{eqnarray}
which in particular implies that
$(h_+(\sigma)\hat g,h_-(\sigma)\hat g)\in {\bf D}_0$, 
and $g$ take values in the
connected component of orbits of $\hat g$ by dressing transformations. 
These orbits are the symplectic leaves of the Poisson-Lie group $G$.

We go on to study the reduced phase space of the model in the open
geometry, $\Sigma={\mathbb R}\times[0,{1}]$. For the moment we take
$g$ free at the boundary. Equivalently, we take a brane which is the
whole target manifold $G$. The $A$ field must then vanish on vectors
tangent to $\partial\Sigma$, so that $h_\pm, {\tilde h}_\pm$ are
constant along the connected components of the boundary. Writing
$\sigma=(t,x)$, we have $h_\pm(t,0)=h_{0\pm}$, ${\tilde
h}_{\pm}(t,0)={\tilde h}_{0\pm}$, $h_\pm(t,{1})= h_{{1}\pm}$, ${\tilde
h}_\pm(t,{1})={\tilde h}_{{1}\pm}$.

Denote $I=[0,{1}]$. The canonical symplectic form on the space of
continuous maps $(g,A):TI\rightarrow G\times {\mathfrak g}$ with
continuously differentiable base map is:
$$\omega =\int_0^{1} \tr(\delta g g^{-1}\wedge \delta g g^{-1}A-\delta
g g^{-1}\wedge \delta A)\dd x.$$

When restricted to the solutions of the equations of motion the
symplectic form $\omega$ becomes degenerate, its kernel given by the
gauge transformations (\ref{gaugesymmetryG}) which vanish at
$ x=0,{1}$. By definition, the reduced phase space ${\bar{\cal P}}(G;G,G)$
is the (possibly singular) quotient of the space of solutions by the
kernel of $\omega$.

If we parametrize the solutions in terms of $h_\pm(\sigma)$ and ${\hat
g}$ we obtain
\begin{eqnarray}\notag
\omega={\frac{1}{2}}\int_0^{1} \partial_ x\Omega((h_+(\sigma)\hat
g,h_-(\sigma)\hat g))\dd x.
\end{eqnarray}
That is, $\omega$ depends only on the values of the fields at the
boundary (i.e. the degrees of freedom of the theory are all at the
boundary, as expected from the topological nature of the model) and is
expressed in terms of the symplectic structure on the Heisenberg
double (\ref{HeiDoubleFact}). Namely,
$$\omega={\frac{1}{2}}[\Omega((h_{{1}+}\hat g,h_{{1}-}\hat g))-
\Omega((h_{0+}\hat g,h_{0-}\hat g))].$$

Or if we take $\sigma_0=(t_0,0)$, i.e. $h_{0\pm}={\tilde h}_{0\pm}=e$
$$\omega={\frac{1}{2}}\Omega((h_{{1}+}\hat g,h_{{1}-}\hat g))$$

The reduced phase space ${\bar{\cal P}}(G;G,G)$ is the set of pairs
$([(h_+,h_-)],\hat g)$ with $[(h_+,h_-)]$ a homotopy class of maps
from $[0,{1}]$ to $G_r$ which are the identity at $0$, are fixed at
${1}$ and such that $(h_+(x)\hat g,h_-(x)\hat g)\in {\bf D}_0$,
$\forall x\in[0,{1}]$. The symplectic form on ${\bar{\cal P}}(G;G,G)$
can be viewed as the pull-back of $\Omega$ by the map
$([h_{+},h_{-}],{\hat g})\mapsto (h_{{1} +}{\hat g},h_{{1} -}{\hat
g})$.

\subsection{The dual factorizable model and G/G theory}

Using the Poisson structure given in (\ref{dualPoissonLie}) the action
of the dual model reads
\begin{eqnarray}\label{dualPLS}
S_{PL}^{*}(g_{+},g_{-},A)&=&\int_\Sigma \tr[({\dd}\gp\gpinv
-{\dd}\gm\gminv)\wedge A+ \cr\cr
&+&{\frac{1}{2}}A\wedge(\Ad_{\gp}-\Ad_{\gm})(r_+\Ad_{\gm}^{-1}
-r_-\Ad_{\gp}^{-1})A].\qquad
\end{eqnarray}

The equations of motion of the model can be written
\begin{eqnarray}\label{eqofmotdual}
&&g_{\pm}^{-1}{\dd}g_\pm+r_{\pm}(Ad_{g_+}^{-1}-Ad_{g_-}^{-1})A=0\cr\cr
&&{\dd}{A}+[{A}, {A}]=0.
\end{eqnarray}

The gauge transformations, for $\beta:\Sigma\rightarrow{\mathfrak g}$,
are:
\begin{eqnarray}
g_{\pm}^{-1}\delta_\beta g_\pm{\hskip -2mm}&=&{\hskip
-2mm}r_{\pm}(Ad_{g_-}^{-1}-Ad_{g_+}^{-1})\beta \notag \\ \notag\\
{\hskip -2mm}\delta_\beta A{\hskip -2mm}&=&{\hskip
-2mm}{\dd}\beta+[A,\beta]+ \notag\\ &+&{\hskip -3mm}{\frac{1}{2}}
(r_+Ad_{g_-}+r_-Ad_{g_+})[g_{+}^{-1}{\dd}g_+-g_{-}^{-1}{\dd}g_-+
(Ad_{g_+}^{-1}-Ad_{g_-}^{-1})A,\tilde\beta]\notag
\end{eqnarray}
where $\tilde\beta:=(r_- Ad_{g_+}^{-1}-r_+ Ad_{g_-}^{-1})\beta$.  
The gauge transformations
close on-shell. Namely, 
$[\delta_{\beta_1},\delta_{\beta_2}]=\delta_{[\beta_1,\beta_2]}$
which corresponds now to the gauge group $G$.

The solutions of the equations of motion can be obtained along
the same lines as before. Locally,
\begin{eqnarray}\notag
A=h{\dd}h^{-1}
\end{eqnarray}
and $(g_+(\sigma),g_-(\sigma))$ is obtained as the solution of
$$
(g_+(\sigma){\tilde h}(\sigma)
,g_-(\sigma){\tilde h}(\sigma))=
({h}(\sigma)\hat g_+,
{h}(\sigma)\hat g_-)
$$ which means that $(g_+,g_-)$ is the dressing-transformed of $(\hat
g_+,\hat g_-)$ by $h$. 
At this point it is evident the symmetry
between both dual models under the exchange of the roles of $G$ and
$G_r$.

In the open geometry, $\Sigma={\mathbb R}\times[0,{1}]$, with free
boundary conditions, $h$ is constant along connected components of the
boundary and one may take $h(t,0)=\tilde h(t,0)=e$, $h(t,{1})=h_{1}$,
$\tilde h(t,{1})=\tilde h_{1}$. The symplectic form, which we denote
by $\omega^*$, can be written
\begin{eqnarray}\notag
\omega^{*}={\frac{1}{2}}\Omega(({h_{1}}\hat g_+,{h_{1}}\hat g_-)).
\end{eqnarray}

The duality between ${\bar{\cal P}}(G;G,G)$ and ${\bar{\cal P}}(G_r;G_r,G_r)$ was
pointed out in \cite{CalFalGar}. The symplectic forms of the two
models coincide upon the exchange of $h_{1}$ with $\hat g^{-1}$ and
$(\hat g_+,\hat g_-)$ with $(h_{{1}+}^{-1},h_{{1}-}^{-1})$. Hence, one
can talk about a bulk-boundary duality between the Poisson-Lie sigma
models for $G$ and $G^*$ since the exchange of degrees of freedom maps
variables associated to the bulk of one model to variables associated
to the boundary of the other one.

\vskip 0.9 cm

At this point we would like to recall the equivalence between the
Poisson-Lie sigma model with target $G^*$ and the $G/G$ theory. Such
equivalence was first discovered in \cite{AleSchStr} in the
Hamiltonian formalism and was given a covariant description in
\cite{FalGaw02}, which we present below.

The $G/G$ theory is the gauged version of the WZW model \cite{Wit84}
with the group $G$ as target. The fields of this model on a
two-dimensional oriented surface $\Sigma$ equipped with a conformal or
a pseudo-conformal structure are $g:\Sigma\rightarrow G$ and a gauge
field $A$, a $\gl g$-valued 1-form on $\Sigma$. The action functional
of the model has the form
\begin{eqnarray}\label{swzw}
S_{\mbox{{\scriptsize WZW}}}(g,A)&=& {1\over4\pi}\int_\Sigma\tr[(g^{-1}\partial_l g)(g^{-1}
\partial_r g)]{\dd}x^l\wedge {\dd}x^r+S_{WZ}(g)\cr\cr
&&\hskip -3mm+{1\over 2\pi}\int_\Sigma\tr[\partial_l g g^{-1}A_r - A_l 
g^{-1}\partial_r g]{\dd}x^l\wedge {\dd}x^r\cr\cr 
&&\hskip -3mm+{1\over 2\pi}\int_\Sigma\tr[A_lA_r  - gA_l g^{-1}A_r]
{\dd}x^l\wedge {\dd}x^r
\end{eqnarray}
where  $x^l=z$, $x^r=\bar z$ are the complex variables in the Euclidean 
signature and the light-cone ones $x^l=x+t$, $x^r=x-t$ in the Minkowski 
signature. The derivatives $\partial_l={\partial\over\partial 
x^l}$, $\partial_r={\partial\over\partial x^r}$ and $A=A_l{\dd}x^l+A_r{\dd}x^r$. 
The Wess-Zumino term $S_{\mbox{{\scriptsize WZ}}}(g)$ is often written as
${1\over 4\pi}\int\limits_\Sigma g^*\,{\dd}^{-1}\chi$ where 
${\dd}^{-1}\chi$ stands for a 2-form whose differential is equal to the 
3-form $\chi$ on $G$. Since such 2-forms do not exist globally, some 
choices are needed. On closed surfaces, one may replace 
$\int\limits_\Sigma g^*{\dd}^{-1}\chi$ by $\int\limits_B{\tilde g}^*\chi$ 
where $B$ is an oriented 3-manifold with $\partial B=\Sigma$ and 
$\tilde g:B\rightarrow G$ extends $g$. This gives a well defined 
amplitude $e^{ikS_{\mbox{{\scriptsize WZW}}}}$ for integer $k$. When $\Sigma$ has 
a boundary, the definition of the amplitude is more problematic and, 
in general, it makes sense only as an element of a product of line bundles
over the loop group $LG$ \cite{Gaw88}. To extract a numerical value of such
an amplitude one has to use sections of the line bundle that are not
globally defined. 

What we shall do here is to restrict the values of the field $g$ of
the gauged WZW model to the subset $\nu(G_r)\subset G$ and to define
\begin{eqnarray}\label{wz}
S_{\mbox{{\scriptsize WZ}}}(g)={1\over 4\pi}\int\limits_\Sigma g^*\rho
\end{eqnarray}
where $\rho$ is given by (\ref{rho}). In the geometric language, 
this corresponds to a particular choice of a section in the (trivial) 
restriction of the line bundle over $LG$ to the loop 
space $L\nu(G_r)$. We shall call the resulting field theory the 
restricted $\,G/G\,$ coset model. 

The model defined above has simple transformation properties under
the gauge transformations of fields
\begin{eqnarray}\notag
{}^h\hspace{-0.05cm}g=h gh^{-1}\,\qquad 
{}^h\hspace{-0.08cm}A=h Ah^{-1}+h \dd h^{-1}
\end{eqnarray}
for $h:\Sigma\rightarrow G$ such that ${}^h\hspace{-0.05cm}g$ takes 
also values in $\nu(G_r)$ (this is always 
accomplished for $h$ sufficiently close to unity). The action transforms
according to 
\begin{eqnarray}\notag
S_{\mbox{{\scriptsize WZW}}}({}^h\hspace{-0.05cm}g,{}^h\hspace{-0.08cm}A)=S_{\mbox{{\scriptsize WZW}}}
(g,A)+\int_\Sigma \Omega(h,g)
\end{eqnarray}
where $\Omega(h,g)$ is the closed 2-form given by (\ref{Omega2}).
In particular, it follows that the action is invariant under infinitesimal 
gauge transformations that vanish on the boundary.

The equations of motion  of the model $\delta S_{\mbox{{\scriptsize WZW}}}=0$ for field
variations vanishing at the boundary are
\begin{subequations} \label{eomWZW}
\begin{align}
&D_l(g^{-1}D_rg)\dd x^l\wedge \dd x^r +F(A)=0 \label{eomWZWa} \\
&g^{-1}D_rg=0\,,\qquad gD_lg^{-1}=0 \label{eomWZWb}
\end{align}
\end{subequations}
where $D$ stands for the covariant derivative
$D_{l,r}=\partial_{l,r} + [A_{l,r},\,\cdot\,]$ 
and $F(A)=\dd A+A\wedge A$ is the field strength of $A$.

Although (\ref{eomWZWa}) is a second order differential equation we can 
write a system of first order equations equivalent to (\ref{eomWZW}),
by simply taking
\begin{eqnarray} \label{eomWZWfo}
&&F(A)=0, \cr
&&g^{-1}Dg=0. 
\end{eqnarray}

Since we have first order equations of motion it should be possible to
get the system from a first order Lagrangian. As mentioned above, the
Poisson-Lie sigma model with $G_r$ target and fields $(g_+,g_-)$ and
$A$ is closely related to the gauged WZW model with the target
$\nu(G_r)$ and fields $g=g_-g_+^{-1}$ and $A$. The action of the
latter is given by equation (\ref{swzw}) with the Wess-Zumino term
defined by (\ref{wz}). More explicitly,
\newpage
\begin{eqnarray}\notag
S_{\mbox{{\scriptsize WZW}}}(g_-g_+^{-1},A){\hskip -2mm}&=&{\hskip
-2mm}{1\over4\pi}\int_\Sigma \tr[(\jlp)(\jrp)+(\jlm)(\jrm) \cr&&\hskip
15mm -2(\jlp)(\jrm)] \dd x^l\wedge \dd x^r\cr &&\hskip -3mm -{1\over
2\pi}\int_\Sigma\tr[\AL g^{-1}\partial_r g+ (g\partial_l g^{-1})\AR
\cr&&\hskip 16mm +g\AL g^{-1}\AR -\AL\AR]\dd x^l\wedge \dd x^r.
\end{eqnarray}
The equations of motion (\ref{eomWZWfo}) read in terms of
$(\gp,\gm)$ as
\begin{eqnarray}\notag
E_{l ,\pm}&\equiv&g_\pm^{-1}\partial_l g_{\pm}+
r_\pm (\Ad_{\gp}^{-1}-\Ad_{\gm}^{-1})\AL = 0,
\cr
E_{r ,\pm}&\equiv&g_\pm^{-1}\partial_r g_{\pm}+ 
r_\pm (\Ad_{\gp}^{-1}-\Ad_{\gm}^{-1})\AR = 0.
\end{eqnarray}
Defining $E=E_l\dd x^l+E_r\dd x^r$, $E_{l,r}=E_{l,r,+}-E_{l,r,-}$ and
after a straightforward calculation one obtains the identity:
\begin{eqnarray}\notag
S_{\mbox{{\scriptsize WZW}}}(g_-g_+^{-1},A)&=& -\frac{1}{8\pi}\int_\Sigma\tr(E\wedge
rE)+\frac{1}{4\pi}\int_\Sigma\tr(E_l E_r)\dd x^l\wedge \dd x^r -\cr
&&{}\cr
&-&\frac{1}{2\pi}S_{PL}^*(g_+,g_-,A)
\end{eqnarray}
where $S_{PL}^*$ is the action (\ref{dualPLS}).

Note that the difference between $S_{\mbox{{\scriptsize WZW}}}$ and
$S_{PL}^*$ is quadratic in the equations of motion for
$S_{\mbox{{\scriptsize WZW}}}$. This implies that neither the
classical solutions nor the phase space structure of both models
(based on first functional derivatives of the action evaluated
on-shell) differ, as long as the boundary conditions coincide.

\section{Triangular Poisson-Lie sigma models}

We now go back to equations (\ref{eqofmotdual}) and assume
$r$ is triangular ($\alpha = 0$). We start noting that whereas $A$ is
a pure gauge of the group $G_r$, ${\frac{1}{2}}rA$ is a pure gauge of
the group $G$, i.e.
$$\dd ({\frac{1}{2}}rA)+[{\frac{1}{2}}rA,{\frac{1}{2}}rA]=0.$$

Now take
$${\bar X}(\sigma) = \begin{pmatrix}e&0\cr X(\sigma)&e\end{pmatrix}h_{X}(\sigma) \in G_{r}.$$

Then, locally,
\begin{eqnarray}\notag
{\bar X}^{-1}\dd{\bar X} = \begin{pmatrix}h_{X}^{-1}\dd h_{X}&0 \cr
h_{X}^{-1}\dd Xh_{X}&h_{X}^{-1}\dd
h_{X}\end{pmatrix}=\begin{pmatrix}rA&0 \cr A&rA\end{pmatrix}.
\end{eqnarray}
${\tilde A}$ is also a pure gauge of the group $G_r$, so ${\tilde
A}=h_{{\tilde X}}^{-1}d{\tilde X}h_{{\tilde X}}$ and the equation of
motion for $g$ reads,
\begin{eqnarray}\notag
\dd gg^{-1}+{\frac{1}{2}}(h_{X}^{-1}\dd h_{X}-\Ad_{g}h_{{\tilde
X}}^{-1}\dd h_{{\tilde X}})=0
\end{eqnarray}
which implies
\begin{eqnarray}\notag
g=h_{X}^{-1}{\hat g}h_{\tilde X}, \ {\hat g}\in G.
\end{eqnarray}
${\tilde A}=\Ad_{g}^{-1}A \Rightarrow {\tilde X}=\Ad_{\hat g}^{-1}X$ and
we can write, locally, the solution as an equation in the double:
\begin{eqnarray} \label{solutiondirect}
\begin{pmatrix}e&0\cr X&e\end{pmatrix}h_{X}\begin{pmatrix}e&0\cr 0&e\end{pmatrix}g = \begin{pmatrix}e&0 \cr 0&e\end{pmatrix}{\hat g}\begin{pmatrix}e&0\cr {\tilde X}&e\end{pmatrix}h_{\tilde X}.
\end{eqnarray}

%\vskip 2mm

The analysis of the reduced phase space when
$\Sigma={\mathbb R}\times[0,{1}]$ and $g\vert_{\partial\Sigma}$ is
free works as in the factorizable case. The field $A$ must vanish on
vectors tangent to ${\partial\Sigma}$ and hence $\bar X$ is constant
along each connected component of the boundary. By using the explicit
solution (\ref{solutiondirect}) we can identify ${\bar{\cal
P}}(G;G,G)$. Notice that we can always choose ${\bar
X}(t,0)={\bar{\tilde X}}(t,0)=e$. With this choice and defining
$X_{{1}}:=X(t,{1}),{\tilde X}_{{1}}:={\tilde
X}(t,{1}),g_{{1}}:=g(t,{1})$, a straightforward calculation yields
\begin{eqnarray}\notag
\omega &=& {\frac{1}{2}}\tr \left(\delta X_{{1}}+[X_{{1}},\delta
h_{X_{1}}h_{X_{1}}^{-1}])\wedge \delta {\hat g} {\hat g}^{-1} +
\Ad_{h_{{\tilde X}_{1}}}^{-1}\delta {{\tilde X}_{1}} \wedge
g^{-1}_{{1}}\delta g_{{1}}\right)\cr\cr
&=&{\frac{1}{2}}\Omega({\bar X}_{1}{\hat g}).
\end{eqnarray}

The reduced phase space ${\bar{\cal P}}(G;G,G)$ turns out to be the
set of pairs $([{\bar X}],{\hat g})$ with $[{\bar X}]$ a homotopy
class of maps from $[0,{1}]$ to $G_r$ which are the identity at $x=0$
and have fixed value at $x = {1}$.

\subsection{The dual triangular model and BF-theory}

Take (\ref{dualPoissonLieTriangular}) and write the action of the
Poisson sigma model with target $G_r$:
\begin{eqnarray}
S^{*}_{PL}({\bar X},A)&=&\int_{\Sigma}\tr \Big((\dd X\wedge A + \dd
X\wedge
\Ad_{g_{{}_X}}r\Ad_{g_{{}_X}}^{-1}[X,A]+{\frac{1}{2}}A\wedge[X,A]-\cr
&-&{\frac{1}{2}}[X,A]\wedge \Ad_{g_{{}_X}}r\Ad_{g_{{}_X}}^{-1}[X,A]\Big)
\end{eqnarray}
with fields $A \in \Omega^{1}(\Sigma)\otimes {\mathfrak g}$, ${\bar
X}:\Sigma \to G_{r}$.

Note that the action is actually determined by
$X$ and $A$, since it is invariant under $g_{X} \mapsto g_{{}_X}g_{0},\
g_{0} \in G_{0}$.

Varying the action with respect to $A$ we get the equation of motion for $X$,
\begin{eqnarray} \label{eomX}
\dd X + [A,X] = 0.
\end{eqnarray}

Taking variations with respect to ${\bar X}$ and after a rather cumbersome
 calculation we obtain the equation of motion for $A$,
\begin{eqnarray} \label{eomAtrian}
\dd A + [A,A] = 0.
\end{eqnarray}

The infinitesimal gauge symmetry for $\beta:\Sigma\rightarrow
{\mathfrak g}$ is:
\begin{subequations} \label{dualgaugesymmetry}
\begin{align}
&\delta_{\beta}X = [X,\beta] \label{dualgaugesymmetrya} \\
&\delta_{\beta}A=\dd \beta + [A,\beta]-r[\dd X+[A,X],\beta+\Ad_{g_{{}_X}}r\Ad_{g_{{}_X}}^{-1}[X,\beta]] \label{dualgaugesymmetryb}
\end{align}
\end{subequations}
which this time corresponds to the vector fields of the infinitesimal form of the right dressing action of $G$ on $G^{*}$. On-shell, $[\delta_{\beta_{1}},\delta_{\beta_{2}}]=\delta_{[\beta_{1},\beta_{2}]}$ .

The solutions of the equations of motion are, locally,
\begin{eqnarray}\label{solutiondual}
&&A=h^{-1}\dd h \cr
&&X=\Ad_{h}^{-1}{\hat X}.
\end{eqnarray}
They may also be rewritten as an equation in ${\bf D}$:
\begin{eqnarray}\notag
\label{solutiondualdouble}
\begin{pmatrix}e&0\cr {\hat X}&e\end{pmatrix}g_{\hat X}\begin{pmatrix}e&0\cr 0&e\end{pmatrix}{\tilde h} = \begin{pmatrix}e&0 \cr 0&e\end{pmatrix}h \begin{pmatrix}e&0\cr X&e\end{pmatrix}g_{{}_X}
\end{eqnarray}
which can be obtained from (\ref{solutiondirect}) taking $X\rightarrow
{\hat X}$, ${\hat g}\rightarrow h$, $g\rightarrow {\tilde h}$,
${\tilde Y}\rightarrow Y$. Now, ${\bar X}={}^{h^{-1}}{\hat{\bar X}}$.

Now consider $\Sigma={\mathbb R}\times[0,{1}]$ and ${\bar
X}\vert_{\partial\Sigma}$ free. $h$ must be constant along each
connected component of the boundary. By choosing $h(t,0)={\tilde
h}(t,0)=e$, defining $h_{{1}}:=h(t,{1}),{\tilde h}_{{1}}:={\tilde
h}(t,{1}),X_{{1}}:=X(t,{1})$ and plugging in (\ref{solutiondual}), we
get
\begin{eqnarray}\notag
\omega^* &=& {\frac{1}{2}}\tr \left((\delta {\hat X}+[{\hat X},\delta {\hat
g_{{}_X}}{\hat g_{{}_X}}^{-1}]) \wedge \delta {\tilde h}_{{1}} {\tilde
h}_{{1}}^{-1} + \Ad_{g_{X_{{1}}}}^{-1}\delta {X_{1}} \wedge {\tilde h}
^{-1}_{{1}}\delta {\tilde h} _{{1}}\right)= \notag\\
&&{}\notag\\
&=&{\frac{1}{2}}\Omega(h_{1} {\hat{\bar X}}).
\end{eqnarray}

The reduced phase space ${\bar{\cal P}}(G_r;G_r,G_r)$ is the set of
pairs $([h],{\hat {\bar X}})$ with $[h]$ a homotopy class of maps from
$[0,{1}]$ to $G$ which are the identity at $x=0$ and have fixed value
at $x = {1}$. Notice the duality between ${\bar{\cal P}}(G;G,G)$ and
${\bar{\cal P}}(G_r;G_r,G_r)$ under the interchange ${\hat g}
\leftrightarrow h_{1},\ X_{1} \leftrightarrow {\hat X}$. This is the
triangular version of the bulk-boundary duality found in
\cite{CalFalGar} and recalled in Section 5.1 for the factorizable
case.

\vskip 1 cm

Now, consider as target of the Poisson sigma model the dual of a
simple complex Lie algebra ${\mathfrak g}^*$ with the Kostant-Kirillov
Poisson bracket. As mentioned in Example \ref{ex:zeroPL}, this is the
dual Poisson-Lie group of the simply connected Lie group $G$ whose Lie
algebra is ${\mathfrak g}$ endowed with the zero Poisson
structure. The action in this particular case is:

\begin{eqnarray}\label{BFtheory}
S_{{\rm BF}}(X,A)=\int_{\Sigma}\tr \left(\dd X \wedge A
-{\frac{1}{2}}[X,A]\wedge A\right)
\end{eqnarray}
with $X:\Sigma \rightarrow {\mathfrak g}$ and $A \in
\Omega^{1}(\Sigma)\otimes {\mathfrak g}$. This is the action of
BF-theory (\cite{Hor}) up to a boundary term.

The equations of motion are
\begin{eqnarray} \label{eomLinear}
&&\dd X+[A,X]=0 \cr
&&\dd A + [A,A] = 0.
\end{eqnarray}

For $\epsilon\in C^\infty(M)\otimes{\mathfrak g}$ the gauge transformation
\begin{subequations}\notag
\begin{align}
&\delta_\epsilon X=[X,\epsilon] \\
&\delta_\epsilon A= \dd \epsilon+ [A,\epsilon]
\end{align}
\end{subequations}
induces the change of the action (\ref{BFtheory}) by a boundary term
\begin{eqnarray}
\delta_\epsilon S_{{\rm BF}}=-\int_\Sigma \dd \tr (\dd X\epsilon).
\end{eqnarray}

Note that in this case the gauge transformations close even off-shell
$$[\delta_\epsilon,\delta_{\epsilon'}]=\delta_{[\epsilon,\epsilon']}$$
and induce the Lie algebra structure of
$C^\infty(\Sigma)\otimes{\mathfrak g}$ in the space of parameters.

The equations of motion (\ref{eomX}), (\ref{eomAtrian}) are the same as
(\ref{eomLinear}). We would like to understand this fact at the level
of the action. A direct computation shows that the following equality
holds:
\begin{eqnarray}\notag
S^*_{PL}({\bar X},A)=S_{{\rm BF}}(X,A)-{\frac{1}{2}}(\dd
X+[A,X])\Ad_{g_{{}_X}}r\Ad_{g_{{}_X}}^{-1}(\dd X+[A,X]).
\end{eqnarray}

Hence, both actions differ by terms quadratic in the equations of
motion. This means that the Poisson-Lie sigma model with target $G_r$
is equivalent, for any triangular $r$-matrix, to the Poisson-Lie sigma
model over $G_{r=0}$, i.e. BF-theory. This is the triangular version
of the connection encountered in the factorizable case
(\cite{FalGaw02}), where every Poisson-Lie sigma model with target
$G_r$ for any factorizable $r$-matrix is (locally) equivalent to the
$G/G$ WZW model with target $G$.

\section{More general Poisson sigma models over Lie groups}

In this section we solve the model with Poisson structure
(\ref{generalPoisson}) following the lines of Section
\ref{FactPLSM}. As remarked in Section \ref{sec:PLstructsimpLG} this
Poisson structure does not make $G$ into a Poisson-Lie group. In the
resolution of the model we shall introduce a generalization of the
dressing transformations and of the Heissenberg double and we shall be
able to identify the symplectic leaves for the Poissson structure of
$G$. For concreteness, we shall consider $G$ factorizable, although an
analogous procedure might be carried out in the triangular case.

The action for the model is
\begin{eqnarray}\notag
S(g,A)=\int_{\Sigma}\tr(\dd gg^{-1}\wedge A)-
{1\over4}\tr(A\wedge (r+Ad_{g}r'Ad_{g}^{-1})A)
\end{eqnarray}
where $r$ and $r'$ are two solutions of the modified Yang-Baxter
equation (\ref{YangBaxter}) with $\alpha=1$. The equations of motion
are
\begin{subequations}
\begin{align} 
&\dd gg^{-1}+{1\over2}(r+Ad_{g}r'Ad_{g}^{-1})A=0 \label{eomPoissonSigmaa}\\
&\dd{\tilde A}-[{\tilde A},{\tilde A}]_{r'}=0,\ {\tilde A}:=Ad_{g}^{-1}A.
\label{eomPoissonSigmab}
\end{align}
\end{subequations}

{}From the previous equations, or performing in the action
variations of the fields that keep $\tilde A$ unchanged, we obtain 
\begin{eqnarray}\notag
%\label{eomA}
&&\dd A+[A,A]_{r}=0.
\end{eqnarray}

Take $\beta:\Sigma \rightarrow {\gl g}$. The gauge symmetry in its
infinitesimal form reads:
\begin{eqnarray}\notag
&&\delta_{\beta}gg^{-1}=-{1\over2}(r+Ad_g r'Ad_{g}^{-1})\beta \cr
&&\delta_{\beta}A=\dd\beta + [A,\beta]_{r}-{1\over2}[dgg^{-1}+
{1\over2}(r+Ad_{g}r'Ad_{g}^{-1})A,\beta].
\end{eqnarray}

The transformation for $g$ corresponds to the right {\it dressing}
vector field that comes from the contraction of the Poisson structure
with the right-invariant forms.  On-shell,
$[\delta_{\beta_{1}},\delta_{\beta_2}]=
\delta_{[\beta_{1},\beta_{2}]_{r}}$. Hence, the symmetry group is the
one corresponding to the matrix $r$, i.e. $G_{r}$.  The reason for the
preferred role of $r$ against $r'$ in the gauge symmetry is simply the
choice of right-invariant one forms to express the Poisson structure.
Had we chosen left-invariant forms (i.e. changing the variable $A$ by
$\tilde A$ in the action) the symmetry algebra would have been that of
$r'$.

We can consider $G_r$ or $G_{r'}$ as different subgroups of the same
double group ${\bf D}$.  We shall denote by $(h_+,h_-)$
(resp. $(h_{+'},h_{-'})$) the elements of $G_r$ (resp. $G_{r'}$).

Using the procedure of Section \ref{FactPLSM}, we can easily solve the
model. As before, we first write locally the solutions for $A$ and
${\tilde A}$,
\begin{eqnarray}\notag
%\label{solutionsAAtilde}
&&A=h_{+}\dd h_{+}^{-1}-h_{-}\dd h_{-}^{-1}\cr
&&{\tilde A}={\tilde h}_{-'}\dd{\tilde h}_{-'}^{-1}-
{\tilde h}_{+'}\dd{\tilde h}_{+'}^{-1}
\end{eqnarray}
with $h_\pm(\sigma_0)=\tilde h_{\pm'}(\sigma_0)=e$.

The equation of motion (\ref{eomPoissonSigmaa}) can be transformed
into
$$\dd gg^{-1}+r_{\pm}A+Ad_{g}r'_{\mp}\tilde A=0$$
and  inserting the solutions for $A$, $\tilde A$
$$\dd gg^{-1}+h_{\pm}\dd h_{\pm}^{-1}- Ad_{g}{\tilde
h}_{\mp'}\dd{\tilde h}_{\mp'}^{-1}=0.$$

Or equivalently,
$$h_+^{-1}g\tilde h_{-'}=h_-^{-1}g\tilde h_{+'}=\hat g$$
with $\hat g=g(\sigma_0)\in G$.

We can write the general solution as an equation in ${\bf D}$:
\begin{eqnarray} \label{solgen}
(g(\sigma)\tilde h_{-'}(\sigma),g(\sigma)\tilde h_{+'}(\sigma))=
(h_+(\sigma)\hat g,h_-(\sigma)\hat g).
\end{eqnarray}

If we now define ${\bf D}'_0 := G_rG_d\cap G_dG_{-r'}$, we see that
for solutions of the equations of motion, $(h_+(\sigma)\hat
g,h_-(\sigma)\hat g)\in {\bf D}'_0$.  The symplectic leaves of
$G_{r,r'}$ are connected components of the orbits of the generalized
dressing transformation of $\hat g$ by $(h_+,h_-)\in G_r$, that comes
from solving equation (\ref{solgen}) in $g(\sigma)$.

In order to describe the presymplectic structure on the space of
solutions we need to introduce a new Poisson bracket on ${\bf D}$. Recall
that the Heisenberg double was defined in Section
\ref{sec:PLstructsimpLG} as ${\bf D}_{R,R}$ whith $R=P_d-P_r$. We can
generalize this construction by introducing another $r$-matrix $r'$
that gives rise to $R'=P_d-P_{-r'}$. The Poisson structure in the
double we are interested in is ${\bf D}_{R,R'}$. It is non-degenerate around
the unit and its main symplectic leaf is denoted by ${\bf D}'_0$.  If we
parametrize the points in ${\bf D}'_0$ by $(\eta_+\xi,\eta_-\xi)=
(\tilde\xi\tilde\eta_{-'},\tilde\xi\tilde\eta_{+'})$, the symplectic
structure in ${\bf D}'_0$ obtained by inverting the Poisson bracket is
$$\Omega'(\eta_+\xi,\eta_-\xi)= \tr[\dd\tilde\xi\tilde \xi^{-1}\wedge
(\dd\eta_{+}{\eta}_{+}^{-1}-\dd{\eta}_{-}{\eta}_{-}^{-1})-\xi^{-1}\dd\xi\wedge(\tilde
\eta_{+'}^{-1}\dd\tilde \eta_{+'}- \tilde \eta_{-'}^{-1}\dd\tilde
\eta_{-'})].
$$

Note again that although for a given point of ${\bf D}'_0$ factors 
$\xi,\eta_{\pm},\tilde\xi,\tilde\eta_{\pm'}$ in general
are not uniquely determined (different choices differ by elements of
the discrete groups $G_0$ or $G_0'$) the form $\Omega'$
is not affected by the ambiguity and is indeed well defined
in ${\bf D}_0'$. 
 
The presymplectic structure of the $r,r'$ Poisson sigma model,
$\omega'$, in the open geometry, $\Sigma={\mathbb R}\times[0,{1}]$,
can then be written in terms of $\Omega'$. It reads
$$\omega'=\Omega'((h_{{1}+}\hat g,h_{{1}-}\hat g)-
\Omega'((h_{0+}\hat g,h_{0-}\hat g)$$  
with $h_{0\pm}=h_\pm(0)$
$h_{{1}\pm}=h_\pm({1})$. And if we take $\sigma_0=(0,t_0)$,
i. e. $h_{0\pm}=e$,
$$\omega'=\Omega'((h_{{1}+}\hat g,h_{{1}-}\hat g)).$$

The discussion of the gauge transformations and the reduced phase
space goes parallel to the previous models. Points of the reduced
phase space are pairs $([(h_+,h_-)],\hat g)$ with $[h_+,h_-]$ the
homotopy class of maps $[0,{1}]\rightarrow G_r$ with fixed boundary
values and such that $(h_{+}(x)\hat g,h_{-}(x)\hat g)\in {\bf
D}_0'$. The symplectic form on the reduced phase space can be obtained
as the pullback of $\Omega'$ by the map $([(h_+,h_-)],\hat g)\mapsto
(h_{{1}+}\hat g,h_{{1}-}\hat g)\in {\bf D}_0'$.

\section{Branes in Poisson-Lie sigma models}

In the previous section we have seen that, if $\Sigma={\mathbb
R}\times[0,{1}]$, the moduli spaces of solutions of the Poisson-Lie
sigma models over $G$ and $G_r\cong G^*$ coincide when $g$ is free at
the boundary (i.e. when the brane is the whole target group). We would
like to find out whether such duality holds for more general boundary
conditions. That is, we address the problem of finding pairs of branes
$N\subset G$ and $N^*\subset G^*$ such that ${\bar{\cal
P}}(G;N,N)\cong {\bar{\cal P}}(G^*;N^*,N^*)$.

Let us restrict $g\vert_{\partial\Sigma}$ to a closed submanifold
(brane) $N\subset M$. It is natural to ask the brane $N$ to respect
the Poisson-Lie structure of $G$ given by the $r$-matrix. To this end
we consider a simple subalgebra ${\mathfrak h}\subset {\mathfrak g}$
such that it is $r$-invariant, i.e. $r{\mathfrak h}\subseteq{\mathfrak
h}$. The restriction of $r$ to ${\mathfrak h}$, $r\vert_{\mathfrak h}$
is an $r$-matrix in ${\mathfrak h}$.  Since ${\mathfrak h}$ is simple,
its Killing form coincides (up to a constant factor) with the
restriction to ${\mathfrak h}$ of the Killing form in $\mathfrak g$.
Let $H\subset G$ be the subgroup with Lie algebra ${\mathfrak
h}$. Then, for $g\in H$,\ $X,Y\in {\mathfrak h}$,
\begin{eqnarray} \label{inducedPoisson}
\pi_{g}^{r\vert_{\mathfrak h}}(X,Y)={\frac{1}{2}}\
\tr(Xr\vert_{\mathfrak h}Y-X\Ad_{g}r\vert_{{\mathfrak h}}\Ad_{g}^{-1}Y)
\end{eqnarray}
defines a Poisson-Lie structure on $H$.

The nice point is that we can realize $H^*$, the dual Poisson-Lie
group of $H$, as a subgroup of $G_r$. $H^*$ is simply identified with
the subgroup $H_r\subset G_r$ corresponding to the Lie subalgebra
$({\mathfrak h},[\cdot,\cdot]_r)$ of $({\mathfrak g},[\cdot,\cdot]_r)$. We claim
that the Poisson-Lie sigma model with target $G$ and brane $H$ is dual
to the Poisson-Lie sigma model with target $G_r$ and brane $H_r$. That
is to say, there is a bulk-boundary duality between ${\bar{\cal
P}}(G;H,H)$ and ${\bar{\cal P}}(G_r;H_r,H_r)$.

Before describing the duality we shall make some general
considerations about the properties of these branes.  The results of
Section \ref{sec:branes} on the boundary conditions of the fields are
written in terms of the field $\eta \in \Gamma(T^*\Sigma \otimes
X^*T^*M)$.  Let us rewrite them in terms of the field $A$ appearing in
the action (\ref{PSgroup}).

When taking variations of (\ref{PSgroup}) with respect to $g$, a
boundary term $-\int_\Sigma\dd\tr(\dd gg^{-1}\wedge A)$ appears. Its
cancellation requires $A_{\bf t} \in {\mathfrak h}^\perp$ ($\perp$
means orthogonal with respect to $\tr(\ )$). On the other hand, the
continuity of (\ref{eomG}) at the boundary imposes $\pi^{\sharp}A_{\bf
t} \in {\mathfrak h}$. Consequently, the boundary condition for
$A_{\bf t}$ is
\begin{equation} \label{BCA}
A_{\bf t}(\sigma) \in {\mathfrak h}^\perp\cap \pi_{g(\sigma)}^{\sharp
-1}{\mathfrak h},\ \sigma \in \partial\Sigma
\end{equation}
and the gauge transformation parameter $\beta$ at the boundary is
restricted by the same condition.

The condition (\ref{PoiDir}) for pointwise Poisson-Dirac branes,
applied to our present situation reads
$${\mathfrak h}\cap \pi_g^{\sharp}{\mathfrak h}^\perp=0,\
\forall g\in H.$$
In particular, if $H$ is Poisson-Dirac the gauge transformations do
not act on $g$ at the boundary.

We have that $H$ is coisotropic if
$$\pi_g^{\sharp}{\mathfrak h}^\perp \subseteq {\mathfrak h},\ \forall
g\in H.$$

We show now that an $r$-invariant, simple, subgroup $H$ is a Poisson-Dirac
submanifold of $G$, and its dual $H_r$ is also Poisson-Dirac in
$G_r$. Denote by ${\mathfrak h}^\perp \subset {\mathfrak g}$ the
subspace orthogonal to ${\mathfrak h}$ with respect to ${\rm
tr}(\ )$. Firstly, the $r$-invariance of ${\mathfrak h}$ implies that
$\pi_g^{r\sharp}{\mathfrak h}\subseteq{\mathfrak h},\forall g\in
H$. Using that $\pi^{r\sharp}$ is antisymmetric we obtain that
$\pi_g^{r\sharp}{\mathfrak h}^\perp\subseteq{\mathfrak h}^\perp,\forall
g\in H$. Finally, recalling that ${\mathfrak h}$ simple
$\Rightarrow{{\mathfrak h}}\cap{\mathfrak h}^\perp=0$, one immediately
deduces that $H$ is Poisson-Dirac. 

Observing that ${\mathfrak h}_r$ (i.e. ${\mathfrak h}$ equipped with
the Lie bracket $[\cdot,\cdot]_{r\vert_{\mathfrak h}}$) is the same as
${\mathfrak h}$ as a vector subspace of ${\mathfrak g}$ and reasoning
as above one shows that $H_r$ is a Poisson-Dirac submanifold in $G_r$.
\vskip 2mm

$H$ and $H_r$ inherit a (smooth) Poisson structure (the Dirac bracket)
from $G$ and $G_r$ respectively. They coincide with the Poisson
structures defined by $r\vert_{\mathfrak h}$ on $H$ and $H_r$ (formula
(\ref{inducedPoisson}) for $H$, and analogously for $H_r$), making
them into a pair of dual Poisson-Lie groups. Notice, however, that
this induced Poisson structure does not make $H$ (resp.  $H_r$) into a
Poisson submanifold of $G$ (resp. $G_r$), so that in general it is not
a Poisson-Lie subgroup, it is so if and only if it is coisotropic.

We now address the issue of the duality of the models with a pair of
branes $H$ and $H_r$ as above. The general picture is as follows.  In
the case of the model over $G$ with brane $H$ the space of solutions,
once reduced by the gauge transformations in the bulk, can be
identified with the universal covering of $G_rH_d\cap H_dG_r\subset
{\bf D}_0$.  The symplectic form $\Omega$ (see (\ref{HeisenbergDouble})) in
${\bf D}_0$ (corresponding to free boundary conditions) becomes degenerate
in $G_rH_d\cap H_dG_r$. This reflects the existence of gauge
transformations at the boundary. Since ${\mathfrak h}$ is
$r$-invariant, there is a natural choice of gauge fixing for these
transformations: $H_rH_d\cap H_dH_r$. The pullback of $\Omega$ to
$H_rH_d\cap H_dH_r$ is nondegenerate and the infinitesimal gauge
transformations span a complementary subspace to $T_p (H_rH_d\cap
H_dH_r)$ in $T_p(G_rH_d\cap H_dG_r)$ for every $p\in H_rH_d\cap
H_dH_r$.

The dual model over $G_r$ with brane $H_r$ behaves in an analogous
way. The space of solutions of the equations of motion can be
identified with the set $G_dH_r\cap H_rG_d$, but there are still gauge
transformations acting on this space. The gauge fixing is given again
by considering the restriction to $H_rH_d\cap H_dH_r$, which makes
this model equivalent by bulk-boundary duality to the previous one.

The considerations of the previous paragraphs did not care about the
existence of singular points or whether the gauge fixing is local or
global.  These subtleties may depend on the concrete model. Let us
work out an example in which all properties of regularity and global
gauge fixing are met.

Take $G=SL(n,{\mathbb C})$ with the Poisson-Lie structure
(\ref{PoissonLie}) given by the standard $r$-matrix (\ref{standardr})
and
\begin{equation}
H=\left\{{\begin{pmatrix}A&0 \cr 0&I \end{pmatrix}}
\in SL(n,{\mathbb C}),\quad {\rm s.t.}\quad
\ A\in SL({k},{\mathbb
C})\right\}
\end{equation}
for a given ${k}<n$.
The dual group $H_r\subset G_r\subset G\times G$ is easily described:
\begin{equation}
H_r=\left\{(g_+,g_-)\in G_r, 
\quad {\rm s.t.}\quad
g_\pm={\begin{pmatrix}A_\pm&0 \cr 0&I \end{pmatrix}},\ A_\pm \in 
SL({k},{\mathbb
C})\right\}.
\end{equation} 

In this case,
\begin{equation}
{\mathfrak h}^\perp=\left\{
\begin{pmatrix}
\lambda I&B\\
C&X
\end{pmatrix}
\in {\mathfrak sl}(n,{\mathbb C})
\right\}.
\end{equation}

An easy calculation shows that $\pi_g^{r\sharp} {{\mathfrak
h}}^\perp=0$, i.e. $H\subset G$ is Poisson-Dirac and coisotropic. In
particular, the inclusion map $i:H\rightarrow G$ is a Poisson map and
$H$ is a Poisson-Lie subgroup of $G$.

The situation is different for the dual model with target $G_r$ and brane
$H_r$. Recall that ${\mathfrak h}_r$ is the same as ${\mathfrak h}$ as
a vector space, and hence also their orthogonal complements. In this
case ${\mathfrak h}^\perp\cap \pi_{(g_+,g_-)}^{r\sharp -1}{\mathfrak
h}\neq {{\mathfrak h}}^\perp$ and $H_r\subset G_r$ is not
coisotropic. In fact, in generic points
\begin{equation}
{\mathfrak h}^\perp\cap
\pi_{(g_+,g_-)}^{r\sharp -1}{\mathfrak h}=\left\{
\begin{pmatrix}
\lambda I&0\\
0&X
\end{pmatrix}
\in {\mathfrak sl}(n,{\mathbb C})\right\}
\end{equation}
so that the brane $H_r$ is classically admissible.

The solutions of the equations of motion for $g$ in the model with
target $G$ are given by
\begin{equation}
g(\sigma)={}^{(h_+(\sigma),h_-(\sigma))}{\hat g}
\end{equation}
where $g(t,0),g(t,{1})\in H$. One can always take
$(h_+(t,{1}),h_-(t,{1}))=(e,e)$ and $(h_+(t,{1}),h_-(t,{1}))\in
{H}_r$. One can fix the gauge freedom for $A_{\bf t}$ at the boundary
imposing $A_{\bf t}=0$. Then, $h_\pm$ are constant at every connected
component of the boundary of $\Sigma$.  Therefore, the reduced phase
space ${\bar{\cal P}}(G;H,H)$ covers the set of pairs
$((h_+,h_-),{\hat g})$ where ${\hat g}\in H$ and $(h_+,h_-)\in H_r$
and $(h_+{\hat g}, h_-{\hat g})\in H_rH_d\cap H_dH_r$.

%The equation of motion (\ref{eomGa}) can be written as
%$$\dd g g^{-1} + 1\over2P^{r\sharp}A=0$$

%Taking into account the boundary condition (\ref{BCA}) and that ${\bar
%G}$ is Poisson-Dirac, we

%Recall (see \cite{CraFer}) that the symplectic leaves of a
%Poisson-Dirac submanifold $N\subset M$ are the connected components of
%the intersections of $N$ with the symplectic leaves of $M$.

For the Poisson-Lie sigma model with target $G_r$ the solutions of the
equations of motion for $(g_+,g_-)$ are
\begin{equation}
(g_+(\sigma),g_-(\sigma))={}^{h(\sigma)}({\hat g}_+,{\hat g}_-)\qquad
\end{equation}
with $(g_+(t,0),g_-(t,0)),(g_+(t,{1}),g_-(t,{1}))\in H_r$.

An analogous argument shows that ${\bar{\cal P}}(G_r;H_r,H_r)$ is (a
covering of) the set of pairs $(h,({\hat g}_+,{\hat g}_-))$ where
$({\hat g}_+,{\hat g}_-)\in H_r$, $h\in H$ and $(h{\hat g}_+,h{\hat
g}_-)$ belongs to $H_dH_r\cap H_rH_d$.

The duality then exchanges degrees of freedom at the boundary with
degrees of freedom in the bulk, exactly in the same way as it does for
the free boundary conditions.

Notice that the duality described so far exists only for very special
branes given by $r$-invariant subalgebras. If one considers more
general situations the result is not that clean and one has, in the dual model,
non-local boundary conditions that relate the fields at both connected
components of $\partial\Sigma$.  A more comprehensive treatment of
this case will be done elsewhere.

\newpage

\pagestyle{empty}
\chapter{Supersymmetric WZ-Poisson sigma model} \label{ch:SusyWZP}
\thispagestyle{empty}
\pagestyle{empty}
\cleardoublepage

\pagestyle{headings}
\renewcommand*{\chaptermark}[1]{\markboth{\small \scshape  \thechapter.\ #1}{}}
\renewcommand*{\sectionmark}[1]{\markright{\small \scshape \thesection. \ #1}}

Generalized complex structures (\cite{Hit},\cite{Gua}) on a manifold
$M$ are objects defined on $TM\oplus T^*M$ which unify complex and
symplectic geometry and interpolate between them. Recently, there has
been much activity related to field theory realizations of generalized
complex structures. Recall that extended supersymmetry in
two-dimensional sigma models is related to complex geometry
(\cite{GatHulRoc}) in the target. On the other hand, in the
first-order formulation the fields of the models take values in
$TM\oplus T^*M$. This led to investigate the conditions for the
existence of extended supersymmetry in first-order actions. The task
was started in \cite{Lin} and systematically carried out in
\cite{LinMinTomZab}, where it was shown that extended supersymmetry is
closely related to generalized complex geometry in the target.

Later, in \cite{Zab}, it was shown that the Hamiltonian formalism
allows to rederive and extend the results of \cite{LinMinTomZab} in a
transparent fashion. In particular, it is proved in a
model-independent way that if the generators of a second supersymmetry
(denoted by ${\bf Q}_2(\epsilon)$) are to satisfy the supersymmetry
algebra, the target must be a (twisted, if a WZ term is present)
generalized complex manifold. Choosing a concrete model (i.e. a
concrete Hamiltonian) and imposing that ${\bf Q}_2(\epsilon)$ be a
symmetry should yield some compatibility conditions relating the
geometrical objects defining the model and the generalized complex
structure ${\cal J}$. In the general case (\cite{LinMinTomZab}) these
conditions include a number of algebraic as well as differential
equations.

Following the approach of \cite{Zab} we work out these compatibility
conditions for the case of vanishing metric in the first-order sigma
model, including a WZ term. When the closed 3-form $H$ defining this
twisting term vanishes, it is the supersymmetric version of the
PSM. For $H\neq 0$ it is the supersymmetric version of the WZ-Poisson
sigma model (\cite{KliStr}), whose target is equipped with a twisted
Poisson structure $\Pi$. We show that the notion of contravariant
connections on twisted Poisson manifolds is the key for unraveling the
differential compatibility conditions between $\Pi$ and ${\cal J}$. We
prove that, remarkably, the differential compatibility conditions are
implied by the algebraic ones, for which we also give a simple
geometrical interpretation.

\section{Twisted Poisson manifolds}\label{sec:twistedPoissonmanifolds}

The effect of closed 3-form fields on Poisson manifolds has recently
received attention in String Theory (see \cite{Par},
\cite{CorSch}). The presence of such closed 3-form leads to a
modification of the Jacobi identity and to the notion of twisted
Poisson manifolds, introduced in \cite{KliStr}. Following
\cite{SevWei}, in this section we introduce these concepts and show
that twisted Poisson structures fit naturally in the framework of
Courant algebroids and Dirac structures.

\vskip 5mm

Let $M$ be an $m$-dimensional manifold. A {\it twisted Poisson
structure} or {\it $H$-Poisson structure} on $M$ is defined by a
bivector field $\Pi$ and a closed 3-form $H$ which in local
coordinates $(x^1,\dots,x^m)$ satisfy a modified Jacobi identity:
\begin{equation} \label{Jacobimodified}
\Pi^{{i}{l}}\partial_{{l}}\Pi^{{j}{k}}+
\Pi^{{j}{l}}\partial_{{l}}\Pi^{{k}{i}}+
\Pi^{{k}{l}}\partial_{{l}}\Pi^{{i}{j}}=
\Pi^{{i}l}\Pi^{{j}n}\Pi^{{k}r}H_{lnr}.
\end{equation}

In the untwisted case, $H=0$, $\Pi$ is a Poisson structure and
(\ref{Jacobimodified}) reduces to the Jacobi identity.

In general, the bracket of two functions $f,g\in C^\infty(M)$,
\begin{equation}\notag
\{f,g\}(p)=\iota(\Pi_p)(\dd f\wedge \dd g)_p,\ p\in M
\end{equation}
is not a Lie bracket anymore. Instead, we have
\begin{equation}\notag
\{\{f,g\},h\}+\{\{g,h\},f\}+\{\{h,f\},g\}-\iota(X_f)\iota(X_g)\iota(X_h)H
\end{equation}
where, for any function $f\in C^\infty(M)$, $X_f=\Pi^\sharp\dd f$.

\vskip 5mm

Recall from Section \ref{sec:Diracstructures} that Poisson structures
are particular examples of the much more general and powerful notion
of Dirac structures. Nicely, the Courant bracket on $TM\oplus T^*M$
can be twisted by a closed 3-form in such a way that the twisted
version of Dirac structures can be consistently defined and twisted
Poisson structures fit in this picture.

 The $H$-{\it twisted Courant bracket} on $TM\oplus T^*M$ is defined
by
\begin{eqnarray}\notag
%\label{Courant}
&&{\hskip -5mm}[({X}_1,\xi_1),({X}_2,\xi_2)]_H=\cr
&&{\hskip
  -5mm}([{X}_1,{X}_2],\iota(X_1)\dd\xi_2-\iota(X_2)\dd\xi_1+{\frac{1}{2}}\dd
(\iota(X_1)\xi_2-\iota(X_2)\xi_1)+\iota(X_1)\iota(X_2)H)
\end{eqnarray}
for $({X}_1,\xi_1),({X}_2,\xi_2)\in \Gamma(TM\oplus T^*M)$.

\vskip 0.4 cm

An $H$-{\it Dirac structure} in $TM\oplus T^*M$ is defined analogously
to the untwisted case of Section \ref{sec:Diracstructures}. It is a
maximally isotropic subbundle of $TM\oplus T^*M$ with respect to the
symmetric pairing
\begin{equation}\label{pairing}
\langle {X}_1+\xi_1,{X}_2+\xi_2 \rangle = \xi_1({X}_2)+\xi_2({X}_1)
\end{equation}
and whose sections are closed under the $H$-twisted Courant bracket.

As expected, the graph of a bivector field $\Pi$,
\begin{equation}\label{Diracstructure}
L_\Pi = \{(\Pi^\sharp\xi,\xi)\in TM\oplus T^*M \vert \xi\in T^*M\}
\end{equation}
is an $H$-Dirac structure if and only if $\Pi$ is an $H$-Poisson
structure. The identification of $L_\Pi$ with $T^*M$ by projection to
the second component makes $T^*M$ into a Lie algebroid with anchor
$\Pi^\sharp$ and Lie bracket defined on exact sections by
\begin{equation}\notag
[\dd f,\dd g]=\dd\{f,g\}+\iota(X_f)\iota(X_g)H.
\end{equation}

The action of two-forms on $TM\oplus T^*M$ will play a very important
role in twisted Poisson geometry and twisted generalized complex
geometry. Let $b$ be a 2-form on $M$ and define its action on
$TM\oplus T^*M$ as
\begin{equation} \label{Btransform}
e^b({X},{\xi}):= (X,\xi + \iota(X)b)
\end{equation}
which is called a {\it $b$-transform}. The interesting point is that
this action is a morphism of Courant algebroids. Concretely,
\begin{equation}\notag
e^b[(X_1,\xi_1),(X_2,\xi_2)]_H=[e^b(X_1,\xi_1),e^b(X_2,\xi_2)]_{H+\dd
b}.
\end{equation}

Under a $b$-transform (\ref{Btransform}), the $H$-Dirac structure
$L_\Pi$ is transformed into $e^b(L_\Pi)$, which is an $(H+\dd
b)$-Dirac structure. However, the subbundle $e^b(L_\Pi)$ is the graph
of a bivector field if and only if $1_m+b\Pi$ is invertible. In this
case, $e^b(L_\Pi)=L_{\Pi_b}$, where
\begin{equation}\notag
{\Pi_b}:=\Pi(1_m+b\Pi)^{-1}
\end{equation}
is an $(H+\dd b)$-Poisson structure. In particular, if $H=\dd B$, a
$b$-transform with $b=-B$ untwists the twisted Poisson structure, so
that $\Pi_b$ is an ordinary Poisson structure (provided that
$1_m-b\Pi$ is invertible, of course).

If $b$ is closed the transformation (\ref{Btransform}) gives an
orthogonal automorphism of the twisted Courant bracket. The most
general orthogonal automorphism of the twisted Courant bracket
(\cite{Gua}) is a semidirect product of a diffeomorphism of $M$ and a
$b$-transform with $b\in\Omega_{{\rm closed}}^2(M)$.

\subsection{Contravariant connections}\label{sec:contravconnections}

Assume we want to endow the twisted Poisson manifold $(M,\Pi,H)$ with
a covariant connection. It is natural to demand this connection to be
compatible with $\Pi$, so that the covariant derivative of $\Pi$
vanishes. Already in the untwisted case one can prove (\cite{Vai94})
that a compatible covariant connection exists if and only if the rank
of the Poisson tensor is constant. Much of the interest of Poisson and
twisted Poisson structures resides in the fact that the rank of $\Pi$
can be non-constant, differing in an essential way from symplectic and
twisted symplectic manifolds. Therefore, the notion of covariant
derivative is not appropriate for twisted Poisson manifolds. The
relevant concepts in Poisson manifolds are those of contravariant
derivatives introduced by Vaisman (\cite{Vai94}) and contravariant
connections developed by Fernandes (\cite{Fer}). Next, we extend their
definitions and some results to the case of twisted Poisson manifolds.

A {\it contravariant derivative} on a vector bundle $E$ over the
twisted Poisson manifold $(M,\Pi,H)$ is an operator $\nabla$ such that
for each $\alpha \in \Omega^1(M)$, $\nabla_\alpha$ maps sections of
$E$ to sections of $E$ satisfying:
\newpage
\begin{enumerate}
\item[(i)]$\nabla_{\alpha_1 + \alpha_2}\psi
=\nabla_{\alpha_1}\psi+\nabla_{\alpha_2}\psi$
\vspace{-0.25cm}
\item[(ii)]$\nabla_{\alpha}(\psi_1 + \psi_2)=\nabla_{\alpha}\psi_1 +
\nabla_{\alpha}\psi_2$
\vspace{-0.25cm}
\item[(iii)]$\nabla_{f\alpha}\psi=f\nabla_\alpha\psi$
\vspace{-0.25cm}
\item[(iv)]$\nabla_\alpha(f\psi)=f\nabla_\alpha\psi+
\Pi^\sharp\alpha(f)\psi$.
\end{enumerate}

A contravariant derivative defines a contravariant connection in an
analogous way to the covariant case. We shall be interested in
defining the contravariant derivative of tensor fields on $M$. To this
end it is enough to have a contravariant derivative on
$E=T^*M$. Take local coordinates $x^i$ on $M$ and define the
Christoffel symbols $\Gamma^{ij}_k$ by
\begin{equation}\notag
\nabla_{\dd x^i}\dd x^j = \Gamma^{ij}_k \dd x^k.
\end{equation}

The contravariant derivative of a tensor field $K$ of type $(p,q)$ is given by
\begin{eqnarray}\notag
\nabla^{n}K^{{i}_1\dots{i}_p}_{{j}_1\dots{j}_q}&=&\Pi^{nl}\partial_l
K^{{i}_1\dots{i}_p}_{{j}_1\dots{j}_q}-\sum_{{\mu}=1}^{p}\Gamma^{n{i}_{\mu}}_l
K^{{i}_1\dots l\dots{i}_p}_{{j}_1\dots{j}_q}+\cr
&+&\sum_{{\nu}=1}^{q}\Gamma^{nl}_{{j}_{\nu}}K^{{i}_1\dots{i}_p}_{{j}_1\dots l\dots{j}_q}.
\end{eqnarray}

A tensor field $K$ is called {\it parallel} if $\nabla K = 0$. The
relevant result for us is given by the following

{\bf Theorem:} {\it Let $(M,\Pi,H)$ be a twisted Poisson
manifold. Then, there exists a contravariant connection such that
$\Pi$ is parallel.}

\vskip 0.3 cm

{\it Proof:} Let $\{{\cal U}_n\}$ be an open cover of $M$. Take local
coordinates $x^i$ on ${\cal U}_n$ and define
\begin{equation} \label{localsymbols}
\Gamma^{ij}_{(n)k}=\partial_k\Pi^{ij}-\frac{1}{2}\Pi^{il}\Pi^{jm}H_{klm}.
\end{equation}
On ${\cal U}_n$, $\Pi$ is parallel for the contravariant connection
$\nabla_{(n)}$ with symbols $\Gamma_{(n)}$. If $\sum_n f_n=1$ is a
partition of unity subordinated to the cover $\{{\cal U}_n\}$, $\nabla=\sum_n
f_n\nabla_{(n)}$ gives a contravariant connection on $M$ such that
$\nabla\Pi=0$.  $\,$\hfill$\Box$\break

\section{Twisted generalized complex structures}

If $x^i$ are local coordinates on $M$ and take $(\partial_i,\dd
x^i)$ as a basis in the fibers of $TM\oplus T^*M$, the bilinear form
(\ref{pairing}) reads
\begin{equation}\notag
%\label{metric}
{\cal I}=\begin{pmatrix}0&1_m\cr 1_m&0\end{pmatrix}.
\end{equation}

\vskip 0.2cm

An {\it almost generalized complex structure} on the manifold $M$ is a
linear map ${\cal J}:TM\oplus{T^*M}\rightarrow TM\oplus{T^*M}$ such
that ${\cal J}^t{\cal I}{\cal J}={\cal I}$ and ${\cal J}^2 =
-1_{2m}$. For such ${\cal J}$ define ${\cal
J}_{\pm}=\frac{1}{2}(1_{2m}\pm i{\cal J})$. The almost generalized
complex structure ${\cal J}$ is an $H$-{\it twisted generalized
complex structure} if
\begin{equation}\label{integrability}
{\cal J}_\mp[{\cal J}_\pm({X}_1+\xi_1),{\cal J}_\pm({X}_2+\xi_2)]_H=0,\ 
\end{equation}
$\forall {X}_1+\xi_1,{X}_2+\xi_2\in \Gamma(TM\oplus{T^*M})$. It follows that the $b$-transform of an $H$-twisted generalized complex
structure ${\cal J}$,
\begin{equation}\notag
{\cal J}_b:=e^b{\cal J}e^{-b}
\end{equation}
is an $(H+\dd b)$-twisted generalized complex structure.

\vskip 0.4 cm

In the coordinate basis $(\partial_i,\dd x^i)$ we can write,
\begin{equation}\notag
{\cal J}=\begin{pmatrix}J&P\cr L&K \end{pmatrix}
\end{equation}
with $J:TM\rightarrow TM,P:T^*M\rightarrow TM,L:TM\rightarrow
T^*M,K:T^*M\rightarrow T^*M$.

The condition ${\cal J}^t{\cal I}{\cal J}={\cal I}$ becomes
\begin{equation}\notag
J^i_{\,\,\,j} + K_{i}^{\,\,\,j} =0,\,\,\,\,\,\,\,\,\,\,\,
 P^{ij} = - P^{ji},\,\,\,\,\,\,\,\,\,\,\,
 L_{ij}=-L_{ji}
\end{equation}
whereas ${\cal J}^2 = -1_{2m}$ translates into
\begin{subequations}\notag
\begin{align}
&J^{i}_{\,\,\,{k}} J^{k}_{\,\,\,{j}} + P^{{i}{k}} L_{{k}{j}} =
-\delta^{i}_{\,\,\,{j}}\\ &J^{i}_{\,\,\,{k}}
P^{{k}{j}} + P^{{i}{k}}K_{k}^{\,\,\,{j}}
=0\\ &K_{i}^{\,\,\,{k}} K_{k}^{\,\,\,{j}} +
L_{{i}{k}} P^{{k}{j}} = -\delta^{i}_{\,\,\,{j}}\\
&K_{{i}}^{\,\,\,{k}} L_{{k}{j}} + L_{{i}{k}} J^{k}_{\,\,\,{j}}
=0.
\end{align}
\end{subequations}

The integrability condition (\ref{integrability}) is equivalent to the
following differential equations:
\begin{eqnarray}\notag
&& J^{j}_{\,\,\,[{l}} J^{i}_{\,\,\,{k}],{j}} + J^{i}_{\,\,\,{j}}
 J^{j}_{\,\,\,[{l},{k}]} + P^{{i}{j}}(L_{[{l}{k},{j}]} +
 J^{n}_{\,\,\,[{l}}H_{{k}]{n}{j}} )=0 \cr\cr
&&P^{[{i}|{j}} P^{|{l}{k}]}_{\,\,\,\,\,\,,{j}} = 0\cr\cr
&& J^{i}_{\,\,\,{j},{k}} P^{{k}{l}} +
 P^{{k}{l}}_{\,\,\,\,\,\,,{j}} J^{i}_{\,\,\,{k}} -
 J^{{l}}_{\,\,\,{k},{j}} P^{{i}{k}} +
 J^{{l}}_{\,\,\,{j},{k}} P^{{i}{k}} -
 P^{{i}{l}}_{\,\,\,\,\,\,,{k}}
 J^{k}_{\,\,\,{j}}-P^{{l}{n}} P^{{i}{k}} H_{{n}{k}{j}}=0
 \cr\cr
&& J^{l}_{\,\,\,{j}} L_{[{l}{k},{r}]} + L_{{j}{l}}
 J^{l}_{\,\,\,[{r},{k}]} +J^{l}_{\,\,\,{k}}
 L_{{r}{j},{l}} +J^{l}_{\,\,\,{r}} L_{{j}{k},{l}} +
 L_{{l}{k}} J^{l}_{\,\,\,{r},{j}} + J^{l}_{\,\,\,{k}}
 L_{{l}{r},{j}} +\cr\cr
&& + H_{{k}{r}{j}} - J^{l}_{\,\,\,[{k}}
 J^{n}_{\,\,\,{r}} H_{{j}]{l}{n}}=0
\end{eqnarray}
where a comma denotes a partial derivative and the brackets stand for
antisymmetrization. Notice that $P^{ij}$ is always a Poisson
structure.

\section{WZ-Poisson sigma model}

In this section we introduce the the twisted Poisson sigma model or
WZ-Poisson sigma model (WZ-PSM) in the Lagrangian formalism and see
how twisted Poisson geometry arises.

Let $(M,\Pi,H)$ be a twisted Poisson manifold. We assume for
simplicity that $H=\dd B$ (which is always locally true). The action
of the WZ-PSM is (\cite{KliStr}):
\begin{equation}\label{eq:actionWZPSM}
S=\int_\Sigma \eta_i\wedge\dd
X^i+\frac{1}{2}\Pi^{ij}(X)\eta_i\wedge\eta_j - \frac{1}{2}B_{ij}\dd
X^i\wedge\dd X^j
\end{equation}
which is obtained from the action of the PSM (\ref{PScoor}) by adding
the WZ term $-\frac{1}{2}\int_\Sigma \dd^{-1}H$. The equations of
motion in the bulk are:
\begin{subequations}\label{eq:eomWZPSM}
\begin{align}
&\dd X^i+\Pi^{ij}\eta_j=0\label{eq:eomWZPSMa}\\
&\dd\eta_i+\frac{1}{2}\partial_i\Pi^{jk}\eta_j\wedge\eta_k-
\frac{1}{2}H_{ijk}\dd X^j\wedge\dd X^k =0\label{eq:eomWZPSMb}
\end{align}
\end{subequations}
with
$H_{ijk}=\partial_iB_{jk}+\partial_jB_{ki}+\partial_kB_{ij}$. Notice
that (\ref{eq:eomWZPSMa}) does not get any contribution from $B$
whereas the additional term in (\ref{eq:eomWZPSMb}) only depends on
$H$. In fact, if $\partial\Sigma=\emptyset$, the model depends on $H$
and not on $B$. Analogously to the case of the ordinary PSM, the
consistency of the equations of motion (\ref{eq:eomWZPSM}) require
that $\Pi$ and $H$ satisfy (\ref{Jacobimodified}), so that $\Pi$ is an
$H$-Poisson structure.

The action (\ref{eq:actionWZPSM}) is invariant (up to boundary terms)
under the local transformations:
\begin{subequations}\notag
\begin{align}
&\delta_\epsilon X^i=\Pi^{ij}(X)\epsilon_j\\
&\delta_\epsilon\eta_i=-\dd\epsilon_i-\partial_i\Pi^{jk}(X)\eta_j\epsilon_k
+H_{ijl}\Pi^{jm}\Pi^{lk}\eta_m\epsilon_k
\end{align}
\end{subequations}
with $\epsilon(\sigma)=\epsilon_i(\sigma)\dd X^i$ the gauge parameter.

The treatment of twisted Poisson structures in terms of Dirac
structures given in Section \ref{sec:twistedPoissonmanifolds} shows
that $b$-transforms act on $\Pi$ connecting it with other twisted
Poisson structures. In particular, if $H=\dd B$ and $1 -B\Pi$ is
invertible, the transformed object ${\tilde
\Pi}:=\Pi_{-B}=\Pi(1-B\Pi)^{-1}$ is an ordinary Poisson structure. It
is then natural to ask whether, in this case, there exists a field
redefinition transforming (\ref{eq:actionWZPSM}) into an untwisted PSM
with Poisson structure $\tilde\Pi$. On-shell this is actually
possible, as we proceed to show.

Take $\tilde\eta =(1-B\Pi)\eta$. The action (\ref{eq:actionWZPSM})
written as a functional of $X$ and $\tilde\eta$ reads:
\begin{equation}\notag
S=\int_\Sigma \tilde\eta_i\wedge\dd
X^i+\frac{1}{2}\tilde\Pi^{ij}(X)\tilde\eta_i\wedge\tilde\eta_j-
\frac{1}{2}B_{ij}(\dd X^i+\tilde\Pi^{ik}\tilde\eta_k)\wedge (\dd
X^j+\tilde\Pi^{jl}\tilde\eta_l)
\end{equation}
which differs from the PSM with Poisson structure $\tilde\Pi$ by terms
quadratic in the equations of motion and therefore it is classically
equivalent to it.

Consequently, if $H=\dd B$ and $1-B\Pi$ is invertible the WZ-PSM is
equivalent to an ordinary PSM. However, if $H$ is not exact or
$1-B\Pi$ is degenerate some mathematical subtleties appear,
particularly when $\Sigma$ has non-empty boundary. These issues are
still to be tackled and will not be pursued further in this
dissertation.

\section{Supersymmetric WZ-Poisson sigma model}

Let $S^{1,1}$ be the supercircle with coordinates
${\bar\sigma}=(\sigma,\theta)$. In local coordinates the cotangent
bundle of the superloop space, $T^*{\bf L}M$, is given by scalar
superfields $\Phi^i(\sigma,\theta)$ and spinorial superfields
${\Psi}_i(\sigma,\theta)$. In components,
\begin{equation}\notag
\Phi^{i}(\sigma) = X^{i}(\sigma)+\theta\lambda^{i}(\sigma),\quad
{\Psi}_{i}(\sigma)=\rho_{i}(\sigma)+\theta \eta_{i}(\sigma)
\end{equation}
where $X^{i}$ and $\eta_{i}$ are bosonic fields.

Assume that $M$ is equipped with a closed 3-form $H$. Then, the
 following 2-form on $T^*{\bf L}M$ is symplectic:
\begin{equation} \label{symplecticform}
\omega=\int_{S^{1,1}}\dd\sigma\dd\theta\left(
\delta\Phi^{i}\wedge\delta {\Psi}_{i}
-\frac{1}{2}H_{{i}{j}{k}}D\Phi^{i}\delta\Phi^{j}\wedge\delta\Phi^{k}
\right)
\end{equation}
where $\delta$ stands for the de Rham differential and
$D=\partial_\theta - \theta\partial_\sigma$. Since $\omega$ is closed
and non-degenerate, it defines a Poisson bracket on functions of the
superfields which we shall denote by $\{\cdot ,\cdot \}$.

The basic Poisson brackets read:
\begin{eqnarray}\notag
&&\{\Phi^{i}({\bar\sigma}),\Phi^{j}({\bar\sigma}')\}=0\cr
&&\{\Phi^{i}({\bar\sigma}),{\Psi}_{j}({\bar\sigma}')\}=\delta^{i}_{j}
\delta({\bar\sigma}-{\bar\sigma}')\cr
&&\{{\Psi}_{i}({\bar\sigma}),{\Psi}_{j}({\bar\sigma}')\}=H_{{i}{j}{k}}D\Phi^{k}\delta({\bar\sigma}-{\bar\sigma}')
\end{eqnarray}
with $\delta({\bar\sigma}-{\bar\sigma}')$ the superspace delta
distribution. Notice that
\begin{equation} \label{canonicaltransformation}
{\Psi}_i\mapsto {\Psi}_i - B_{ij}D\Phi^j
\end{equation}
is a canonical transformation for closed $B$.

The Hamiltonian formulation of the $N=1$ supersymmetric WZ-PSM is as
follows. The phase space of the theory, denoted by ${\cal P}$, is the
set of points of $T^*{\bf L}M$ satisfying the constraints:
\begin{equation}\notag
%\label{constraintWZ}
D\Phi^i(\sigma,\theta) + \Pi^{ij}(\Phi){\Psi}_j(\sigma,\theta)
= 0,\ i = 1,\dots,m.
\end{equation}
which are a consequence of the singular nature of the Lagrangian of
the WZ-PSM (\cite{KliStr}). The Hamiltonian of the WZ-PSM can be
written:
\begin{equation}\notag
%\label{Hamiltonian}
{\cal H} =
\int_{S^{1,1}}F_i(\sigma,\theta)\left(D\Phi^i(\sigma,\theta) +
\Pi^{ij}(\Phi){\Psi}_j(\sigma,\theta) \right) \dd\sigma\dd\theta
\end{equation}
where the fields $F_i$ act as Lagrange multipliers. As we already
know, the consistency of the model requires $\Pi$ to be an $H$-Poisson
structure. In the Hamiltonian formalism this is obtained as the
condition for the dynamics to preserve the submanifold ${\cal P}$,
i.e. $\{D\Phi^i + \Pi^{ij}{\Psi}_j,{\cal H}\}\vert_{\cal P}=0$.

By construction, the WZ-PSM is invariant under the supersymmetry
transformation generated by:
\begin{equation}\notag
{\bf Q}_1(\epsilon)=\int_{S^1}\dd\sigma\epsilon \left(
\eta_{i}\lambda^{i}-\rho_{i}\partial X^{i}
-\frac{1}{3}H_{{i}{j}{k}}\lambda^{i}\lambda^{j}\lambda^{k} \right)
\end{equation}
where $\epsilon$ is a constant anticommuting parameter and $\partial
X^i := \frac{\partial X^i}{\partial\sigma}$.

We address now the issue of extended supersymmetry in the WZ-PSM. The
most general ansatz for a second supersymmetry transformation is given
by (\cite{Zab})
\begin{equation} \label{Q2}
{\bf Q}_2(\epsilon) = -\frac{1}{2}\int_{S^{1,1}}\dd\sigma\dd\theta \epsilon
(2D\Phi^{i} {\Psi}_{j} J^{j}_{\,\,\,{i}}+D\Phi^{i} D\Phi^{j}
L_{{i}{j}}+{\Psi}_{i} {\Psi}_{j} P^{{i}{j}}).
\end{equation}

As shown in \cite{Zab} in a model-independent way, the generators
${\bf Q}_2(\epsilon)$ satisfy the supersymmetry algebra if and only if
\begin{equation}\notag
{\cal J}=\begin{pmatrix}J&P\cr L&-J^t \end{pmatrix}
\end{equation}
is an $H$-twisted generalized complex structure. In this context, the
canonical transformation (\ref{canonicaltransformation}) is identified
with a $b$-transform for closed $B$, i.e. an automorphism of the
Courant algebroid. If $\dd B\neq 0$, ${\Psi}_i\mapsto {\Psi}_i - B_{ij}D\Phi^j$
is not a canonical transformation and it changes the twisting term in
(\ref{symplecticform}). In particular, if $H=\dd B$ it untwists the
symplectic structure.

It remains to find out when ${\bf Q}_2(\epsilon)$ generates a symmetry
transformation of the WZ-PSM. That is, when
\begin{equation} \label{symmetryWZ}
\{{\bf Q}_2(\epsilon),{\cal H}\}\vert_{\cal P}=0.
\end{equation}

At this stage one expects that (\ref{symmetryWZ}) holds only if some
compatibility conditions relating ${\cal J}$ and $\Pi$ are
satisfied. These conditions were worked out in the Lagrangian
formalism for a general (untwisted) first-order sigma model in
\cite{LinMinTomZab}. It turns out that in the general case some
algebraic as well as differential conditions must be imposed. Our aim
is to prove that in the WZ-PSM the differential conditions are
automatically implied by the algebraic ones. We shall see that the
contravariant connections introduced in Section
\ref{sec:contravconnections} are extremely helpful in the derivation
of this result.

A direct calculation shows that (\ref{symmetryWZ}) holds if and only if
the following two conditions are met:

{\it Algebraic condition:}
\begin{equation} \label{algebraic}
P^{{i}{j}}+J^{i}_{\,\,\,{k}}\Pi^{{k}{j}}+\Pi^{{i}{k}}J^{j}_{\,\,\,{k}}-
\Pi^{{i}{k}}L_{{k}{l}}\Pi^{{l}{j}}=0.
\end{equation}

{\it Differential condition:}
\begin{eqnarray} \label{differential}
&&\frac{1}{2}\Big[\Pi^{{i}{l}}\partial_{l}
  P^{{j}{k}}-\partial_{l}
  \Pi^{{i}{j}}P^{{l}{k}}-\partial_{l}\Pi^{{i}{k}}P^{{j}{l}}+\cr
  &&+(\Pi^{{i}{l}}\partial_{l}
  J^{j}_{\,\,\,{n}}-\partial_{l}\Pi^{{i}{j}}J^{l}_{\,\,\,{n}}+\partial_{n}\Pi^{{i}{l}}J^{j}_{\,\,\,{l}})\Pi^{{n}{k}}+\cr
  &&+\Pi^{{j}{n}}(\Pi^{{i}{l}}\partial_{l}
  J^{k}_{\,\,\,{n}}-\partial_{l}\Pi^{{i}{k}}J^{l}_{\,\,\,{n}}+\partial_{n}\Pi^{{i}{l}}J^{k}_{\,\,\,{l}})-\cr
  &&-\Pi^{{j}{n}}(\Pi^{{i}{l}}\partial_{l}
  L_{{n}{r}}+\partial_{n}\Pi^{{i}{l}}L_{{l}{r}}+\partial_{r}\Pi^{{i}{l}}L_{{n}{l}})\Pi^{{r}{k}}\Big]+\cr
  &&+(P^{{r}{j}}+J^{r}_{\,\,\,\tau}\Pi^{\tau{j}})\Pi^{{i}{n}}\Pi^{{l}{k}}H_{{l}{r}{n}}=0.
\end{eqnarray}

The untwisted version of the differential condition
(\ref{differential}) was deduced in \cite{LinMinTomZab} (see equation
(6.35) therein) for a general untwisted sigma model. In the case of
non-vanishing metric it was interpreted as a condition of constancy
with respect to a covariant derivative compatible with the metric
tensor. The WZ-PSM is the limit of vanishing metric of the twisted
first-order sigma model and there is no such covariant derivative at
hand. We have learnt in Section \ref{sec:contravconnections} that the
natural objects which allow to compare tangent spaces at different
points of a twisted Poisson manifold are contravariant connections. In
fact, after some manipulations we can rewrite condition
(\ref{differential}) in terms of the local symbols
(\ref{localsymbols})\footnote{We omit the subscript $(n)$ referring to
an open set of a cover of $M$.}:
\begin{eqnarray}\notag
&&\nabla^{i} P^{{j}{k}}+(\nabla^{i}
J^{j}_{\,\,\,{n}})\Pi^{{n}{k}}+\Pi^{{j}{n}}(\nabla^{i}
J^{k}_{\,\,\,{n}})-\Pi^{{j}{n}}(\nabla^{i}
L_{{n}{r}})\Pi^{{r}{k}}+\cr
&&+(P^{{r}{j}}+J^{r}_{\,\,\,{p}}\Pi^{{p}{j}}-J^{j}_{\,\,\,{p}}\Pi^{{p}{r}}-\Pi^{{r}\delta}L_{\delta{p}}\Pi^{{p}{j}})\Pi^{{i}{n}}\Pi^{{l}{k}}H_{{l}{r}{n}}=0.
\end{eqnarray}

Using that $\nabla\Pi=0$ we can go one step further and write:
\begin{eqnarray}\notag
&&\nabla^{i} \left( P^{{j}{k}}+
J^{j}_{\,\,\,{n}}\Pi^{{n}{k}}+\Pi^{{j}{n}}J^{k}_{\,\,\,{n}}
-\Pi^{{j}{n}}L_{{n}{r}}\Pi^{{r}{k}} \right)+\cr
&&+(P^{{r}{j}}+J^{r}_{\,\,\,{p}}\Pi^{{p}{j}}-J^{j}_{\,\,\,{p}}\Pi^{{p}{r}}-\Pi^{{r}s}L_{s{p}}\Pi^{{p}{j}})\Pi^{{i}{n}}\Pi^{{l}{k}}H_{{l}{r}{n}}=0
\end{eqnarray}
which is obviously satisfied if the algebraic condition
(\ref{algebraic}) holds. Hence, we obtain the remarkable result that
the WZ-PSM has a second supersymmetry (\ref{Q2}) if and only if ${\cal
J}$ is a generalized complex structure and the algebraic condition
(\ref{algebraic}) is satisfied. This gives an enormous simplification
with respect to the general sigma model, in which the differential
condition analogous to (\ref{differential}) is not necessarily implied
by the algebraic ones.

Now, we would like to give a geometrical interpretation of
(\ref{algebraic})\footnote{The algebraic condition (\ref{algebraic})
was already deduced in \cite{Ber}, but a geometrical interpretation
was lacking.}, which imposes a compatibility condition between the
$H$-Poisson structure $\Pi$ and the Poisson structure $P$. Define two
endomorphisms of $TM\oplus T^*M$ by
\begin{equation}\notag
\tau_1:=\begin{pmatrix}0 & \Pi \cr 0 & 1_m \end{pmatrix},\quad
\tau_2:= \begin{pmatrix}1_m & -\Pi \cr 0 & 0 \end{pmatrix}.
\end{equation}

Notice that $\tau_i^2=\tau_i,i=1,2$. $\tau_1$ projects onto the Dirac
structure $L_\Pi$ (see definition (\ref{Diracstructure})) along $TM$. $\tau_2$
projects onto $TM$ along $L_\Pi$. In particular, ${\rm
Im}(\tau_1)={\rm Ker}(\tau_2)=L_\Pi$, which will be the important
point for us.

In matrix notation the condition (\ref{algebraic}) can be expressed as
\begin{equation}\label{interpretation}
\tau_2 \begin{pmatrix}J & P \cr L & -J^t \end{pmatrix} \tau_1 = 0
\end{equation}
which says that ${\cal J}$ can be restricted to an endomorphism of
$L_\Pi$. That is, the WZ-PSM supports extended supersymmetry if and
only if
\begin{equation}\notag
{\cal J}(L_\Pi)\subset L_\Pi.
\end{equation}

Note that the canonical transformation (\ref{canonicaltransformation})
can be viewed as a $b$-transform with $b=-B\in\Omega^2_{{\rm
closed}}(M)$ acting on ${\cal J}$ and $\Pi$. Since a canonical
transformation does not modify the Poisson brackets, one would expect
that (\ref{algebraic}) hold for the transformed objects
\begin{equation}\notag
{\cal J}_{-B} = \begin{pmatrix}J_{-B} & P_{-B} \cr L_{-B} & -J_{-B}^t
\end{pmatrix},\quad \Pi_{-B}=\Pi(1_m-B\Pi)^{-1}.
\end{equation}
This is not evident from (\ref{algebraic}). However, the result
follows easily by using that the expression (\ref{interpretation}) is
manifestly invariant under a $b$-transform. But the argument is purely
algebraic so that the result holds even if $B$ is not closed.

The fact that in the WZ-PSM there is only one algebraic compatibility
condition between ${\cal J}$ and $\Pi$, given by (\ref{algebraic}),
should make easier the search of backgrounds admitting extended
supersymmetry. Notice that when trying to solve (\ref{algebraic}) one
can use that it is invariant under a $b$-transform. Therefore, if
$H=\dd B$ and $1_m-B\Pi $ is invertible one can look for solutions of
(\ref{algebraic}) in terms of the untwisted objects and twist back at
the end of the day.

\newpage

\pagestyle{empty}
\chapter*{Conclusions} \label{ch:Conclusions}
\addcontentsline{toc}{chapter}{Conclusions}
\thispagestyle{empty}
\pagestyle{empty}
\cleardoublepage

\pagestyle{plain}

We end this dissertation by collecting and summarizing our main
results:
\begin{itemize}
\item[--]We have presented a purely algebraic approach to the
reduction of Poisson manifolds. We describe the reduced Poisson
algebra and give sufficient conditions for the submanifold to inherit
a Poisson structure.
\item[--]We give a natural way of producing new Dirac subbundles from
a given one. The case in which the original Dirac subbundle is
actually a Dirac structure is especially interesting. We apply our
procedure to the reduction and projection of Dirac structures,
generalizing previous results in the literature. The situation in
which the original or the induced Dirac structure are Poisson is also
investigated.
\item[--]A detailed study of the classical PSM defined on a surface
with boundary has been performed. We have identified very general
sufficient conditions ({\it weak regularity}) under which a
submanifold is an admissible brane for the model.
\item[--]The Hamiltonian analysis of the PSM on the strip with two
arbitrary branes has been carried out. We have described how the phase
space encodes the Poisson geometry of the branes:
\begin{itemize}
\item[--]We showed that the presymplectic structure on the phase space
is related to the Poisson bracket induced on each brane through a dual
pair structure.
\item[--]When the brane is second-class, we have proven that there
exists a connection between the reduced phase space and the symplectic
groupoid integrating the brane.
\end{itemize}
\item[--]At the quantum level, we have shown that the PSM on the disk
with a brane satisfying the {\it strong regularity condition} (a
slightly stronger version of the previous one) can be perturbatively
quantized. When second-class constraints are present, the standard
Feynman expansion valid for coisotropic branes breaks down and a
redefinition of the perturbative series is needed. In the particular
case of a second-class brane we prove that the perturbative
quantization of the model is related to the Kontsevich formula for the
Dirac Poisson structure on the brane.
\item[--]We have tackled the study of the PSM when the target is a
Poisson-Lie group. We gave a thorough analysis for any Poisson-Lie
structure associated to a simple Lie group $G$ (its dual group being
denoted by $G^*$). Such structures are divided in two classes:
factorizable and triangular.
\begin{itemize}
\item[--]We provide explicit realizations of any factorizable or
triangular structure and solve the models over $G$ and $G^*$. We also
solve some related PSMs whose targets are Lie groups but not
Poisson-Lie groups.
\item[--]In the Hamiltonian formalism, in the open geometry with
free boundary conditions, we discovered a duality transformation
relating the reduced phase space of the models corresponding to a pair
of dual groups $G$ and $G^*$.
\item[--]Then, we identified a broad family of pairs of branes which
preserve the duality.
\item[--]We have shown that, if the Poisson-Lie structure is
triangular, the model over $G^*$ is equivalent to BF-theory.
\end{itemize}
\item[--]We have worked out the conditions for extended supersymmetry
in the twisted Poisson sigma model. They reduce to an algebraic
equation with a beautiful geometrical interpretation relating the
twisted Poisson structure and the twisted generalized complex
structure on the target.
\begin{itemize}
\item[--]In the derivation of this result the notion of contravariant
connection is very useful. We have proven that any twisted Poisson
manifold has a compatible contravariant connection, extending the
existing theorems on (standard) Poisson manifolds.
\end{itemize}
\end{itemize}

\newpage

\thispagestyle{empty}

\pagestyle{headings}
\renewcommand*{\chaptermark}[1]{\markboth{\small \scshape  \thechapter.\ #1}{}}
\renewcommand*{\sectionmark}[1]{\markright{\small \scshape \thesection. \ #1}}

%%\bibliography{bibliography_PhDthesis_arxiv}
%%\bibliographystyle{amsplain}

\end{document}